%% file: main-g3x.tex
\documentclass[%
 reprint,longbibliography,
showpacs,preprintnumbers,
bibnotes,
 amsmath,amssymb,
prb,
]{revtex4-2}

\usepackage{graphicx}
\usepackage[english]{babel}
\usepackage{microtype}       
\usepackage[utf8]{inputenc}
\usepackage[breaklinks=true,colorlinks=true,linkcolor=blue,urlcolor=blue,citecolor=blue]{hyperref}

\usepackage[dvipsnames]{xcolor}      
\usepackage{mathtools}
\mathtoolsset{showonlyrefs=true}
\usepackage{braket}
\usepackage{tikz}

 \usepackage{longtable}
  \usepackage{booktabs}
 \usepackage{siunitx}
 \usepackage{csquotes}
 
 \usepackage{multirow}
\usepackage{amssymb}

\newcommand{\nn}{\nonumber}

\newcommand{\refeqn}[1]{Eq.~\eqref{#1}}
\newcommand{\refeqns}[2]{Eqs.~\eqref{#1} and \eqref{#2}}

\newcommand{\reffigs}[2]{Figs. (\ref{#1}) and (\ref{#2})}

\newcommand{\eqrefs}[2]{Eqs.~\eqref{#1} and \eqref{#2}}

\newcommand{\eqrefrno}[3]{Eqs.~\eqref{#1} -- \eqref{#2}\noeqref{#3}}

\newcommand{\fref}[1]{Fig.~\ref{#1}}

\newcommand{\GKBA}{{}}

\newcommand{\lowerbossphantom}{\vphantom{\bar{\bar{x}}}}
\newcommand{\upperbossphantom}{\vphantom{\dagger}}
\newcommand{\tempop}[3][\textstyle]{\settowidth{\dimen1}{$#1\hat{#2}$}\makebox[\dimen1][l]{$#1\hat{#2\mspace{#3}}$}}
\newcommand{\xop}[1]{{\mathchoice{\tempop[\displaystyle]{#1}{3.5mu}}{\tempop{#1}{3.5mu}}{\tempop[\scriptstyle]{#1}{3.5mu}}{\tempop[\scriptscriptstyle]{#1}{3mu}}}}
\newcommand{\chat}[1]{\ensuremath{\xop{#1}}}
\newcommand{\aop}[2]{\ensuremath{\chat{c}_{#1#2\lowerbossphantom}^{\upperbossphantom}}}
\newcommand{\cop}[2]{\ensuremath{\chat{c}_{#1#2\lowerbossphantom}^{\dagger\upperbossphantom}}}
\newcommand{\cbar}[1]{\ensuremath{\xbar{#1}}}

\newcommand{\tempbar}[3][\textstyle]{\settowidth{\dimen1}{$#1\bar{#2}$}\makebox[\dimen1][l]{$#1\bar{#2\mspace{#3}}$}}
 
\newcommand{\xbar}[1]{{\mathchoice{\tempbar[\displaystyle]{#1}{3.5mu}}{\tempbar{#1}{3.5mu}}{\tempbar[\scriptstyle]{#1}{3.5mu}}{\tempbar[\scriptscriptstyle]{#1}{3mu}}}} 
\renewcommand{\i}{\mathrm{i}}
\renewcommand{\d}{\mathrm{d}}
\newcommand{\tn}[1]{\textnormal{#1}}

\usepackage{dcolumn}
\usepackage{bm}

\makeatletter
\g@addto@macro\bfseries{\boldmath}
\newcommand*{\balancecolsandclearpage}{%
   \close@column@grid
   \clearpage
   \twocolumngrid
 }
\makeatother

\begin{document}
\preprint{APS/123-QED}

\title{The dynamically screened ladder approximation: Simultaneous treatment of strong electronic correlations and dynamical screening out of equilibrium}

\author{Jan-Philip Joost, Niclas Schl\"unzen, Hannes Ohldag, and Michael Bonitz
 \email{bonitz@theo-physik.uni-kiel.de}}
\affiliation{
Institut f\"ur Theoretische Physik und Astrophysik, 
Christian-Albrechts-Universit\"{a}t zu Kiel, D-24098 Kiel, Germany \\ and
Kiel Nano, Surface and Interface Science KiNSIS, Kiel University, Germany
}
\author{Fabian Lackner and Iva B\v rezinov\' a}
\affiliation{Institute for Theoretical Physics, Vienna University of Technology, Wiedner Hauptstrasse 8-10/136, 1040 Vienna, Austria, EU}

\date{\today}

\begin{abstract}
Dynamical screening is a key property of charged many-particle systems. Its theoretical description is based on the $GW$ approximation that is extensively applied for ground-state and equilibrium situations but also for systems driven out of equilibrium. The main limitation of the $GW$ approximation is the neglect of strong electronic correlation effects that are important in many materials as well as in dense plasmas. Here we derive the dynamically screened ladder (DSL) approximation that selfconsistently includes, in addition to the $GW$ diagrams, also particle--particle and particle--hole $T$-matrix diagrams. The derivation is based on reduced-density-operator theory and the result is equivalent to the recently presented G1--G2 scheme [Schlünzen \textit{et al.}, Phys. Rev. Lett. \textbf{124}, 076601 (2020); Joost \textit{et al.}, Phys. Rev. B \textbf{101}, 245101 (2020)]. We perform extensive time-dependent DSL simulations for finite Hubbard clusters and present tests against exact results that confirm excellent accuracy as well as total energy conservation of the approximation.
At strong coupling and for long simulation durations, instabilities are observed. These problems are solved by enforcing contraction consistency and applying a purification approach.
\end{abstract}

\maketitle

\section{Introduction}\label{s:intro}
The ultrafast dynamics of many-particle systems following a rapid excitation are of high interest in many fields, including dense plasmas, correlated electrons in solids, femtosecond laser pulse excited atoms and molecules, or fermionic atoms in optical lattices. Among the key properties of these systems---most importantly, in case of long range Coulomb interaction between the particles---are dynamical screening, plasmonic and excitonic effects. A theoretical treatment of these effects is possible within the $GW$ approximation \cite{hedin_pr_65} that has allowed one to achieve excellent ground-state and equilibrium results for model systems and real materials, e.g. Refs.~\cite{aryasetiawan_rpp_98, onida_rmp_02}.
In situations where the system is out of equilibrium, $GW$ simulations are much more challenging. Early approaches have been derived in kinetic theory of plasmas by Balescu \cite{balescu_60}, Lenard \cite{lenard_60} and others who replaced, in the collision integral, the pair potential, $V(q)$ by a dynamically screened interaction $V(q)/\epsilon(q,\omega)$, where the dielectric function $\epsilon$ takes into account the screening behavior of the surrounding charged particles. These results were extended to optically excited semiconductors by Binder \textit{et al.} \cite{binder_prb_92}.
However, the Balescu--Lenard collision integral does not conserve total energy and neglects the formation of the plasmon spectrum.

These problems can be solved within nonequilibrium Green functions (NEGF) theory with the $GW$ selfenergy, and first selfconsistent time-dependent $GW$ simulations were reported by Banyai \textit{et al.} \cite{banyai_prl_98} who applied, in addition, the generalized Kadanoff--Baym ansatz (GKBA) \cite{lipavsky_generalized_1986}, see below. These results were extended to plasmas in strong laser fields in Ref.~\cite{bonitz_99_cpp}.
However, NEGF simulations with the $GW$ approximation exhibit a unfavorable cubic scaling of the computation time with the 
number of time steps $N_\tn{t}$, both, for two-time and for GKBA simulations which restricts the simulations to very short times. The situation radically changed with the introduction of the G1--G2 scheme by Schlünzen \textit{et al.} \cite{schluenzen_19_prl} which solves coupled time-local equations for the one-particle and two-particle Green functions. This scheme eliminates all memory integrals and, therefore, scales linearly with $N_\tn{t}$. Interestingly, this favorable scaling is achieved already after a small number of time steps and for all common selfenergies, including the second-order Born approximation, the $T$-matrix approximation and $GW$, as was demonstrated by Joost \textit{et al.} \cite{joost_prb_20}.
The G1--G2-scheme was recently applied to the photoionization of organic molecules \cite{pavlyukh_prb_21} and ultrafast electron--boson dynamics \cite{karlsson_prl21}.
In particular, G1--G2 simulations with the $GW$ selfenergy were reported for the simulation of ultrafast carrier and exciton dynamics in 2D materials by Perfetto \textit{et al. }\cite{perfetto_prl_22}.

However, $GW$ simulations apply only to weakly and moderately coupled many-particle systems. The reason is that the selfenergy is only of first order in the screened potential and, therefore, neglects multiple scattering effects that become increasingly important in strongly correlated materials, a modern example being transition-metal dichalcogenides (TMDCs) or twisted bilayers of graphene or TMDCs, e.g. \cite{wu_prl_18, li_nat_21, smolenski_nat_21, bonitz_pj_21}. On the other hand, strong-coupling effects are well captured with the particle--particle and particle--hole $T$-matrix selfenergies, e.g. \cite{schluenzen_jpcm_19}. But these approximations are available only in combination with a static pair potential. Therefore, a selfconsistent combination of strong coupling and dynamical screening effects remains a major open problem which is in the focus of the present paper. 

There have been various approximate methods to combine strong coupling and dynamical screening. These include the fluctuating-exchange approximation (FLEX), e.g.  \cite{schluenzen_jpcm_19, stahl_prb_21} and the Gould--DeWitt approximation \cite{gould-dewitt-67, gericke_pngf1}. A perturbation theory approach is the third-order approximation (TOA) \cite{schluenzen_prb17} which we discuss below.
We note that, more systematically, dynamical screening effects have been analyzed in detail for electron--hole plasmas and excitons in equilibrium within the Bethe--Salpeter equation by Zimmermann \textit{et al.} \cite{zimmermann_pss_78, haug_78} where also the dynamically screened ladder approximation (DSL) was introduced, for a text book discussion see Ref.~\cite{kremp-springer}. Especially the plasma effects on excitonic and atomic bound states as well as exciton--plasmon coupling remain a topic of high current interest for semiconductors, TMDCs, and dense plasmas, see e.g. \cite{kremp_jpcs_10, glazov_pssb_18, semkat_prb_19, kremp-springer, van-tuan_PhysRevX.7.041040, steinhoff_PhysRevB.98.045304} and references therein.

However, until now a selfconsistent treatment of dynamical screening and strong correlations under general nonequilibrium conditions has not been reported within nonequilibrium Green functions theory.
DSL-type equations for the pair-correlation operator have been presented within a reduced-density-operator approach in Refs. \cite{bonitz_qkt} and were analyzed in Ref. \cite{joost_prb_20}. However, the resulting equations did not include all exchange contributions, and the relation to NEGF remained unclear.
Here we re-analyze the G1--G2 scheme on the DSL level, starting from a density operator approach \cite{bonitz_qkt}.

In fact, the theory of single-time reduced density operators (RDO, BBGKY-hierarchy) has emerged independently of the NEGF approach in a variety of fields including quantum gases \cite{boercker_ap_79}, nuclear matter \cite{wang-cassing-85, lacroix_epja_14, schuck_epja_16}, 
dense plasmas \cite{bonitz-etal.96pla}, semiconductor optics (semiconductor Bloch equations) and transport \cite{lindberg_prb88,axt_zphysb_94,bonitz-etal.96jpcm, kuhn_rmp_02}. The DSL approximation emerges naturally in this approach when three-particle correlations are neglected. 
Extensive developments for the ground state of correlated electrons have also occured in  atomic and molecular physics \cite{valdemoro_93,MAZZIOTTI_98,lackner_propagating_2015} where the approach is known under the name two-particle reduced density matrix (2RDM) method. Recently, extensions to time-dependent electron dynamics have lead to the time-dependent 2RDM  (TD2RDM) method \cite{lackner_propagating_2015, lackner_pra_17} which is conceptionally equivalent to the BBGKY-hierarchy for the reduced density operators.
A particular problem, when applied to finite systems such as atoms, is that the solution of the RDO (2RDM) equations may become unstable during the time propagation. A solution was presented by Lackner \textit{et al.} \cite{lackner_propagating_2015, lackner_pra_17} by enforcing contraction consistency, e.g.~\cite{coleman2000reduced,mazzioti_reduced_2007} and applying a purification scheme.

In this paper present a detailed derivation of the DSL--G1--G2 equation within RDO (2RDM) theory paying particular attention to a complete account of the exchange contributions. 
The resulting DSL-G1--G2 equations are then compared to the results following from known selfenergy approximations of NEGF theory. Furthermore, we present numerical DSL results for the ultrafast electron dynamics in finite Hubbard clusters and demonstrate excellent agreement with exact results. We verify time-linear scaling and demonstrate that long propagation times can be achieved by enforcing contraction consistency and applying an improved purification scheme for the two-particle Green function.

The structure of this paper and its main goals are as follows: 
\begin{enumerate}
    \item In section~\ref{s:theory} we recall the second quantization scheme and introduce the NEGF approach.
    \item In Sec.~\ref{s:sigmas} we introduce the selfenergy approximations that are of interest for our analysis and for comparison with the DSL approximation.
    \item Section~\ref{s:g1-g2} is devoted to the G1--G2 scheme that was introduced in Refs.~\cite{schluenzen_19_prl, joost_prb_20}. 
    We present the explicit form of the G2 equations for the selfenergies of Sec.~\ref{s:sigmas} and
    pay special attention to the correct treatment of the exchange diagrams in the $T$-matrix and $GW$ approximations.
    \item In Sec.~\ref{s:do-test} we introduce the alternative approach to many-particle dynamics that is based on reduced density operators. There we derive the DSL approximation and present a detailed term by term comparison to the G1--G2 equations that were derived from nonequilibrium Green functions above, which is summarized in table~\ref{tab:negf-rdo}. 
    \item In Sec.~\ref{s:hubbard} we apply the G1--G2 scheme to the ultrafast dynamics of finite Hubbard clusters. There we also discuss the issues of contraction consistency and discuss how to deal with intrinsic instabilities of the dynamical equations via a purification scheme.
    \item Numerical benchmarks of the DSL-G1--G2 approximation for finite Hubbard clusters against exact results are presented in Sec.~\ref{s:numerics}.
\end{enumerate}

\section{Theoretical framework}\label{s:theory}
The goal of this section is to provide the basis to link the NEGF formalism to the G1--G2 scheme. The equations of motion for the NEGF are the Keldysh--Kadanoff--Baym equations that contain a single input quantity---the selfenergy $\Sigma$. For each approximation to $\Sigma$ we will identify a counterpart in the G1--G2 scheme below.

\subsection{Keldysh--Kadanoff--Baym Equations}
\label{ss:framework}
Even though we will present numerical results for a Hubbard system below, it is instructive to start from a formulation of the nonequilibrium many-body problem with the  general Hamiltonian in second quantization
\begin{align}
    \chat{H}(t) &= \sum_{ij} h^{(0)}_{ij}(t) \cop{i}{} \aop{j}{} + \frac{1}{2} \sum_{ijkl} w_{ijkl}(t) \cop{i}{} \cop{j}{} \aop{l}{} \aop{k}{} \, .
    \label{eq:h}
\end{align}
Here, $h^{(0)}$ is the single-particle contribution and $w$ the pair interaction. Note the two-fold time dependencies of the Hamiltonian. The time dependence of the single-particle contribution $h^{(0)}$ accounts for the interaction of the particles with external electromagnetic fields, e.g.~\cite{kremp_99_pre,haberland_01_pre}, charged particle impact (stopping) \cite{balzer_prb16, balzer_prl_18,schluenzen_cpp_18}, or the rapid variation (quench) of system parameters such as the confinement potential  \cite{schneider_fermionic_2012, schluenzen_prb16,schluenzen_prb17}. Similarly, quenches of the pair interaction $w$ have been studied \cite{gericke_jpa03,moeckel_prl08,schollwoeck_prb_19}.
There is a second type of time dependence in $w$ that is not related to the coupling to an external excitation but that results from the numerical preparation of a correlated initial state. In many cases an efficient procedure is to start from an uncorrelated initial state and to build up correlations dynamically via ``adiabatic switching'', e.g. \cite{schluenzen_jpcm_19}. This approach will also be used in some of our simulation results below. Therefore, in the derivations, we will retain the full time dependence of $w(t)$ throughout this paper.

\begin{figure}[h]
\includegraphics[width=\columnwidth]{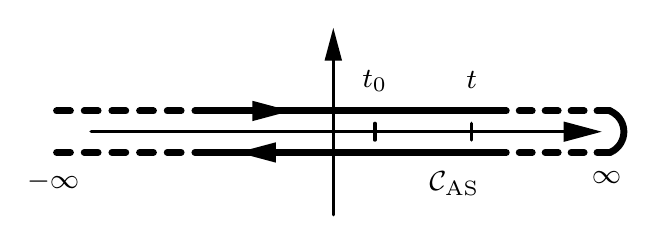}
\caption{Keldysh ``round-trip'' time contour that is used in NEGF theory to treat initial correlations via ``adiabatic switching'' of the pair interaction, starting from an uncorrelated state in the remote past, for more details, see Refs. \cite{stefanucci_nonequilibrium_2013,balzer-book,bonitz_pss_19_keldysh,joost_prb_20}.
}
\label{fig:contour}
\end{figure}

The matrix indices and summations in the Hamiltonian of \refeqn{eq:h}
refer to an arbitrary complete orthonormal system of single-particle orbitals $|i\rangle$ for which we define
 creation ($\chat c^\dagger_i$) and annihilation ($\chat c_i$) operators that obey Bose or Fermi statistics. Using the standard Heisenberg procedure, these operators are made time dependent and are used to  define the one-body nonequilibrium Green function where all time arguments $z, z'$ are defined on the Keldysh contour $\mathcal{C}$~\cite{schluenzen_jpcm_19} (see Fig.~\ref{fig:contour}),
\begin{align}
    G_{ij}(z,z')=\frac{1}{\i\hbar}\left\langle \mathcal{T}_\mathcal{C} \left\{\chat{c}_i(z)\chat{c}^\dagger_j(z')\right\} \right\rangle\,,
\end{align}
where, $\mathcal{T}_\mathcal{C}$ is the time-ordering operator on the contour, and the averaging is performed with the correlated unperturbed $N$-particle density operator of the system.

The equations of motion for the NEGF are the Keldysh--Kadanoff--Baym  equations (KBE)~\footnote{Throughout this work, ``$\pm$'' refers to bosons/fermions.}\cite{kadanoff-baym-book,keldysh64}
\begin{align}\label{eq:KBE1}
    \sum_k & \left[ \i\hbar \frac{\d}{\d z} \delta_{ik} - h^{(0)}_{ik}(z) \right] G_{kj}(z,z') - \delta_{ij} \delta_\mathcal{C}(z,z') \\ 
    &=\pm \i\hbar \sum_{klp} \int_\mathcal{C} \d \cbar{z}\, w_{iklp}(z,\cbar{z})G^{(2)}_{lpjk}(z,\cbar{z},z',\cbar{z}^+)\\
    & = \, \sum_k \int_\mathcal{C} \d \cbar{z}\, \Sigma_{ik}(z,\cbar{z})  G_{kj}(\cbar{z},z')\, , 
    \label{eq:kbe-sigma-form1} \\
   \sum_k & G_{ik}(z,z') \left[- \i\hbar \frac{\overset{\leftarrow}{\d}}{\d z'} \delta_{kj} - h^{(0)}_{kj}(z') \right] - \delta_{ij} \delta_\mathcal{C}(z,z') \label{eq:KBE2}\\ 
   &=\pm \i\hbar \sum_{klp} \int_\mathcal{C} \d \cbar{z}\, G^{(2)}_{iklp}(z,\cbar{z}^-,z',\cbar{z}) w_{lpjk}(\cbar{z},z') \\
    &= \sum_k \int_\mathcal{C} \d \cbar{z}\, G_{ik}(z,\cbar{z}) \Sigma_{kj}(\cbar{z},z') 
   \,,
   \label{eq:kbe-sigma-form2}
\end{align} 
where ${z^\pm \coloneqq z \pm \epsilon}$ with $\epsilon \to +0$, and
we introduced a two-time version of the interaction potential using the delta function on the Keldysh contour, $w_{ijkl}(z,z') = \delta_\mathcal{C}(z,z') w_{ijkl}(z)$, see, e.g. Refs.~\cite{schluenzen_jpcm_19,schluenzen_cpp16,stefanucci_nonequilibrium_2013}.

Note that we have presented two forms of the r.h.s. of the KBE. The first lines contain the two-particle Green function $G^{(2)}$, Eq.~(\refeq{eq:g2-def}), that will be discussed in detail below. The second form of the r.h.s. contains the selfenergy $\Sigma$ which is introduced in NEGF theory to eliminate the two-particle Green function. Below, in Sec.~\ref{s:sigmas}, we will consider several approximations for $\Sigma$. Here we already notice that the dependence of the single-particle Green function on two time arguments, combined with the time integral on the r.h.s. of Eqs.~(\refeq{eq:kbe-sigma-form1}) and (\refeq{eq:kbe-sigma-form2}), gives rise to a cubic scaling, $N_t^3$, of the computing time with the number of time steps $N_t$. It is the main achievement of the G1--G2 scheme that this scaling can be reduced to $N^1_t$, regardless of the choice of the selfenergy \cite{schluenzen_19_prl,joost_prb_20}. In this scheme, the two-particle Green function is restored and propagated. This will be introduced in Sec.~\ref{s:g1-g2}. But first, we introduce and briefly discuss the relevant approximations for the selfenergy.

\section{Selfenergy approximations}\label{s:sigmas}
A graphical overview of the most important selfenergy approximations in terms of Feynman diagrams is presented in \reffigs{fig:diagrams_order}{fig:diagrams_resummation}. A main selection criterion is that each of the approximations is conserving, i.e. conserves particle number, momentum and total energy.
To shorten the presentation we only provide the results for the greater and less component, $\Sigma^\gtrless(t,t')$ which follows from $\Sigma(z,z')$ by taking the time arguments on different branches of the contour in Fig.~\ref{fig:contour}, for details see Refs.~\cite{stefanucci_nonequilibrium_2013,schluenzen_jpcm_19}.

\subsection{Single-particle Green functions}\label{ss:g1}
Equations (\refeq{eq:KBE1}) and (\refeq{eq:KBE2}) for the one-particle NEGF are formulated on the Keldysh contour, cf. Fig.~\ref{fig:contour} and are equivalent to equations for Keldysh Green function matrices of real-time arguments, where the matrix components differ by the location of the time arguments on the contour \cite{keldysh64,bonitz_pss_19_keldysh}. This gives rise to the correlation functions, $G^\gtrless(t,t')$, and the retarded and advanced functions, $G^{\rm R/A}(t,t')$,
\begin{align}
    G^<_{ij}(t,t') &=\pm \frac{1}{\i\hbar}\left\langle \chat c_j^\dagger(t') \chat c_i(t)\right\rangle\,,
    \\
    G^>_{ij}(t,t') &= \frac{1}{\i\hbar}\left\langle \chat c_i(t) \chat c_j^\dagger(t')\right\rangle  \,,
    \\
    G^{\rm R/A}_{ij}(t,t') &=
    \pm \Theta[\pm(t-t')]\left\{G^>_{ij}(t,t')-G^<_{ij}(t,t')\right\}\,,
    \label{eq:gra-def}
\end{align}
for details see the text books \cite{stefanucci_nonequilibrium_2013,balzer-book}.

Let us summarize a few important properties of the correlation functions.
First, on the time diagonal the less component of the NEGF can be written as 
\begin{align}
G_{ij}^<(t,t) \equiv  G_{ij}^<(t) &= G_{ij}^>(t) - \frac{1}{\i\hbar}\delta_{ij} = \pm \frac{1}{\i\hbar} n_{ij}(t)\,,
\label{eq:g-glsymm}
\end{align}
where $n_{ij}$ is the single-particle density matrix. Thus, $G^\gtrless$ have a clear physical meaning and are directly related to observables.
We, therefore, provide the selfenergy approximations in terms of these functions.

\subsection{Hartree--Fock selfenergy}\label{ss:hf}
The first-order terms of the selfenergy describe particle interaction on the mean-field level. They are combined in the so-called Hartree--Fock (HF) selfenergy, which for a time-diagonal interaction tensor only has a single-time-dependent (delta) component for the real time $t$,
\begin{align}
     \Sigma^{\tn{HF},\delta}_{ij}(t) =& \pm \i \hbar \sum_{kl} w^\pm_{ikjl}(t) G^<_{lk}(t,t) \, . \label{eq:sigma_hf}
    \end{align}
For this reason, the first-order terms are easily accounted for by including $\Sigma^{\tn{HF},\delta}$ into an effective single-particle Hamiltonian of the form
\begin{align}
 h^\mathrm{HF}_{ij}(t) = h^{(0)}_{ij}(t) \pm \mathrm{i}\hbar \sum_{kl} w^\pm_{ikjl}(t) G^<_{lk}(t,t)\,. \label{eq:h_HF}
\end{align}
In \eqrefs{eq:sigma_hf}{eq:h_HF} we introduced the (anti-)symmetrized matrix element of the pair potential,

\begin{align}
 w^\pm_{ijkl}(t) &\coloneqq w_{ijkl}(t) \pm w_{ijlk}(t)\,,
\label{eq:wpm-def}
\\
&=w_{ijkl}(t) \pm w_{jikl}(t)\,,
\end{align}
which has the symmetries
\begin{align}
    w^\pm_{ijkl}(t) &= \pm w^\pm_{ijlk}(t) = \pm w^\pm_{jikl}(t)\,. \label{eq:wpm-symm}
\end{align}
and, for fermions, in particular, $w^\pm_{ijkk}(t)=  w^\pm_{iikl}(t)= 0$. 

\subsection{Second-order Born selfenergy (SOA)}\label{ss:soa}
The simplest selfenergy that includes correlations and, thus, allows to describe dissipation and relaxation effects (selfenergy beyond Hartree--Fock) is given by the second-order Born approximation~\cite{schluenzen_cpp16}, 
\begin{align}
 \Sigma^\gtrless_{ij}\left(t,t'\right) 
 &= \pm\left(\i\hbar\right)^2 \sum_{klpqrs} \, w_{iklp}\left(t\right) w^\pm_{qrjs}\left(t'\right) 
 \label{eq:sigma-soa} \\ \nonumber
 &\qquad \qquad \times G^\gtrless_{lq}\left(t,t'\right) G^\gtrless_{pr}\left(t,t'\right) G^\lessgtr_{sk}\left(t',t\right)\, .
\end{align}
We will use the notation ``SOA'' for the selfenergy that includes all terms up to second order (including HF).
SOA provides the starting point for all following approximations.
Note that the potential $w^\pm$ gives rise to two contributions---the direct SOA and the associated exchange diagram which are shown in Fig.~\ref{fig:diagrams_order}.

\subsection{Third-order approximation (TOA)}\label{ss:toa}
\begin{figure}[t]
\includegraphics[width=\columnwidth]{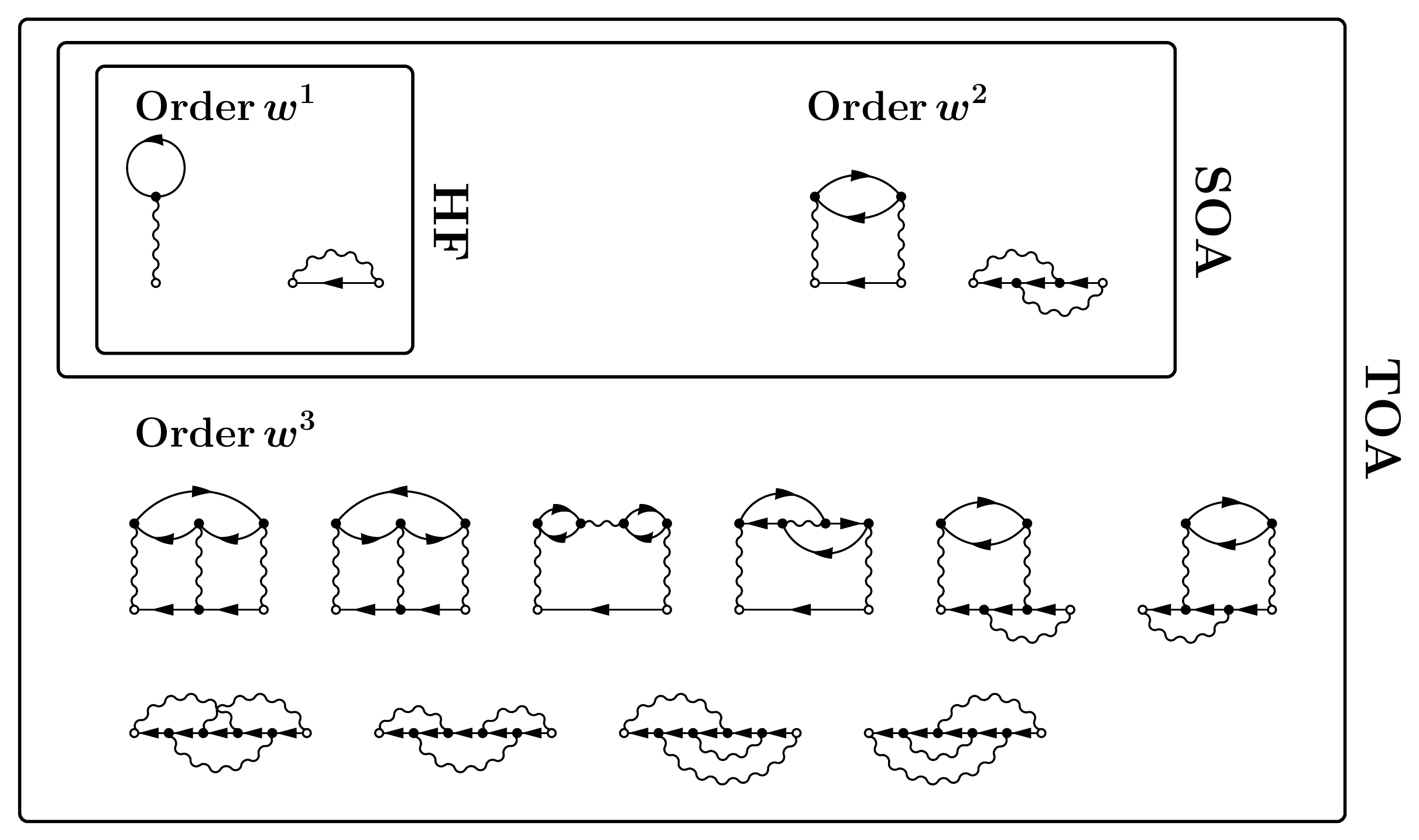}
\caption{Selfenergy diagrams for the perturbative (with respect to powers of the interaction) approach up to order three. Hartree--Fock (HF) contains contributions of first order in $w$, the second-order selfenergy (SOA) contains second-order diagrams (together with the first order), whereas the third-order approximation (TOA) contains all diagrams up to  order $w^3$. Note that the second diagrams of order $w^1$ and $w^2$ describe exchange processes, respectively. For order $w^3$, exchange processes are included in the latter six diagrams.}
\label{fig:diagrams_order}
\end{figure}

\begin{figure}[t]
\includegraphics[width=\columnwidth]{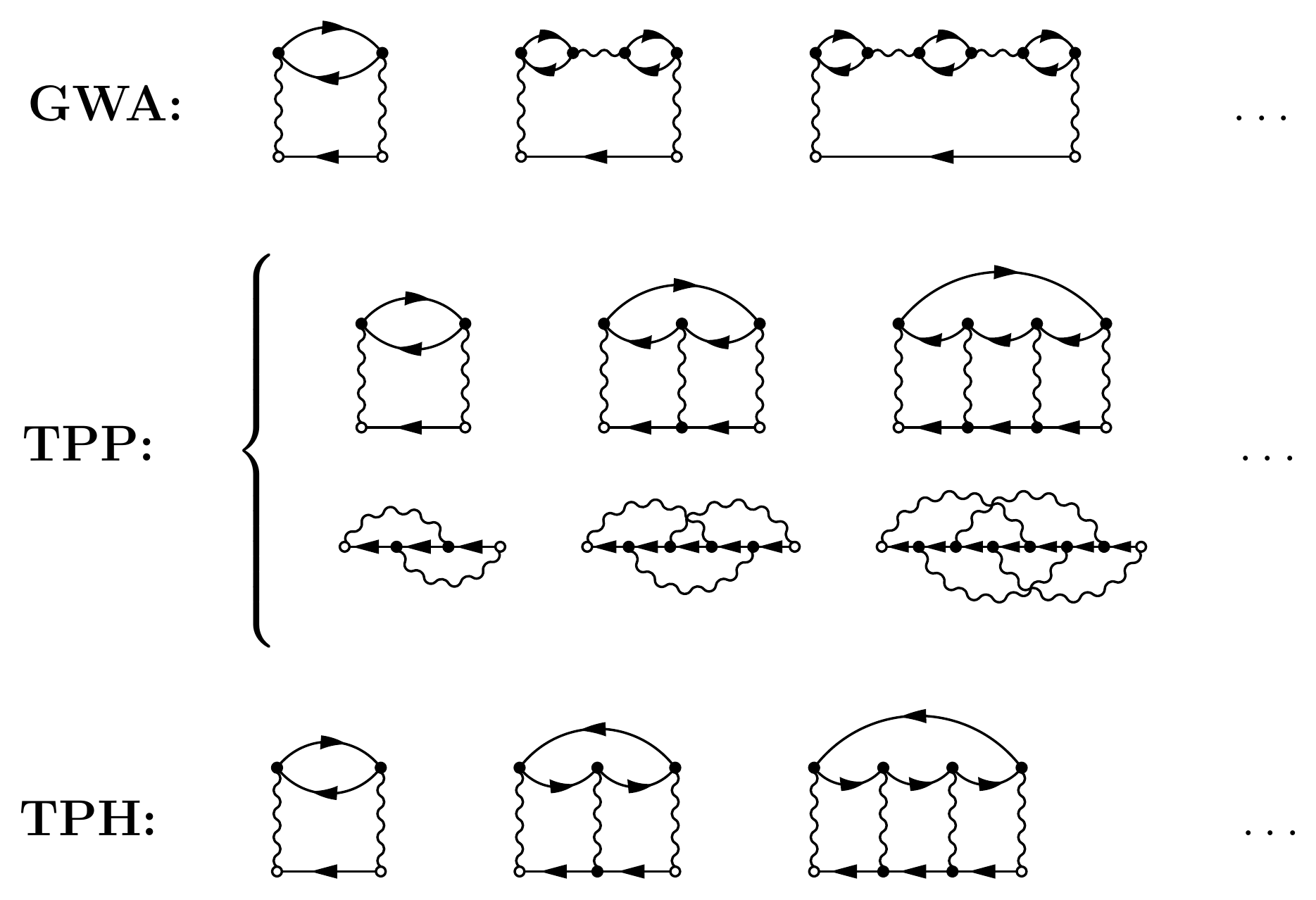}
\caption{Selfenergy diagrams for the three resummation approaches starting from the second-order contributions. Dots indicate continuation of the sums to infinite order. Note that for TPP it is possible to also include the corresponding exchange diagrams (second line). }
\label{fig:diagrams_resummation}
\end{figure}

The third-order approximation for the selfenergy allows to significantly improve the accuracy of simulations, compared to SOA. It contains all diagrams that include up to three interaction lines, cf. Fig.~\ref{fig:diagrams_order}. There exist 10 skeletonic diagrams that are of order $w^3$ and which are also part of the GW, TPP and TPH approximations, cf. Fig.~\ref{fig:diagrams_resummation}. Thus, TOA contains the starting terms of the ladder and bubble sums.
TOA was first introduced and tested in Ref.~\cite{schluenzen_prb17} and was found to be very accurate for weak and moderate coupling, independently of the filling (density) \cite{schluenzen_jpcm_19}.

\subsection{Infinite series summations. Dynamical screening and strong coupling}\label{ss:negf-gf-gh}
After considering perturbation theory results for the selfenergy we now turn to another class of approximations that result from summation of an infinite series of diagrams. The first example is the polarization approximation (summation of bubble diagrams, $GW$ approximation) that allows to include dynamical screening effects which is important, in particular, for charged particles in plasmas, condensed matter or in macromolecules. $GW$ is a weak coupling approximation, but includes a selfconsistently screened pair potential $W$. To account for strong coupling effects, the second example is the particle--particle $T$ matrix that results from summing up the entire Born series. Finally, we will consider the second flavor of the $T$ matrix---the particle--hole $T$-matrix approximation. These approximation use a static potential as an input. Since these are standard approximations, a derivation is not necessary, e.g. \cite{schluenzen_cpp16,schluenzen_jpcm_19}. Instead, we list the compact final result, together with the associated diagrams in Fig.~\ref{fig:diagrams_resummation}.

A compact notation is achieved by introducing the following products of single-particle Green functions
\begin{align}
    \mathcal{G}^{\tn{H},\gtrless}_{ijkl}(t,t') &\coloneqq G^\gtrless_{ik}(t,t') G^\gtrless_{jl}(t,t')\,,
    \label{eq:g2h-def}\\
    \mathcal{G}^{\tn{F},\gtrless}_{ijkl}(t,t') &\coloneqq G^\gtrless_{il}(t,t') G^\lessgtr_{jk}(t',t)\,.
    \label{eq:g2f-def}
\end{align}
We will see in Sec.~\ref{ss:g2} that $\mathcal{G}^{\tn{H},\gtrless}_{ijkl}$ is just the Hartree part of the two-particle Green function whereas $\mathcal{G}^{\tn{F},\gtrless}_{ijkl}$ is the Fock part.

\subsection{Dynamical screening. $GW$ selfenergy}\label{ss:negf-gw}
The selfenergy in $GW$ approximation (GWA) is defined in terms of the dynamically screened potential $W$ 
\begin{align}
 \Sigma_{ij}^{\rm{GWA},\gtrless}(t,t') = \i \hbar \sum_{kl} W^\gtrless_{ilkj}(t,t') G^\gtrless_{kl}(t,t')\, ,
\label{eq:sigma-gw}
\end{align}
that obeys the following integral equation (Dyson equation)
\begin{align}
\label{eq:W-def}
    & W^\gtrless_{ijkl}(t,t') = \\&\pm \i \hbar \sum_{pqrs} w_{ipkq}(t) w_{rjsl}(t') \mathcal{G}^{\tn{F},\gtrless}_{qspr}(t, t') \pm \i\hbar \sum_{pqrs} w_{ipkq}(t)\times\\ &\qquad\Bigg\{ \int_{t_0}^t \d\cbar{t} \, \Big[ \mathcal{G}^{\tn{F},>}_{qspr}(t, \cbar t) - \mathcal{G}^{\tn{F},<}_{qspr}(t, \cbar t) \Big] W^\gtrless_{rjsl}(\cbar{t},t') \nonumber 
    \\&\qquad+\int_{t_0}^{t'} \d \cbar{t} \, \mathcal{G}^{\tn{F},\gtrless}_{qspr}(t, \cbar t) \left[W^<_{rjsl}(\cbar{t},t') - W^>_{rjsl}(\cbar{t},t') \right]\Bigg\}\,. \nonumber
\end{align}
Here, the first term on the right coincides with the second order direct Born diagram (SOA) whereas the integral term gives rise to an infinite sum of additional diagrams that follow iteratively, starting by inserting the second-order terms for $W$, under the integral. The first diagrams are sketched in Fig.~\ref{fig:diagrams_resummation}. Note that here we did not use the (anti-)symmetrized potential $w^\pm$. Finally, the dependence of $W$ on two times and the time integral on the r.h.s. imply that the computational effort for evaluating the $GW$ selfenergy scales cubically with the simulation duration.

\subsection{Strong coupling. Particle--particle $T$-matrix selfenergy (TPP)}\label{ss:negf-tpp}
The definition of the TPP selfenergy has a similar structure as $GW$, but the screened potential is replaced by the particle--particle $T$ matrix,
\begin{align}
 \Sigma_{ij}^{{\rm TPP},\gtrless}(t,t') = \i \hbar \sum_{kl} T^{\tn{pp},\gtrless}_{ikjl}(t,t') G^\lessgtr_{lk}(t',t)\, ,
 \label{eq:sigma-tpp}
\end{align}
which obeys a slightly different integral equation,
\begin{align}
  & T^{\tn{pp},\gtrless}_{ijkl}(t,t') = \\&\pm \i \hbar \sum_{pqrs} w_{ijpq}(t) \mathcal{G}^{\tn{H},\gtrless}_{pqrs}(t, t') w^\pm_{rskl}(t') + \i\hbar \sum_{pqrs} w_{ijpq}(t) \times \\&\qquad\Bigg\{ \int_{t_0}^t \d \cbar t\, \Big[ \mathcal{G}^{\tn{H},>}_{pqrs}(t, \cbar t) - \mathcal{G}^{\tn{H},<}_{pqrs}(t, \cbar t) \Big] T^{\tn{pp},\gtrless}_{rskl}(\cbar t, t') \\&\qquad+\int_{t_0}^{t'} \d \cbar t\, \mathcal{G}^{\tn{H},\gtrless}_{pqrs}(t, \cbar t) \Big[ T^{\tn{pp},<}_{rskl}(\cbar t, t') - T^{\tn{pp},>}_{rskl}(\cbar t, t')\Big] \Bigg\}\, .\nonumber
\end{align}
The main difference to the Dyson equation for the screened potential $W$ is the replacement of the Fock Green function by the Hartree Green function, $\mathcal{G}^{\tn{F},\gtrless}\to \mathcal{G}^{\tn{H},\gtrless}$. Furthermore, the first term (which again reproduces the SOA diagram) here contains the potential $w^\pm$, thus it includes the exchange diagram. As a consequence, each diagram of the iteration series is complemented by an exchange diagram, cf. Fig.~\ref{fig:diagrams_resummation}, second and third line, respectively.

\subsection{Particle--hole $T$-matrix selfenergy}\label{ss:negf-teh}
The particle--hole $T$ matrix is defined analogously to the particle--particle $T$ matrix,
\begin{align}
 \Sigma_{ij}^{\rm{TPH},\gtrless}(t,t') = \i \hbar \sum_{kl} T^{\tn{ph},\gtrless}_{ikjl}(t,t') G^\gtrless_{lk}(t,t')\, ,
 \label{eq:sigma-tph}
\end{align}
with the main difference given by the appearance of $\mathcal{G}^{\tn{F},\gtrless}$ in the Lippmann--Schwinger equation, 
 \begin{align}
     &T_{ijkl}^{\tn{ph},\gtrless}(t,t') = \\&\pm \i \hbar \sum_{pqrs} w_{iqpl}(t) \mathcal{G}^{\tn{F},\gtrless}_{psqr}(t,t') w_{rjks}(t') + \i\hbar \sum_{pqrs} w_{iqpl}(t)\times\\ &\qquad\bigg\{ \int_{t_0}^t \d \cbar t\, \Big[ \mathcal{G}^{\tn{F},>}_{psqr} (t,\cbar t) - \mathcal{G}^{\tn{F},<}_{psqr} (t,\cbar t) \Big] T^{\tn{ph},\gtrless}_{rjks}(\cbar t, t') \\&\qquad +\int_{t_0}^{t'} \d \cbar t \, \mathcal{G}^{\tn{F},\gtrless}_{psqr} (t,\cbar t) \Big[ T^{\tn{ph},<}_{rjks}(\cbar t, t') - T^{\tn{ph},>}_{rjks}(\cbar t, t')\Big]\bigg\} \, .
     \nonumber
 \end{align}
 Note that, in contrast to the particle--particle $T$ matrix, in this definition no exchange contributions are included.  
 While it is possible to sum up an additional diagram series by using the (anti-)symmetrized $w^\pm$, instead of $w$, this would lead to a violation of physical conservation laws \cite{schluenzen_phd_21}.

\subsection{Combining strong coupling and\\ dynamical screening}\label{ss:negf-dsl}
An important task of many-body theory, in particular for systems with long-range Coulomb interaction,  is to combine strong coupling and dynamical screening effects. There exist several approximate solutions. One is the third-order approximation (TOA) that was discussed above in Sec.~\ref{ss:toa} which contains diagrams from both approximations, up to the third order. Another approximate solution is provided by the FLEX (fluctuating exchange) scheme, e.g. \cite{schluenzen_jpcm_19, stahl_prb_21}. A combination of strong coupling and dynamical screening (the dynamically screened ladder approximation, DSL) for the case of excitons in thermal equilibrium has been formulated in terms of a Bethe-Salpeter equation 
 \cite{zimmermann_pss_78, haug_78}. However, a fully selfconsistent nonequilibrium expression for the DSL selfenergy is still missing. At the same time, as we will show in Sec.~\ref{s:do-test}, a G1--G2 scheme on the level of the  nonequilibrium DSL approximation is straightforwardly derived using reduced density operator theory. This also allows to incorporate exchange diagrams into the particle--hole $T$-matrix and $GW$ approximation in a conserving manner.


\section{The G1--G2 scheme}\label{s:g1-g2}
Instead of solving the full two-time equations (\refeq{eq:kbe-sigma-form1}, \refeq{eq:kbe-sigma-form2}), from now on we will concentrate on an approximation scheme that considers the solution along the time diagonal only. This is based on the generalized Kadanoff--Baym ansatz (GKBA), that will be introduced in Sec~\ref{ss:gkba}, and its reformulation in terms of coupled time local equations for the single-particle and two-particle Green functions, leading to the G1--G2 scheme, cf. Secs.~\ref{ss:g1-g2-soa} and \ref{ss:g1g2-beyond-soa}. But first we introduce two-particle Green functions on the real time axis which we will need, in addition to the the single-particle Green functions, $G^\gtrless$, that we discussed in Sec.~\ref{ss:g1}.

\subsection{The two-particle Green function}\label{ss:g2}
We start with formulating the two-particle NEGF on the Keldysh contour,
\begin{align}\label{eq:g2-def}
    &G^{(2)}_{ijkl}(z_1,z_2,z_3,z_4)\\
    &\qquad=\frac{1}{\left(\i\hbar\right)^2}\left\langle \mathcal{T}_\mathcal{C} \left\{\chat{c}_i(z_1)\chat{c}_j(z_2)\chat{c}^\dagger_l(z_4)\chat{c}^\dagger_k(z_3)\right\} \right\rangle\,,\nonumber
\end{align}
which---similar to the selfenergy---can be divided into a mean-field [Hartree (H) plus Fock (F)] and a correlation contribution,
\begin{align}\label{eq:G2_parts}
    G^{(2)}_{ijkl}(z_1,z&_2,z_3,z_4) \\ &=G^{(2),\tn{H}}_{ijkl}(z_1,z_2,z_3,z_4) \pm G^{(2),\tn{F}}_{ijkl}(z_1,z_2,z_3,z_4) \nonumber\\
    &\quad+ G^{(2),\tn{corr}}_{ijkl}(z_1,z_2,z_3,z_4)\,.\nonumber
    \end{align}
Our G1--G2 scheme involves the special case of two-particle functions that depend either on one or two times and their real-time components,
that we define as follows
    \begin{align}
    \mathcal{G}^{\tn{H}}_{ijkl}(z,z') &\coloneqq G^{(2),\tn{H}}_{ijkl}(z,z,z',z') = G_{ik}(z,z') G_{jl}(z,z')\,,\nonumber\\
    \mathcal{G}^{\tn{F}}_{ijkl}(z,z') &\coloneqq G^{(2),\tn{F}}_{ijkl}(z,z',z,z') = G_{il}(z,z') G_{jk}(z',z)\,,\nonumber\\
    \mathcal{G}^{\tn{corr}}_{ijkl}(z,z') &\coloneqq G^{(2),\tn{corr}}_{ijkl}(z,z,z',z^+)\,.\nonumber
\end{align}
For the derivation of the G1--G2 scheme it will be sufficient to consider the greater/less components of the Hartree, Fock and correlated parts of $G^{(2)}$ on the real time diagonal,
\begin{align}
    \mathcal{G}^{\tn{H},\gtrless}_{ijkl}(t) &\coloneqq \mathcal{G}^{\tn{H},\gtrless}_{ijkl}(t,t)\,,\nonumber\\
    \mathcal{G}^{\tn{F},\gtrless}_{ijkl}(t) &\coloneqq \mathcal{G}^{\tn{F},\gtrless}_{ijkl}(t,t)\,,\nonumber\\
    \mathcal{G}_{ijkl}(t) &\coloneqq \mathcal{G}^{\tn{corr},<}_{ijkl}(t,t)\, .
    \label{eq:gscor-def}
\end{align}
where $\mathcal{G}^{\tn{H},\gtrless}_{ijkl}(t,t)$ and $\mathcal{G}^{\tn{F},\gtrless}_{ijkl}(t,t)$ were defined in Eqs.~\eqref{eq:g2h-def} and \eqref{eq:g2f-def}.

The time-diagonal correlated two-particle Green function, $\mathcal{G}(t)$, defined by \refeqn{eq:gscor-def}, is the central quantity of the G1--G2 scheme. The exact solution, $\mathcal{G}(t)$, and the one corresponding to the selfenergy approximations considered in this work, obey the following (pair-) exchange symmetries,
\begin{align}
    \mathcal{G}_{ijkl}(t) &= \mathcal{G}_{jilk}(t)\, , \label{eq:Gsymm_a}  \\
    \mathcal{G}_{ijkl}(t) &= \Big[\mathcal{G}_{klij}(t)\Big]^*\, , \label{eq:Gsymm_b}\\
    \mathcal{G}_{ijkl}(t) &= \pm \mathcal{G}_{jikl}(t)\, 
= \pm \mathcal{G}_{ijlk}(t)\, , 
\label{eq:Gsymm_a2}
\end{align}
which exactly agree with the symmetries of the (anti-) symmetrized potential $w^\pm$, cf. Eqs.~(\refeq{eq:wpm-symm}).

\subsection{Time-diagonal KBE for $G^<(t)$}
\label{ss:diagonal-kbe}
In the following we concentrate on the dynamics of the correlation function ${G_{ij}^\gtrless(t) \coloneqq G_{ij}^\gtrless(t,t)}$ on the real-time diagonal. The corresponding equation of motion follows from adding the two KBE and taking the limit of equal times, e.g.~\cite{schluenzen_cpp16}~\footnote{The commutator of two single-particle quantities $A(t)$ and $B(t)$ is defined as $\left[A,B\right]_{ij}(t) = \sum_{k} \left[A_{ik}(t)B_{kj}(t) - B_{ik}(t)A_{kj}(t)\right]$.}
\begin{align}
 \i\hbar \frac{\d}{\d t}G_{ij}^<(t) &- \left[ h^\mathrm{HF},G^< \right]_{ij}(t) = \big[I + I^\dagger\big]_{ij}(t)\,,\quad \label{eq:eom_gone}
 \\\mbox{with}\;
 \left[ h^\mathrm{HF},G^< \right]_{ij} &=\sum_k \left\{ h^\mathrm{HF}_{ik} G^<_{kj} - G^<_{ik}h^\mathrm{HF}_{kj}
 \right\}\,,
 \label{eq:commutator-matrix}
\end{align}
where 
$I(t)= \mathcal{I}(t) + \mathcal{I}^\tn{IC}(t)\,,$ is the collision 
integral of the kinetic equation
that, in general, consists of the dynamical collision integral $\mathcal{I}$ and the initial-correlation contribution $\mathcal{I}^\tn{IC}$ which includes pair correlations existing in the system at the initial time $t=t_0$. The treatment of initial correlations in the G1--G2 scheme was discussed in detail in Ref.~\cite{joost_prb_20}. Therefore, these results will not be repeated here, hence, we consider the case $\mathcal{I}^\tn{IC}(t)=0$. For a general discussion of initial correlations in NEGF theory, see Refs.~\cite{stefanucci_nonequilibrium_2013,bonitz_pss_18} and references therein. In the numerical applications, below, initial correlations will be properly included. 

The collision integral in Eq.~(\refeq{eq:eom_gone}) has the following general form:
\begin{align}
 \label{eq:definition_D}
 I_{ij}(t) 
  &= \pm\i\hbar \sum_{klp} w_{iklp}(t) \mathcal{G}_{lpjk}(t) \,
\\\label{eq:tiagonal-i-sigmas}
&= 
 \sum_{k} \int_{t_0}^t \mathrm{d}\cbar{t} \left[ \Sigma_{ik}^>(t,\cbar{t}) G_{kj}^<(\cbar{t},t) - \Sigma_{ik}^<(t,\cbar{t}) G_{kj}^>(\cbar{t},t) \right]\,,
\end{align}
where, the first line follows directly from the r.h.s. of  \eqrefs{eq:KBE1}{eq:KBE2}, (the time integral has been taken with the help of the delta function in the two-time potential). In the second line the two-particle Green function has been eliminated by introducing the correlation selfenergy functions $\Sigma^\gtrless$ (we retain the notation $\Sigma$ for the correlated part). Note that for the approximations studied in this paper, the correlation selfenergies $\Sigma^\gtrless(t,t')$ are non-singular functions. This means, the collision integral $I_{ij}(t)$ in \refeqn{eq:definition_D} vanishes for $t\to t_0$ (at this point only the initial correlation term may be present).

\subsection{The generalized Kadanoff--Baym ansatz (GKBA)}\label{ss:gkba}
Even in the time diagonal case, the collision integral (\ref{eq:tiagonal-i-sigmas})  involves Green functions and selfenergies away from the real-time diagonal. Thus, the equation for $G^<(t)$, Eq.~(\refeq{eq:eom_gone}) is not closed.
The generalized Kadanoff--Baym ansatz \cite{lipavski_prb_86} provides a useful approximation for the reconstruction of the 
time-off-diagonal elements of the less and greater NEGF  from their time-diagonal value via~\cite{schluenzen_cpp16}
\begin{align}
 G_{ij}^\gtrless(t, t') = \i \hbar \sum_k \left[G_{ik}^\mathrm{R}(t,t') G_{kj}^\gtrless(t')-G_{ik}^\gtrless(t) G_{kj}^\mathrm{A}(t,t')\right]\, , \label{eq:GKBA}
\end{align}
where the retarded and advanced Green functions were defined in Eq.~(\refeq{eq:gra-def})
with $G^\mathrm{R}(t,t')$ [$G^\mathrm{A}(t,t')$] being nonzero only for $t\ge t'$ ($t\le t'$). Thus, $G_{ij}^\gtrless(t, t')$ are expressed via their values on the time diagonal, i.e. the density matrix [cf. Eq.~(\refeq{eq:g-glsymm})]. 
Alternatively, the individual functions $G^{\mathrm{R/A}}$ can be eliminated in favor of their difference
\begin{align}
 \mathcal{U}_{ij}(t,t') &= G_{ij}^\mathrm{R}(t,t') - G_{ij}^\mathrm{A}(t,t')\,, \label{eq:prop_b}
\end{align}
with the following value on the time diagaonal 
\begin{align}
 \mathcal{U}_{ij}(t,t) &= 
 G_{ij}^>(t) - G_{ij}^<(t)
 = \frac{1}{\i\hbar} \delta_{ij}\, . \label{eq:unity_prop}\
\end{align}
$\mathcal{U}(t,t')$ is a time evolution operator that does not contain a $\Theta$-function and allows us to rewrite the GKBA in the following form
 \begin{align}
      G_{ij}^\gtrless(t' \leq t) &=\i \hbar \sum_k G_{ik}^\gtrless(t') \, \mathcal{U}_{kj}(t',t)\,, \label{eq:GKBA_propa}\\
     G_{ij}^\gtrless(t\geq t') &=\i \hbar \sum_k \mathcal{U}_{ik}(t,t')\, G_{kj}^\gtrless(t')\,. \label{eq:GKBA_propb}
 \end{align}

On the other hand, the retarded and advanced functions in \refeqn{eq:GKBA} and  $\mathcal{U}$ are still depending on two time arguments. We solve this problem by using the Hartree--Fock approximation

\begin{align}
    \mathcal{U}_{ij}(t,t') = \frac{1}{\i\hbar}\exp{\left\{-\frac{1}{\i\hbar}\int_{t'}^t d{\bar t} \,h^{\rm HF}(\bar t)\right\}}\bigg|_{ij}\,,
    \label{eq:u-hf}
\end{align}
which gives rise to the Hartree--Fock-GKBA \cite{balzer-book,hermanns_psc_12, hermanns_prb14}.
On the other hand, using the Hartree--Fock result (\refeq{eq:u-hf}), expressions (\refeq{eq:GKBA_propa}) and (\refeq{eq:GKBA_propb}) solve the time-diagonal KBE (\refeq{eq:eom_gone}) in the collisionless limit, $I\to 0$. Using the HF-GKBA allows us to solve the time-diagonal KBE (\refeq{eq:eom_gone}) with collisions included in a perturbative manner: in the collision integral all functions $G^\gtrless(t,t')$ are replaced by $G^{\gtrless}(t,t')|_{\rm HF-GKBA}$, using Eqs.~(\refeq{eq:GKBA_propa}, \refeq{eq:GKBA_propb}, \refeq{eq:u-hf}).

The HF-GKBA has a number of attractive properties \cite{bonitz_qkt,hermanns_prb14}. It retains total energy conservation and time reversibility. Correlation effects are fully included in the time-diagonal values of the Green function, $G^<(t)$, but cannot be recovered from the off-diagonal components, $G^<(t\ne t')$. Finally, restricting the time propagation to the diagonal reduces the computational effort to $N_{\tn t}^2$, because the memory integration in the collision integral (\refeq{eq:tiagonal-i-sigmas}) has still to be carried out at each time step. 

In the following we demonstrate how this memory integration can be eliminated which leads to the G1--G2 scheme that scales as $N_{\tn t}^1$.

\subsection{G1--G2 scheme for the Second-order Born selfenergy}\label{ss:g1-g2-soa}
We start by considering the simplest selfenergy beyond Hartree--Fock---the second-order Born approximation (SOA), cf. Sec.~\ref{ss:soa}. Then, the collision integral of the time-diagonal equation \eqref{eq:definition_D} transforms into:
\begin{align}
 I^{\rm SOA}_{ij}(t) =
 & \pm\left(\i\hbar\right)^2 \sum_{klpqrsu} \, w_{iklp}\left(t\right) \int_{t_0}^t \d\cbar t\, w^\pm_{qrsu}\left(\cbar t\right) \times \\
 & \times\Big[ G^>_{lq}\left(t,\cbar t\right) G^>_{pr}\left(t,\cbar t\right) G^<_{uk}\left(\cbar t,t\right) G^<_{sj}\left(\cbar t,t\right) \nonumber \\
 &\quad - G^<_{lq}\left(t,\cbar t\right) G^<_{pr}\left(t,\cbar t\right) G^>_{uk}\left(\cbar t,t\right) G^>_{sj}\left(\cbar t,t\right) \Big] \nonumber \\
 = \pm & \left(\i\hbar\right)^2 \sum_{klpqrsu} \, w_{iklp}\left(t\right) \int_{t_0}^t \d\cbar t\, w^\pm_{qrsu}\left(\cbar t\right) \times \\
 & \times \Big[ \mathcal{G}^{\tn{H},>}_{lpqr}(t,\cbar t) \mathcal{G}^{\tn{H},<}_{sujk}(\cbar t, t) - \mathcal{G}^{\tn{H},<}_{lpqr}(t,\cbar t) \mathcal{G}^{\tn{H},>}_{sujk}(\cbar t, t) \Big] \nonumber \\
 = \pm & \left(\i\hbar\right)^2 \sum_{klpqrsu} \, w_{iklp}\left(t\right) \int_{t_0}^t \d\cbar t\, w^\pm_{qrsu}\left(\cbar t\right) \times \\
 & \times \Big[ \mathcal{G}^{\tn{F},>}_{ljqs}(t,\cbar t) \mathcal{G}^{\tn{F},<}_{urkp}(\cbar t, t) - \mathcal{G}^{\tn{F},<}_{ljqs}(t,\cbar t) \mathcal{G}^{\tn{F},>}_{urkp}(\cbar t, t) \Big] \, ,
\end{align}
where, in the second and third expressions we used the two-particle Hartree and Fock Green functions defined in Eqs.~\eqref{eq:g2h-def} and \eqref{eq:g2f-def}, respectively.

Using \refeqn{eq:definition_D} we can identify the correlated part of the two-particle Green function, $\mathcal{G}$  in SOA,
\begin{align}
 \mathcal{G}^{\rm SOA}_{ijkl}(t) =& \i\hbar\sum_{pqrs} \int_{t_0}^t \d\cbar t\, w^\pm_{pqrs}\left(\cbar t\right) \times \label{eq:g2-integral}
\\
 & \times\Big[ \mathcal{G}^{\tn{H},>}_{ijpq}(t,\cbar t) \mathcal{G}^{\tn{H},<}_{rskl}(\cbar t, t) - \mathcal{G}^{\tn{H},<}_{ijpq}(t,\cbar t) \mathcal{G}^{\tn{H},>}_{rskl}(\cbar t, t) \Big] \, . \nonumber
\end{align}


Now we apply the HF-GKBA, using  \refeqns{eq:GKBA_propa}{eq:GKBA_propb}, for the Green functions (taking into account that in the collision integral only $G^>(t\geq \cbar{t})$ and $G^<(\cbar{t}\leq t)$ appear), we reformulate \refeqn{eq:g2-integral} for the two-particle Green function on the time diagonal within the HF-GKBA \cite{joost_prb_20}
\begin{align}\label{eq:dtilde-solution-short}
 \mathcal{G}^{\rm SOA}_{ijkl}(t) &= \left(\i\hbar\right)^3 \sum_{pqrs}\int_{t_0}^t \mathrm{d}\cbar{t}\, 
  \mathcal{U}_{ijpq}^{(2)}(t,\cbar{t}) 
  \Psi^\pm_{pqrs}(\cbar{t})
  \mathcal{U}_{rskl}^{(2)}(\cbar{t},t)\,,\phantom{...}
\end{align}
where we introduced the short notations for the two-particle evolution operators $\mathcal{U}^{(2)}$ and the occupation factors $\Psi^\pm$, for which we give two equivalent expressions,
\begin{align}
  \mathcal{U}_{ijkl}^{(2)}(t,t') &= \mathcal{U}_{ik}(t,t') \mathcal{U}_{jl}(t,t') = \mathcal{U}_{jilk}^{(2)}(t,t')\,, \label{prop_two}
  \\
    \Psi^\pm_{ijkl}(t) &= \left(\i\hbar\right)^2\sum_{pqrs} w^\pm_{pqrs}(t)\left\{
    \mathcal{G}^{\tn{H},>}_{ijpq} \mathcal{G}^{\tn{H},<}_{rskl}
    - (>\leftrightarrow <)
    \right\}_t\,,
    \label{eq:psi-pm-def}
\\
&= \left(\i\hbar\right)^2\sum_{pqrs} w^\pm_{pqrs}(t)\left\{
    \mathcal{G}^{\tn{F},>}_{irkp} \mathcal{G}^{\tn{F},>}_{jslq}
    - (>\leftrightarrow <)
    \right\}_t\,.
\end{align}
The superscript ``$\pm$'' indicates that exchange effects are included which enter via the (anti-)symmetrized potential $w^\pm$.
The function $\Psi^\pm(\bar t)$ has the meaning of pair correlations produced in the system at time $\bar t$ via two-particle scattering per unit time. These correlations are time evolved from $\bar t$
 to $t$ by the evolution operators $\mathcal{U}_{ijkl}^{(2)}$, cf. expression (\refeq{eq:dtilde-solution-short}).
  The appearance of two propagators indicates that $\mathcal{G}^{\rm SOA}_{ijkl}(t)$ does not obey a Schrödinger-type equation but a commutator (Heisenberg--von Neumann) equation, that we present in the following.
  
  Indeed, a straightforward calculation reveals \cite{bonitz_qkt, schluenzen_19_prl,joost_prb_20} that the time-diagonal two-particle Green function (\refeq{eq:dtilde-solution-short}) in SOA obeys the following ordinary differential equation 
\begin{align}
 \i\hbar  \frac{\d}{\d t} \mathcal{G}^{\rm SOA}_{ijkl}(t) &- \Big[ h^{(2),\tn{HF}}(t),\mathcal{G}^{\rm SOA}(t) \Big]_{ijkl} = \Psi^\pm_{ijkl}(t)
 \, .\quad  \label{eq:G1-G2_soa}
 \\
 h^{(2),\tn{HF}}_{ijkl}(t) &= h^\tn{HF}_{ik}(t)\delta_{jl} + h^\tn{HF}_{jl}(t)\delta_{ik}\,.
 \label{eq:h2-hf}
\end{align}
%
Equations \eqref{eq:G1-G2_soa} and \eqref{eq:eom_gone}  constitute 
a closed system of time-local differential equations, for which the computational effort of a numerical implementation scales linearly with time. The present result is obtained for a general single-particle basis. Special cases, such as the Hubbard basis or a momentum basis were discussed in detail in Ref.~\cite{joost_prb_20}.

In similar manner as for the SOA selfenergy, a time-local equation for $\mathcal{G}^\GKBA$ corresponding to more advanced selfenergies can be derived for which the speedup of the G1--G2 scheme is even larger. This is discussed in the next section.

\subsection{G1--G2 equations for selfenergies beyond the second-order Born approximation}\label{ss:g1g2-beyond-soa}
Starting from improved correlation selfenergies beyond SOA, as were presented in Sec.~\ref{s:sigmas}, the derivation of the G1--G2 scheme can be repeated. This was done in great detail in Ref.~\cite{joost_prb_20}, so here we summarize the results for the particle--particle $T$ matrix (including exchange diagrams), the particle--hole $T$ matrix and for the $GW$ approximation. For the latter two no conserving exchange selfenergies are available. At the same time, this problem can be solved within the alternative reduced density operator formalism, as will be demonstrated in Sec.~\ref{s:do-test}.

\subsubsection{Particle--particle $T$-matrix selfenergy}\label{sss:tpp}
Starting from the $T$-matrix approximation for the selfenergy, Eq.~(\refeq{eq:sigma-tpp}), the equation for the time-diagonal two-particle Green function becomes (we suppress the time dependencies) \cite{joost_prb_20}
\begin{align}
\label{eq:G1-G2_tpp_compact}
 \i\hbar  \frac{\d}{\d t} \mathcal{G}^{\rm TPP}_{ijkl} &- \Big[ h^{(2),\tn{HF}},\mathcal{G}^{\rm TPP} \Big]_{ijkl} = \Psi^\pm_{ijkl} + L_{ijkl}
 \\
 L_{ijkl} &\coloneqq \sum_{pq}\Big\{ \mathfrak{h}^{L}_{ijpq} \mathcal{G}^{\rm TPP}_{pqkl} 
- \mathcal{G}^{\rm TPP}_{ijpq} \Big[\mathfrak{h}^{L}_{klpq}\Big]^* \Big\}\,,
\label{eq:l-definition}
\\
 \mathfrak{h}_{ijkl}^{L} &\coloneqq 
\left(\i\hbar\right)^2 \sum_{pq}   \left[ \mathcal{G}^{\tn{H},>}_{ijpq}  - \mathcal{G}^{\tn{H},<}_{ijpq}  \right]w_{pqkl}  \,,
\label{eq:h_omega-pp} 
\end{align}
This equation differs from the SOA case by the ladder term $L$ which can be further transformed, using the identities \begin{align}
(\pm\i\hbar)^2    \left(\mathcal{G}^{\tn{H},>}_{ijrs}  -
\mathcal{G}^{\tn{H},<}_{ijrs}\right) &= (\delta_{ir}\pm n_{ir})(\delta_{js}\pm n_{js})-n_{ir}n_{js}
\nn\\
&=\delta_{ir}\delta_{js}\pm \delta_{js}n_{ir}\pm \delta_{ie}n_{js} \coloneqq N_{ijrs}
\label{eq:g2h-transform}
\end{align}
with the result for the ladder term
\begin{align}
    L_{ijkl} &= 
    \sum_{rs} \sum_{pq}\Big\{
N_{ijrs} w_{rspq}\mathcal{G}_{pqkl} 
\nn\\
&\qquad\qquad 
- \mathcal{G}_{ijpq}w_{pqrs}N_{rskl} 
 \Big\}\,.
\label{eq:ladder-final}
\end{align}
Note that, while the inhomogeneity, $\Psi^\pm$, involves the (anti-)symmetric potential, $w^\pm$, the ladder term contains the bare interaction potential $w$.

\subsubsection{Particle--hole $T$-matrix selfenergy}\label{sss:tph}
Starting from the particle--hole $T$-matrix approximation for the selfenergy, Eq.~(\refeq{eq:sigma-tph}), the equation for the time-diagonal two-particle Green function becomes \cite{joost_prb_20}
\begin{align}
 \i\hbar  \frac{\d}{\d t} \mathcal{G}^{\rm TPH}_{ijkl} &- \Big[ h^{(2),\tn{HF}},\mathcal{G}^{\rm TPH} \Big]_{ijkl} = \Psi_{ijkl} 
 +\Lambda^{\rm PH}_{ijkl}
\label{eq:G1-G2_tph_compact}
 \\
     \Lambda^{\rm PH}_{ijkl} &\coloneqq \sum_{pq}
     \Big\{ 
     \mathfrak{h}^{\Lambda}_{ipql}\, \mathcal{G}^{\rm TPH}_{qjkp} -
     \mathcal{G}^{\rm TPH}_{ipql}\, \Big[\mathfrak{h}^\Lambda_{kpqj}\Big]^*
     \Big\}
     \,,
\label{eq:lambda-definition}
\\
 \mathfrak{h}_{ijkl}^{\Lambda} &\coloneqq 
\left(\i\hbar\right)^2 \sum_{pq}   \left[ \mathcal{G}^{\tn{F},>}_{iqlp}  - \mathcal{G}^{\tn{F},<}_{iqlp}  \right]w_{pjkq}  \,.
\label{eq:h_omega-ph} 
\end{align}
Compared to the particle--particle $T$-matrix case, the ladder term $L$ is replaced by the term $\Lambda^{\rm PH}$ and the inhomogeneity $\Psi$ is defined with the original interaction matrix element [as opposed to the (anti-)symmetrized $w^\pm$ of Eq.~\eqref{eq:psi-pm-def}].

\subsubsection{$GW$ selfenergy}\label{sss:gw}
The equation for the time-diagonal Green function in $GW$ approximation is \cite{joost_prb_20}
\begin{align}
 \i\hbar  \frac{\d}{\d t} \mathcal{G}^{\rm GW}_{ijkl} &- \Big[ h^{(2),\tn{HF}},\mathcal{G}^{\rm GW} \Big]_{ijkl} = \Psi_{ijkl} +  \Pi_{ijkl}\,,
\label{eq:G1-G2_gw_compact} 
 \\
 \Pi_{ijkl} &\coloneqq \sum_{pq}
 \Big\{ 
 \mathfrak{h}^{\Pi}_{qjpl} \,\mathcal{G}^{\rm GW}_{ipkq} 
 - \mathcal{G}^{\rm GW}_{qjpl} \,\left[\mathfrak{h}^{\Pi}_{qkpi}\right]^*
\Big\}
\,,
\label{eq:pi-definition}
\\
 \mathfrak{h}_{ijkl}^{\Pi} &\coloneqq 
\pm \left(\i\hbar\right)^2 \sum_{pq}  w_{qipk} \left[ \mathcal{G}^{\tn{F},>}_{jplq}  - \mathcal{G}^{\tn{F},<}_{jplq}  \right]  \,,
\label{eq:h_pi} 
\end{align}
where $\Pi$ denotes the polarization terms (ring diagrams). 
Note that, in contrast to the SOA case, this equation contains the inhomogeneity $\Psi$ without (anti-)symmetrization (as in the case of TPH), which is a consequence of the starting diagrams in Fig.~\ref{fig:diagrams_resummation}.
It is interesting to note that \refeqns{eq:pi-definition}{eq:h_pi} can be reformulated by transforming the difference of Fock Green functions:
\begin{align}
(\pm\i\hbar)^2    \left(    \mathcal{G}^{\tn{F},>}_{jplq}  -
\mathcal{G}^{\tn{F},<}_{jplq}\right) &= (\delta_{jq}\pm n_{jq}) n_{pl}-n_{jq}(\delta_{pl}\pm n_{pl})
\nn\\
&=\delta_{jq}n_{pl} -\delta_{pl}n_{jq} \coloneqq M_{jplq}  \,.
\label{eq:g2f-transform}
\end{align}
Using \refeqn{eq:g2f-transform} the effective Hamiltonian, \refeqn{eq:h_pi}, and the polarization term, \refeqn{eq:pi-definition}, become
\begin{align}
     \mathfrak{h}^{\Pi}_{ijkl} & = 
     \sum_{pq} w_{iqkp} 
      M_{jplq}
     \nonumber \\
    \Pi_{ijkl} &=
\sum_{pqrs}
 \Big\{ 
 w_{qspr}M_{jrls} \,\mathcal{G}^{\rm GW}_{ipkq} 
 - \mathcal{G}^{\rm GW}_{qjpl} \,M_{slrk}w_{prqs}
\Big\}.\quad
\label{eq:pipm-m}
\end{align}

\subsubsection{Combining $GW$ and $T$-matrix selfenergy contributions}\label{sss:dsl}
An important task in many-body physics is to combine strong coupling ($T$-matrix contributions) and dynamical screening ($GW$ with exchange) which corresponds to the dynamically screened ladder approximation (DSL), as we discussed in Sec.~\ref{ss:negf-dsl}. Within nonequilibrium Green functions this would require to combine the respective selfenergies. For this problem, a perturbative solution has been reported: the third-order approximation (TOA) that includes ladder and polarization diagrams up to third order in the interaction, as was discussed in Sec.~\ref{ss:toa}.
However, beyond this perturbative result, at the moment no closed selfenergy approximation that corresponds to DSL and includes diagrams of all orders, is known. Furthermore, there is significant asymmetry in the approximations: while for the particle--particle $T$ matrix exchange diagrams are known, no such terms are available for the particle--hole and $GW$ selfenergies.

We, therefore, will proceed differently and approach this problem from the side of the reduced density operator theory which provides an independent approach to the G1--G2 scheme. As we will see this approach provides answers to the questions above. 

\section{Comparison of the G1--G2 scheme to reduced density operator theory}\label{s:do-test}
In this section we recall the basic ingredients of nonequilibrium reduced density operator (density matrix, TD2RDM) theory \cite{bonitz_qkt}. Thereby we will retain the compact operator notation as it allows one to more easily understand the physical meaning of the individual terms in the equations of motion. We start by introducing the $N$-particle nonequilibrium density operator, $\hat \rho_N$, and by defining the reduced density operators, $\hat F_1$, $\hat F_{12}$, $\hat F_{123}$ etc., in Sec.~\ref{ss:rdm-definition}. There we summarize the equations of motion of the latter---the BBGKY-hierarchy. After this, in Sec.~\ref{ss:cluster-expansion} we introduce the cluster (or cumulant) expansion of the density operators and the binary (ternary) correlation operators $\hat g_{12}$ ($\hat g_{123}$), and we present the equation of motion of $\hat g_{12}$. The resulting equation allows us to directly identify the terms that correspond to the selfenergy approximations of NEGF theory that were introduced in Sec.~\ref{s:sigmas}, including the 
$T$-matrix and $GW$ approximations.
Finally, we perform the (anti-)symmetrization of the reduced density and correlation operators and of their equations of motion, cf. Sec.~\ref{ss:bbgky-antisymmetrization}.
In the concluding Sec.~\ref{ss:g1-g2-rdm-comparison} we establish the correspondence between the quantities of reduced density matrix theory and of NEGF and compare their equations of motion.

\subsection{Definitions of reduced density operators. The case of spinless particles}\label{ss:rdm-definition}
We consider a generic quantum many-particle system with the Hamiltonian $\hat H_N$ that is subject to external baths and, thus requires a mixed state description. This is done with the $N$-particle density operator that is composed from solutions of the $N$-particle Schrödinger equation
\begin{align}
    \i\hbar\frac{\partial}{\partial t} |\psi_N^{(a)}\rangle &= \hat H_N  |\psi_N^{(a)}\rangle\,,
\quad \hat H_N = \sum_{i=1}^N \hat H_i +\frac{1}{2}\sum_{i\ne j}\hat V_{ij}\\\
    \hat \rho_N &= \sum_a p_a |\psi_N^{(a)}(t)\rangle \langle \psi_N^{(a)}(t)|\,,\quad \mbox{Tr}_{1\dots N} \hat \rho_N = 1\,,
    \label{eq:rho-n-def}
    \\
    0 &\le p_a\le 1, \qquad \sum_a p_a = 1\,,
\end{align}
where $p_a$ are real non-negative probabilities.
From the $N$-particle density operator one computes $s$-particle reduced density operators ($s=1, 2, \dots N-1$)
\begin{align}
    \hat F_{1\dots s} = \frac{N!}{(N-s)!}\mbox{Tr}_{s+1\dots N} \hat \rho_N\,,\quad \mbox{Tr}_{1\dots s} \hat F_{1\dots s} = \frac{N!}{(N-s)!}\,,
    \label{eq:fs-def}
\end{align}
where, in the following, we will skip the ``hat'' of the operators. 
From the equation of motion for $\hat \rho_N$---the von Neumann equation, with the initial condition $\rho_N^0$, 
\begin{align}\label{eq:von-neumann-1}
    \i\hbar \frac{\d}{\d t}\rho_N &- [H_N,\rho_N] = 0\,,\\
    \rho_N(t_0) &= \rho_N^0 \equiv \sum_k p_a\,| \psi^{(a)}(t_0)\rangle \langle \psi^{(a)}(t_0)|\,,
    \nonumber
\end{align}
one readily derives the equations for the reduced density operators---the quantum BBGKY-hierarchy \cite{bonitz_qkt},
\begin{align}
\label{eq:bbgky_finite}
\i\hbar\frac{\d}{\d t} F_{1\dots s}-[H_{1\dots s},F_{1\dots s}] &=
 \mbox{Tr}_{s+1}[V^{(1\dots s),s+1},F_{1\dots s+1}]\,,\\
  V^{(1\dots s),s+1} &= \sum_{\alpha=1}^s V^{\alpha,s+1},\;
  \label{eq:V1s-s+1}
 \\
 F_{1\dots s}(t_0) = F_{1\dots s}^0 &= \frac{N!}{(N-s)!}\mbox{Tr}_{s+1\dots N}\,\rho^0\,.
\end{align}
Here,  $H_{1\dots s}$ is the $s$-particle Hamilton operator which
follows from $H_{1\dots N}$ by substituting $N\rightarrow s$.
The equations of the hierarchy differ from the von Neumann equation
due to the terms on the r.h.s, which contain the coupling of the
$s$ particles to the remainder of the system via all possible binary
interactions. The complete hierarchy is, obviously, equivalent to the von Neumann equation and, therefore, has the same properties. In particular, the system (\refeq{eq:bbgky_finite}) is time reversible and conserves the total energy.

\subsection{Cluster expansion. Correlation operators.}\label{ss:cluster-expansion}
To derive approximations and decouple the hierarchy, it is useful to introduce correlation operators that are defined via the cluster expansion,
\begin{align}
F_{12}&=F_{1}F_{2}+g_{12}\,,
\label{g12-def}
\\
F_{123}&=F_{1}F_{2}F_{3}+g_{23}F_{1}+g_{13}F_{2}+g_{12}F_{3}+ \dots + g_{123}\,,
\label{g123-def}
\\
& \dots &
\nn
\end{align}
In Eq.~(\refeq{g12-def}) we have introduced the pair correlation operator, $g_{12}$, which is the deviation of the two-particle density operator from its uncorrelated (Hartree) part, $F_1\, F_2$. Note that exchange corrections will be recovered from an (anti-)symmetrization procedure in Sec.~\ref{ss:bbgky-antisymmetrization}. Similarly, $g_{123}$ is the three-particle correlation operator. 

We now rewrite the equation of motion for $F_1$ in terms of $g_{12}$ and also derive the equation of motion for $g_{12}(t)$, starting from the BBGKY-hierarchy---\refeqn{eq:bbgky_finite} for $s=2$---and subtracting the first equation for $F_1\, F_2$ \cite{bonitz_qkt},
\begin{align}
\i \hbar\frac{\d}{\d t} F_{1} - [{\bar H_1},F_1]
=&
I^0_{1, \rm DO}\,,
\label{eq:f1-equation-g}
\\
\i\hbar \frac{\d}{\d t}
g_{12} - [{\bar H}^0_{12}, g_{12}]
=& \Psi^0_{12,\rm DO} + L^0_{12,\rm DO}
\label{eq:g12-equation}\\
&+ \Pi^0_{12, \rm DO} +
\mbox{Tr}_{3}[V^{(12),3},g_{123}]
\,.
\nonumber
\end{align}
Here we have introduced the one-particle and two-particle mean-field Hamilton operators, $\bar H_1$ and $\bar H^0_{12}$ and the operator of the Hartree mean-field energy $U_1^{\rm H}$,
\begin{align}
    \bar H_1 &= H_1 + U_1^{\rm H}\,,
    \label{eq:hbar1}\\
    U_1^{\rm H} &= \mbox{Tr}_2 V_{12}F_2\,,
    \label{eq:uhartree}\\
    \bar H_{12} &= \bar H^0_{12} + V_{12}\,,
    \label{eq:hbar12-0}\\
    \bar H^0_{12} &= \bar H_1 + \bar H_2\,.
    \label{eq:hbar12}
\end{align}
The other notations are as follows:
\begin{align}
    I^0_{1,\rm DO} &= \mbox{Tr}_{2}[V_{12},g_{12}]\,,
    \label{eq:col-int-rdo-0}\\
    \Psi^0_{12,\rm DO} &= [V_{12},F_{1}F_{2}]\,,
    \label{eq:psi-rdo-0}\\
    L^0_{12,\rm DO} &= [V_{12},g_{12}]\,,
    \label{eq:l-rdo-0}\\
    \Pi^0_{12,\rm DO} &= \Pi^{0(1)}_{12,\rm DO } +  \Pi^{0(2)}_{12,\rm DO}\,,
    \label{eq:pi-rdo-0}
    \\
    \Pi^{0 (1)}_{12,\rm DO} &= \mbox{Tr}_{3}
    [V_{13},F_1 g_{23}]\,,
    \label{eq:pi-rdo-01}
\end{align}
where $\Pi_{12,\rm DO}^{0 (2)}$ follows from $\Pi_{12,\rm DO}^{0 (1)}$ by exchanging $(1 \leftrightarrow 2)$.
Furthermore, $I^0_{\rm DO}$ denotes the operator form of the collision integral, $\Psi^0_{\rm DO}$ is the inhomogeneity in the $g_{12}$-equation, $L^0_{\rm DO}$ denotes the ladder terms, and $\Pi^0_{\rm DO}$ the polarization terms. 
A detailed discussion of the physics behind each of these terms can be found in Ref.~\cite{bonitz_qkt}.

In expressions (\refeq{eq:col-int-rdo-0}), (\refeq{eq:psi-rdo-0}), (\refeq{eq:l-rdo-0}), and (\refeq{eq:pi-rdo-0}) we deliberately chose notations resembling those of the G1--G2 scheme. The subscript ``DO'' stands for density operator approach and superscript ``0'' for the spinless limit.

\subsection{Many-body approximations}\label{ss:do-many-body}
Equations (\refeq{eq:f1-equation-g}) and (\refeq{eq:g12-equation}) have a one to one correspondence to the G1--G2 equations that were derived within NEGF theory for given selfenergy approximations, if three-particle correlations are neglected, in the limit of spinless particles. In particular, 
\begin{description}
\item[i.)] $I^0_{\rm DO} \to 0$ recovers the time-dependent Hartree approximation;
\item[ii.)] $L^0_{12,\rm DO}=\Pi^0_{12,\rm DO}\to 0$ is equivalent to the second-order approximation, Eq.~\eqref{eq:G1-G2_soa};
\item[iii.)] $\Pi^0_{12,\rm DO} \to 0$ and $L^0_{12,\rm DO} \ne 0$ is equivalent to the particle--particle $T$-matrix approximation, Eq.~\eqref{eq:G1-G2_tpp_compact}.
\item[iv.)] $L^0_{12,\rm DO} \to 0$ and $\Pi^0_{12,\rm DO} \ne 0$ is equivalent to the $GW$ approximation, Eq.~\eqref{eq:G1-G2_gw_compact}.
\item[v.)] $L^0_{12,\rm DO} \ne 0$ and $\Pi^0_{12,\rm DO} \ne 0$ is equivalent to the DSL approximation.
\end{description}
Note that, without (anti-)symmetrization, density operator theory does not yield the particle--hole $T$-matrix terms of NEGF theory. On the other hand, including both the ladder and polarization terms, i.e. $\Pi^0_{12,\rm DO} \ne 0$ and $L^0_{12,\rm DO} \ne 0$ simultaneously, yields the DSL approximation (so far, without exchange terms) for which no NEGF correspondence exists. 
These issues will be discussed in detail for the (anti-)symmetrized RDO equations. There we also will establish an exact correspondence between the 
respective terms of the two approaches.

\subsection{Energy conservation and trace consistency}\label{ss:econs-cc}
The exact solution of the BBGKY-hierarchy \refeqn{eq:bbgky_finite} conserves density, momentum and total energy. Approximate solutions to the hierarchy should, therefore, also satisfy these constraints, so we briefly discuss this issue in the following. The approximations that were discussed in Sec.~\ref{ss:do-many-body} correspond to the Hartree, second order Born, T-matrix and GW approximation of NEGF theory (this correspondence will be shown explicitly in Sec.~\ref{ss:g1-g2-rdm-comparison}) which are all known to be conserving, e.g. \cite{baym_kadanoff_conservation}. Upon reducing the two-time equations of NEGF theory to single-time equations via the Hartree-Fock GKBA these conservation properties are maintained, as was shown by Hermanns \textit{et al.} \cite{hermanns_prb14}. Therefore, the independent reduced density operator approach, within approximations i)--iv) of Sec.~\ref{ss:do-many-body} should also be conserving. Indeed, this is straightforward to show \cite{bonitz_qkt}, and we reproduce the criterion for a conserving approximations in Appendix~\ref{app:econs}. This criterion consists in the permutation symmetry of the three-particle reduced density operator, 
\begin{align}
F_{123}(t)=F_{132}(t)=\dots\,, \quad \mbox{for all times.}
\label{eq:f123-symmetry}
\end{align}
From this it is easy to see that also approximation v.)---the dynamically screened ladder approximation---is conserving as well, because it is equivalent to neglecting three-particle correlations, $g_{123} \to 0$ which obey the symmetry \refeqn{eq:f123-symmetry}.

Another important property of the above approximations is that they preserve time reversal invariance of the exact BBGKY-hierarchy \cite{bonitz_cpp18} and of the underlying hierarchy of the nonequilibrium Green functions \cite{scharnke_jmp17}. Reversibility is lost only if the Markov limit is enforced (Fermi's golden rule). For more details and a discussion of the relevant time scales, we refer to refs. \cite{bonitz_cpp18, bonitz96pla, bonitz_qkt}.

Aside from conservation properties there exist additional constraints on the solution of the many-body problem. If this solution is expressed via the reduced density operators $F_1$, $F_{12}$, $F_{123}$ and so on there exist consistency constraints between them. The reason is that they are all derived from the same $N-$particle density operator $\rho_N$ obey the BBGKY-hierarchy \refeqn{eq:bbgky_finite} that is equivalent to the von Neumann equation for $\rho_N$. From the definition of the reduced density operators, \refeqn{eq:fs-def} immediately follow relations between any pair of RDO, in particular \begin{align}
    F_1 &= \frac{1}{N-1}\mbox{Tr}_2 F_{12}\,, \label{eq:cc-f1-f12}\\
    F_{12} &= \frac{1}{N-2}\mbox{Tr}_3 F_{123}\,. \label{eq:cc-f12-f123}
\end{align}
The related issues of contraction consistency and N-representability have been studied in detail in Refs.~\cite{lacknerphd,coleman2000reduced,mazzioti_reduced_2007,akbari_prb_12}.
One readily verifies that relation \refeqn{eq:cc-f1-f12} is a consequence of \refeqn{eq:cc-f12-f123}, if, in addition, the three-particle density operator obeys permutation symmetry, i.e., \refeqn{eq:f123-symmetry} is fullfilled.

It is straightforward to show that approximations i)--v) are not trace consistent. Consider, as an example, the Hartree approximation, $F_{12}=F_1\, F_2$. Inserting this into~\refeqn{eq:cc-f1-f12} yields $\frac{N}{N-1}F_1$, on the r.h.s., violating this relation. Even though the correct results is recovered if exchange is restored (Hartree--Fock approximation), cf. Sec.~\ref{ss:bbgky-antisymmetrization}, this holds only for fermions in the ground state when the one-particle density matrix is idempotent. Similar behavior is observed for the other many-particle approximations. To restore trace consistency, we will consider modified approximations in Sec.~\ref{ss:cc}.

\subsection{(Anti-)Symmetrization of the reduced density operators}\label{ss:bbgky-antisymmetrization}
Until now, we have used, as a starting point for constructing the density operators, the solution of the Schrödinger equation without (anti-)symmetrization. Of course, this completely neglects spin and exchange effects, including Pauli blocking.

We now restore the spin statistics, in all expressions, by applying an (anti-)symmetrization procedure to all density operators that was proposed by Boercker and Dufty \cite{dufty-etal.97}, for details see Ref.~\cite{bonitz_qkt}.
The result is that all two- and three-particle operators are (anti-)symmetrized according to
\begin{align}
    F_{12} &\to F_{12}^\pm = F_{12}\lambda^\pm_{12}\,,
    \nn\\
    g_{12} &\to g_{12}^\pm = g_{12}\lambda^\pm_{12}\,,
    \nn\\
    F_{123} &\to F_{123}^\pm = F_{123}\lambda^\pm_{123}\,,
    \nn\\
    g_{123} &\to g_{123}^\pm = g_{123}\lambda^\pm_{123}\,,
\end{align}
and so on,
where the (anti-)symmetrization operators $\lambda^\pm$ are expressed in terms of pair permutation operators $\hat P_{ij}$, with (1, 2, 3, i, j are particle indices)
\begin{align}
    \hat P_{12} |12\rangle &= |21\rangle \,,
    \nn\\
    \lambda^\pm_{12} &= \hat 1 + \epsilon \hat P_{12}\,,
\nn\\
    \lambda^\pm_{123} &=     \lambda^\pm_{12} ( \hat 1 + \epsilon \hat P_{13} + \epsilon \hat P_{23})\,,
\end{align}
where $\epsilon=+1$, for bosons, $\epsilon=-1$, for fermions and $\epsilon=0$, for spinless particles. A number of important properties of the operators $\hat P_{ij}$ and $\lambda^\pm$ are presented in appendix~\ref{app:g12-equation}.

We now perform the (anti-)symmetrization of the first 
 hierarchy equation with the result \cite{bonitz_qkt,dufty-etal.97}
\begin{align}
    \i\hbar \frac{\d}{\d t} &F_{1} - \left[{H}_{1}+U_1^{\rm HF},F_{1}\right] =
    \mbox{Tr}_2 [V_{12},g^\pm_{12}]=I_{1,\rm DO}\,.\quad
    \label{eq:bbgky1-as}
\end{align}
Here we introduced the (anti-)symmetrized versions of the collision integral and of the Hartree mean field, i.e. the Hartree--Fock potential energy operator,
\begin{align}
    U_1^{\rm H} \to U_1^{\rm HF} &= \mbox{Tr}_2 V_{12}^\pm F_2\,,
    \label{eq:uhf-def}\\
    V_{12} \to V_{12}^\pm &= V_{12}\lambda_{12}^\pm\,.
    \label{eq:vpm-def}
\end{align}
As we will see below, the operator (\refeq{eq:uhf-def}) is directly related to the Hartree--Fock selfenergy whereas the (anti-)symmetrized pair potential (\refeq{eq:vpm-def}) coincides with the previously introduced $w^\pm$.

Let us now turn to the (anti-)symmetrization of the equation of motion for the pair correlation operator $g^\pm_{12}$. We start by introducing the following definitions
\begin{align}
    \bar{H}^0_{12} &= H_1 + H_2 + U^\tn{HF}_1 + U^\tn{HF}_2\,,\\
    \hat V_{12} &= N_{12} \, V_{12}\,,
    \label{eq:vhat}\\
    \hat V^\pm_{12} &= N_{12} \, V^\pm_{12}\,,
    \\
    N_{12} &= \hat 1 +\epsilon F_1 +\epsilon F_2\,.
    \label{eq:n12}
\end{align}
$\bar H_{12}^0$ is the (anti-)symmetrized generalization of the previous definition, and $N_{12}$ is the familiar two-particle Pauli blocking (Bose enhancement) factor. Note that, while the operators $V_{12}, F_1, F_{12}, g_{12}$ and so on are hermitian,  the potential $\hat V_{12}$ is not: $\hat V_{12}^\dagger = V_{12}N_{12}$.

To derive the equation of motion for $g_{12}^\pm$ we start from the (anti-)symmetrized version of the second hierarchy equation where we introduced the cluster expansions of $F_{12}$ and $F_{123}$ and factorized the operator $\lambda_{123}^\pm$:
\begin{align}
\nn
&\i\hbar \frac{\d}{\d t}\Big(F_{1}F_{2}\lambda^{\pm}_{12}+g^\pm_{12}\Big)
-[H^0_{12}+V_{12},F_{1}F_{2}\lambda^{\pm}_{12}+g^\pm_{12}]
\\
&=
\mbox{Tr}_{3}\Big\{[V_{13}+V_{23},F_{1}F_{2}F_{3}]
+ [V_{13}+V_{23},F_{1}g_{23}]
\nonumber\\
&\quad +
[V_{13}+V_{23},F_{2}g_{13}]
+ [V_{13}+V_{23},F_{3}g_{12}]
\nonumber\\
&\quad +
[V_{13}+V_{23},g_{123}]\Big\}(1+\epsilon P_{13}+\epsilon P_{23})\lambda^{\pm}_{12}\,.
\label{eq:g12-as-1}
\end{align}
The transformation of Eq.~(\refeq{eq:g12-as-1}) was performed in Refs.~\cite{dufty-etal.97, bonitz_qkt} where it was observed that each term contains a factor $\lambda^\pm_{12}$ that was cancelled. However, this is incorrect, as it neglects important terms. Here, we restore this factor and present the complete derivation.

From Eq.~(\refeq{eq:g12-as-1}) we subtract the equation of motion for $F_1F_2\lambda_{12}^\pm$, using the first hierarchy equation (\refeq{eq:bbgky1-as}), with the result (the main steps involved in the derivation of Eq.~(\refeq{eq:g12-equation-as}) are provided in appendix \ref{app:g12-equation}.)
\begin{align}
\i\hbar \frac{\d}{\d t}
g^\pm_{12} - [{\bar H}^0_{12}, g^\pm_{12}]
=& \Psi^\pm_{12,\rm DO} + L_{12,\rm DO}
\label{eq:g12-equation-as}\\
&+ P^\pm_{12, \rm DO} +
\mbox{Tr}_{3}[V^{(12),3},g^\pm_{123}]
\,,\nn
\\
\Psi^\pm_{12,\rm DO} =& 
\hat{V}^\pm_{12}F_{1}F_{2} - F_{1}F_{2}\hat{V}^{\pm\dagger}_{12}\,,
\label{eq:psi-pm-do}
\\
L_{12,\rm DO} =& \hat{V}_{12}\,g^\pm_{12} - g^\pm_{12}\,\hat{V}^\dagger_{12}\,,
\label{eq:l-pm-do}
\\
P^\pm_{12, \rm DO} =&
\left(\Pi_{12,\rm DO}^{\pm (1)} + 
\Pi_{12,\rm DO}^{\pm (2)}\right)\lambda_{12}^\pm \,,
\label{eq:ppm-do}
\\
\Pi_{12,\rm DO}^{\pm (1)} =& \mbox{Tr}_{3}
    [V^\pm_{13},F_1 g^\pm_{23}]\,,
\label{eq:ppm-do1}
\end{align}
where $\Pi_{12,\rm DO}^{\pm (2)}$ follows from $\Pi_{12,\rm DO}^{\pm (1)}$ by exchanging $(1 \leftrightarrow 2)$.

Let us discuss the terms in this equation.
The operators $\Psi^\pm_{12,\rm DO}$ and $L_{12,\rm DO}$ are generalizations of the spinless results for the inhomogeneity ($\Psi_{12,\rm DO}$) and the ladder terms ($L^0_{12,\rm DO}$), respectively. Instead of the bare potential, $V_{12}$, of the spinless case, they now involve the pair potential (\refeq{eq:vhat}, \refeq{eq:n12}), that is modified by Pauli blocking effects of the surrounding medium. Furthermore, $\Pi^{\pm (1)}_{12, \rm DO}$, Eq.~(\refeq{eq:ppm-do1}), is the polarization contribution that generalizes the previous spinless result, $\Pi^{0(1)}_{12, \rm DO}$, Eq.~(\refeq{eq:pi-rdo-01}). This generalization is two-fold: first, the pair correlation operator $g_{23}$ is replaced by its (anti-)symmetrized generalization, $g_{23}^\pm$. Second, the pair potential $V_{13}$ is replaced by the (anti-)symmetrized potential, $V^{\pm}_{13}$ where the additional contribution involving $\hat P_{13}$ gives rise to exchange corrections to the polarization terms which have no classical counterpart. Finally, note that the full polarization term, \refeqn{eq:ppm-do}, contains an additional factor $\lambda_{12}^\pm=\hat 1 + \epsilon \hat P_{12}$. While the contribution of the $\hat 1$ directly yields \refeqn{eq:ppm-do1} which has the same form as in the spinless case, as discussed above, the additional contributions  $\Pi_{12,\rm DO}^{\pm (1)}\epsilon \hat P_{12}$ and $\Pi_{12,\rm DO}^{\pm (2)}\epsilon \hat P_{12}$ are new, without a counterpart in the BBGKY-hierarchy of classical or spinless quantum systems.

The physical nature of these new terms will become clear from a comparison to the G1--G2 scheme, in Sec.~\ref{ss:g1-g2-rdm-comparison}: taking advantage of the connection of all terms in the G1--G2 scheme to selfenergy diagrams of NEGF theory we will establish that these additional terms correspond to particle--hole ladder diagrams.

The conserving many-body approximations that can be used to decouple the (anti-)symmetrized BBGKY-hierarchy are the same as discussed in Sec.~\ref{ss:do-many-body} where only now the properly (anti-)symmetrized terms have to be used:
\begin{description}
\item[i.)] $I_{\rm DO} \to 0$ recovers the time-dependent Hartree--Fock approximation;
\item[ii.)] $L_{12,\rm DO}=P^\pm_{12,\rm DO}\to 0$ is equivalent to the second-order approximation with exchange, Eq.~\eqref{eq:G1-G2_soa};
\item[iii.)] $P^\pm_{12,\rm DO} \to 0$ and $L_{12,\rm DO} \ne 0$ is equivalent to the particle--particle $T$-matrix approximation with exchange, Eq.~\eqref{eq:G1-G2_tpp_compact};
\item[iv.)] $L_{12,\rm DO} \to 0$ and $P^\pm_{12,\rm DO} \ne 0$ corresponds to a combination of the $GW$ approximation with exchange and the particle--hole $T$-matrix with exchange that also includes the respective cross-coupling terms.
\item[v.)] $L_{12,\rm DO} \ne 0$ and $P^\pm_{12,\rm DO} \ne 0$ is equivalent to the DSL approximation with exchange contributions. 
\end{description}

\subsection{Comparison of the density operator results to the G1--G2 scheme}\label{ss:g1-g2-rdm-comparison}
Let us start the comparison of the two approaches by relating the definitions of the one-particle reduced density operator, $F_1$, and the (anti-)symmetrized pair correlation operator, $g^\pm_{12}$, \cite{bonitz_qkt} to the Green functions (correlation functions) on the time diagonal. Since the Green functions are given in a matrix representation, we also transform the reduced density operators into matrix form with respect to a complete orthonormal set of single-particle orbitals, $\{|i\rangle\}$. For two-particle states we use a product form, $\{|ij\rangle\}=\{|i\rangle |j\rangle\}$.

The relevant relations between the quantities in reduced density operator theory and nonequilibrium Green functions on the time diagonal are
\begin{align}
    (F_1)_{ij} &= F_{ij}=F^*_{ji} = n_{ij}=\pm \i\hbar G^<_{ij}\,,
    \label{eq:f1-gless}
    \\
    (g^\pm_{12})_{ijkl} &=  (\i\hbar)^2 \mathcal{G}_{ijkl} \,,
    \label{eq:g12-g2less}
    \\
    (V_{12})_{ijkl} &=
    w_{ijkl} = w_{jilk} = w^*_{klij}\,,
    \label{eq:v12-wijkl}\\
    (V^\pm_{12})_{ijkl} &= (V_{12}\lambda_{12}^\pm)_{ijkl} =  w^\pm_{ijkl} = w_{ijkl} + \epsilon\, w_{ijlk}\,,
    \label{eq:g12-g2}
    \\
    w^\pm_{ijkl} &= w^\pm_{jilk} = w^{\pm*}_{klij}=\pm w^\pm_{ijlk}\,,
    \label{eq:wpmsym}
    \\
    \mathcal{G}_{ijkl} &= \mathcal{G}_{jilk} =  \mathcal{G}^*_{klij} =\pm \mathcal{G}_{ijlk}=\pm \mathcal{G}_{jikl}\,.
    \label{eq:gsym1}
\end{align}
Thus, the matrix of the single-particle density operator, \refeqn{eq:f1-gless}, coincides with the one-particle density matrix, $n_{ij}$ and, up to the factor $\pm i\hbar$, with the single-particle Green function. Similarly, the matrix of the correlation operator, \refeqn{eq:g12-g2less} coincides, up to the factor $(i\hbar)^2$ with the time-diagonal correlated part of the two-particle Green function.

Note that the (anti-)symmetrized pair potential and, with it also the two-particle functions $\mathcal{G}$ and $g^\pm_{12}$, obey a number of additional symmetries, compared to the original pair potential $V_{12}$ and the non-symmetrized operator $g_{12}$, that are listed in \refeqns{eq:wpmsym}{eq:gsym1}.

In the following we compare, term by term, the first and second equations of the (anti-)symmetrized BBGKY hierarchy to the equations of the G1--G2 scheme.

\subsubsection{First equation}
Comparison of \refeqns{eq:bbgky1-as}{eq:eom_gone} reveals that both are identical, up to a factor $(\pm i\hbar)$ that follows from relation \refeqn{eq:f1-gless}. This follows from the equivalence of the mean field terms and collision integrals, in \refeqns{eq:bbgky1-as}{eq:tiagonal-i-sigmas}, 
\begin{align}
    (U_1^{\rm HF})_{ij} &= \sum_{kl} V^\pm_{ikjl}\,F_{lk} = \pm \i\hbar \,\Sigma^{\rm HF}_{ij}\,,\\
    (H_1+U_1^{\rm HF})_{ij} &= h^{\rm HF}_{ij}
    \,,\\
    (I_{1,\rm DO})_{ij} &= \pm (\i\hbar) (I+I^\dagger)_{ij}\,.
\end{align}
We now turn to a comparison of the second hierarchy equation, \refeqn{eq:g12-equation-as}, and the equations for the two-particle Green function that were derived, separately, for the SOA, TPP, TPH, and GW selfenergies. Due to the relation \refeqn{eq:g12-g2less} both versions of the equations differ by an overall factor $(i\hbar)^2$. 

\subsubsection{Inhomogeneity $\Psi^\pm$ in the second equation}
We start from the comparison of the two inhomogeneities, $\Psi^\pm_{12,\rm DO}$, \refeqn{eq:psi-pm-do}, in the equation for the pair correlation operator to $\Psi^\pm_{ijkl}$, in the equation of $\mathcal{G}$, for SOA and TPP selfenergies. Straightforward transformations confirm that both are identical,
\begin{align}
(\Psi^\pm_{12,\rm DO})_{ijkl} &=
\left\{ \hat{V}^\pm_{12}F_{1}F_{2} - F_{1}F_{2}\hat{V}^{^\pm\dagger}_{12} \right\}_{ijkl} = 
\nn\\
\left\{(1\pm F_1)(1\pm F_2)\right.&V^\pm_{12}F_1F_2 - \left.F_1F_2V^\pm_{12}(1\pm F_1)(1\pm F_2) \right\}_{ijkl} 
\nn\\
 &\equiv(\i\hbar)^2\Psi^\pm_{ijkl}
\end{align}
\subsubsection{Particle--particle ladder term $L$}
Let us consider now the particle--particle ladder contributions, $L_{12,\rm DO}$, \refeqn{eq:l-pm-do}, and $L_{ijkl}$, \refeqn{eq:ladder-final}.
With this we can transform the ladder term of the G1--G2 scheme
\begin{align}
L_{ijkl} &= 
\sum_{rs} \sum_{pq}\Big\{
N_{ijrs} w_{rspq}\mathcal{G}_{pqkl} 
\nn\\
&\qquad\qquad 
- \mathcal{G}_{ijpq}w_{pqrs}N_{rskl} 
 \Big\}\,.
\nn
\end{align}
Using the relation between the two-particle Pauli blocking factors, in the G1--G2 scheme, \refeqn{eq:g2h-transform}, and, in the density operator equations, \refeqn{eq:n12}, $(N_{12})_{ijkl}=N_{ijkl}$,
and the relation of the two bare interaction potentials, \refeqn{eq:v12-wijkl},
the matrix summations can be performed giving rise to products of operators,
\begin{align}
L_{ijkl} &= 
    \sum_{pq}\Big\{
\hat V_{ijpq}
\mathcal{G}_{pqkl} 
- \mathcal{G}_{ijpq}\hat V^\dagger_{pqkl}  \Big\}
\nn\\
&=\frac{1}{\left(\i\hbar\right)^{2}}
    \left\{ 
    {\hat V}_{12}\,g^\pm_{12} - g^\pm_{12}\,{\hat V}^\dagger_{12}
    \right\}_{ijkl} = 
    \frac{1}{\left(\i\hbar\right)^{2}}(L_{12,\rm DO})_{ijkl}\,.
\label{eq:ladder-qkt}
\end{align}
Furthermore, we  identify the effective ladder Hamiltonian with the screened potential,
\begin{align}
    \mathfrak{h}^L_{12} \equiv \hat V_{12} \,.
    \nn
\end{align}
In both cases, the screened potential $\hat V$ appears, without exchange corrections.

\subsubsection{Polarization terms. $GW$ and exchange corrections}\label{sss:gw-comparison}
Let us now transform the polarization term, $\Pi^\pm$. It is convenient to start from Eq.~(\refeq{eq:pipm-m}) where we replace $w\to w^\pm$ and perform the summations
\begin{align}
    \Pi^\pm_{ijkl} &= \sum_{pqr}\bigg\{ \left[w^\pm_{jqrp} n_{rl}  - n_{jr} w^\pm_{rqlp}  \right] \mathcal{G}_{ipkq} \nonumber \\
    &\qquad\quad - \mathcal{G}_{jqlp} \Big[n_{ir} w^\pm_{rpkq}   - w^\pm_{iprq} n_{rk} \Big] \bigg\} \nonumber \\
& = \sum_{pq}\bigg\{ \left[w^\pm_{23}, n_{2}  \right]_{jqlp} \mathcal{G}_{ipkq} 
    +  \Big[w^\pm_{13}, n_{1}   \Big]_{ipkq}\mathcal{G}_{jqlp} \bigg\}
    \nn\\
    &=\left(\i\hbar\right)^{-2}\mbox{Tr}_3\left\{[V^\pm_{13}, F_{1}]g^\pm_{23} + [V^\pm_{23}, F_{2}]g^\pm_{13}\right\}_{ijkl}\,,
 \label{eq:P_polarization}
 \\
    & = \left(\i\hbar\right)^{-2}\left( \Pi^{\pm(1)}_{12,\rm DO}+\Pi^{\pm(2)}_{12,\rm DO}\right)_{ijkl}\,.
\end{align}
Thus, we have demonstrated agreement with the density operator result, \refeqn{eq:ppm-do} if exchange diagrams are taken into account in the GW selfenergy, i.e. $\Pi \to \Pi^\pm$. This means, also the Green functions result can be brought into a compact, basis-independent operator notation. Finally, we note another useful connection of the two approaches. The effective polarization Hamiltonian, \refeqn{eq:h_pi}, can be rewritten using the definition, \refeqn{eq:g2f-transform}, in the following way, again including exchange diagrams,
\begin{align}
     \mathfrak{h}^{\Pi\,\pm}_{ijkl} & = 
     \pm \sum_{pq} w^\pm_{iqkp} \left[  \delta_{jq} n_{pl} - \delta_{pl} n_{jq} \right] \nonumber \\
     &= \pm \sum_{p} \left[w^\pm_{ijkp}  n_{pl} - n_{jp} w^\pm_{ipkl}  \right] \nonumber \\
     &= \pm \left[V^\pm_{12}, F_{2}  \right]_{ijkl}\, ,
\end{align}
where, in the last line, we again introduced the density operator theory notation.\\

\subsubsection{Particle--hole ladder terms with exchange}

What is left is to analyze the correspondence  between the particle--hole $T$-matrix contributions and the density operator result. We consider 
 the effective particle--hole $T$-matrix Hamiltonian, and again include exchange terms via the replacement $w\to w^\pm$. This leads to the replacement  $\mathfrak{h}^\Lambda \to \mathfrak{h}^{\Lambda \pm}$, which are related to the polarization Hamiltonian with exchange, $\mathfrak{h}^{\Pi \pm}$, by an exchange of the first two matrix indices. 
 \\
We rewrite this relation in terms of a permutation operator acting either on the left or right pair of matrix indices,
\begin{align}
    \Lambda^{\rm PH\,\pm}_{ijkl} &= \pm \Pi^\pm_{jikl} \equiv \pm (\Pi^\pm_{12})_{jikl} = \pm \hat P_{12}(\Pi^\pm_{12})_{ijkl} \,,
\\
&= \pm \Pi^\pm_{ijlk} \equiv \pm (\Pi^\pm_{12})_{ijlk} = \pm (\Pi^\pm_{12})_{ijkl}\hat P_{12} \,.
\end{align}

On the other hand, consider the term, \refeqn{eq:ppm-do}, in the density operator equation (\refeq{eq:g12-equation-as}). It contains a factor $\lambda_{12}^\pm = \hat 1 +\epsilon \hat P_{12}$. The term with the $\hat 1$ is just the polarization term (with exchange) for which we established agreement with the G1--G2 scheme in Sec.~\ref{sss:gw-comparison}. Thus, the remaining terms for which we have not yet established a counterpart in the G1--G2 scheme are
\begin{align}
\left(\Pi^{\pm (1)}_{12,\rm DO}+\Pi^{\pm (2)}_{12,\rm DO}\right)_{ijkl}    \epsilon\hat P_{12}\,.
\end{align}
Since the term in the parantheses was shown to be equal to $(i\hbar)^2 \Pi^\pm_{ijkl}$, we immediately conclude that 
\begin{align}
\left(\Pi^{\pm (1)}_{12,\rm DO}+\Pi^{\pm (2)}_{12,\rm DO}\right)_{ijkl}    \epsilon\hat P_{12} = (\i\hbar)^2 \Lambda^{\rm PH\,\pm}_{ijkl}\,.
\end{align}
Thus, we have identified these remaining terms in the density operator equation with the terms derived from the particle--hole ladder selfenergy including exchange, in the G1--G2 scheme.

\subsubsection{Dynamically screened 
ladder approximation. Neglect of three-particle correlations}
\label{sss:dsl-comparison}
So far we have obtained perfect agreement between the G1--G2--scheme for two of the considered selfenergy approximations---SOA with exchange and TPP with exchange---with the corresponding density operator results. We further have identified the particle--hole $T$-matrix terms with exchange and $GW$ terms with exchange which appear in the density operator equations only simultaneously.

On the other hand, the density operator result confirms that all these terms can also be included at the same time, as follows from equation, \refeqn{eq:g12-equation-as} which, therefore, selfconsistently contains strong coupling (TPP) and dynamical screening (polarization and TPH) effects, including the proper exchange diagrams. 
The corresponding many-body approximation, thus, corresponds to the nonequilibrium dynamically screened ladder approximation (DSL). 

Finally \refeqn{eq:g12-equation-as}
contains an additional term involving $g_{123}^\pm$. Via this term, the second hierarchy equation couples to the rest of the hierarchy and, with its full inclusion, the equation would be exact. In this paper we will only consider the case $g_{123}^\pm \to 0$ focusing on extensive numerical tests of the DSL approximation, cf. Sec.~\ref{s:numerics}.

\subsection{Summary of conserving many-body approximations in the G1--G2 scheme}\label{ss:g1-g2-summary-approximations}
Let us summarize the many-body approximations that are available in the G1--G2 scheme. The first group of approximations is derived from 
conserving ($\Phi$-derivable)  approximations of Green functions theory, i.e. from common selfenergy approximations that were introduced in Sec.~\ref{s:sigmas}, to which subsequently the Hartree--Fock GKBA is applied, and which coincide with standard approximations of reduced density operator theory:
\begin{description}
\item[i.)] The Hartree--Fock selfenergy,  $\Sigma=\Sigma^{\rm HF}$, Eq.~\eqref{eq:sigma_hf}, is equivalent to setting  $I_{\rm DO} \to 0$ in the RDO approach;
\item[ii.)] The second-order  selfenergy with exchange (SOA), i.e. $\Sigma=\Sigma^{\rm SOA}+\Sigma^{\rm SOA}_x$, Eq.~\eqref{eq:sigma-soa}, is equivalent to setting $L_{12,\rm DO}=P^\pm_{12,\rm DO}\to 0$, in the RDO approach; 
\item[iii.)] The particle--particle $T$-matrix selfenergy with exchange, $\Sigma=\Sigma^{\rm TPP}$, Eq.~\eqref{eq:sigma-tpp}, is equivalent to setting 
$P^\pm_{12,\rm DO} \to 0$ and $L_{12,\rm DO} \ne 0$, in the RDO approach; 
\end{description}
In addition, there are a number of conserving many-body approximations  which straightforwardly follow from one of the two approaches whereas their correspondence in the other approach is not known or not straightforward:
\begin{description}
\item[iv.)] 
Setting $L_{12,\rm DO} \to 0$ and $P^\pm_{12,\rm DO} \ne 0$ in the RDO approach leads to a consistent combination of all particle--hole ladder contributions [Eq.~\eqref{eq:sigma-tph}] and dynamical-screening terms [Eq.~\eqref{eq:sigma-gw}] from Green functions theory that also includes mixed terms going beyond the direct sum.
\item[v.)] Setting $L_{12,\rm DO} \ne 0$ and $P^\pm_{12,\rm DO} \ne 0$, in the RDO approach is equivalent to the DSL-approximation with exchange contributions. At the same time, this approximation cannot be derived from a direct combination of $\Sigma^{\rm GW}$, $\Sigma^{\rm TPP}$ and $\Sigma^{\rm TPH}$, as one could have expected. At the moment, no nonequilibrium DSL selfenergy that includes all orders in the selfenergy is known.
\item[vi.)] The third-order selfenergy, $\Sigma=\Sigma^{\rm TOA}$, Fig.~(\ref{fig:diagrams_order}), corresponds to an iteration of the DSL selfenergy that includes all diagrams up to third order in the interaction  In the RDO approach this approximation is not straightforward. It can be recovered by treating the ladder and polarization terms iteratively, by using $g^{\pm}\to g^{\rm SOA}$
\cite{schluenzen_phd_21}.
\item[vii.)] Neglecting exchange diagrams in common $\Phi$-derivable selfenergies also gives rise to conserving approximations. In the RDO approach this amounts to replacing the anti-symmetrized terms by the spinless versions, e.g.  $\Psi^\pm \to \Psi_0$ and, similarly, for the other terms, for details see Tab.~\ref{tab:negf-rdo}.
\end{description}
The correspondence between NEGF selfenergies and the RDO approximations is summarized in Tab.~\ref{tab:negf-rdo}.

Our main result is the dynamically screened ladder approximation (DSL) because it selfconsistently combines dynamical screening and strong coupling effects. It arises naturally in the RDO approach by simultaneously including particle--particle ladder, polarization and particle--hole ladder terms. This means, it contains all two-particle correlation contributions and neglects only three-particle correlations, $g^\pm_{123} \to 0$.

This approximation was also studied in detail for nuclear matter by Wang and Cassing \cite{wang-cassing-85}. 
It has also been derived by Valdemoro \textit{et al.} \cite{valdemoro_93} without introducing correlation operators, by exploiting symmetries between the two-particle and two-hole reduced density matrix.
The present RDO approach that uses the cluster expansion including the pair correlation operators has the advantage that it allows one to separate ladder and polarization terms that describe very different physical effects which are important in different situations.
Note that all these approximations conserve total energy, as was discussed in Sec.~\ref{ss:econs-cc}.

\begin{table}[]
    \centering
    \begin{tabular}{c|c|c|c|c}
    approximation & RDO notation & Def.&Selfenergy & Def.\\
    \hline
    SOA&$\Psi^0$ & \eqref{eq:psi-rdo-0} & $\Sigma^{\rm SOA}$
         &\eqref{eq:sigma-soa} \\
    &$\Psi^\pm$& \eqref{eq:psi-pm-do} & $\Sigma^{\rm SOA}+\Sigma_x^{\rm SOA}$ & \eqref{eq:sigma-soa} \\
         \hline    
    TPP&$\Psi^0+ L^0$ & \eqref{eq:l-rdo-0} & $\Sigma^{\rm TPP}$
         &\eqref{eq:sigma-tpp} \\
    &$\Psi^{\pm} + L$ & \eqref{eq:l-pm-do} & $\Sigma^{\rm TPP}+\Sigma_x^{\rm TPP}$
         &\eqref{eq:sigma-tpp} \\
         \hline
    $GW$&$\Psi^0+ \Pi^0$ & \eqref{eq:pi-rdo-0} & $\Sigma^{\rm GW}$
         &\eqref{eq:sigma-gw}\\
    &$\Psi^\pm + \Pi^\pm$ & \eqref{eq:ppm-do1} & --
         &--
         \\
         \hline
        DSL& $\Psi^0 + \Pi^0 + L^0$ &--& --
         & --\\
        & $\Psi^\pm + \Pi^\pm + L$ &-- & --
         & --\\
         \hline    
    TOA&$\Psi^\pm +\Pi^{\pm}[G^{\rm SOA}_2]+$&-- & $\Sigma^{\rm TOA}$
    & Fig.~\ref{fig:diagrams_order}\\
     &$+L[G^{\rm SOA}_2]$ & & &\\
         
    \end{tabular}
    \caption{Correspondence of many-body approximations of Green functions (correlation selfenergies) and reduced density operators (terms in $G_2$-equation) and their defining equations. 
    }
    \label{tab:negf-rdo}
\end{table}

\section{Solving the G1--G2 equations for Hubbard systems. Contraction consistency and purification}\label{s:hubbard}

\subsection{G1--G2 equation in the Hubbard basis}\label{ss:hubbard}

The numerical tests of the time-linear DSL approximation in Sec.~\ref{s:numerics} are performed within the Hubbard model \cite{hubbard_1963}. This model is well established in condensed-matter physics, as it allows for semi-quantitative analysis of strong electronic correlations and phase transitions in solids. Besides, it has been used extensively in experiments with ultracold fermionic and bosonic atoms in optical lattices \cite{bloch_probing_2014} in particular, to study time-dependent correlation phenomena, see, e.g, Refs.~\cite{kajala_expansion_2011,schneider_fermionic_2012,von_friesen_kadanoff-baym_2010-1, hermanns_prb14}.
For the Fermi--Hubbard model, the general pair-interaction matrix element becomes ($\cbar{\delta}_{\alpha \beta} \coloneqq 1 - \delta_{\alpha \beta}$)
\begin{align}
 w_{ijkl}^{\alpha\beta\gamma\delta}(t) = U(t) \delta_{ij} \delta_{ik} \delta_{il} \delta_{\alpha \gamma} \delta_{\beta \delta} \cbar{\delta}_{\alpha \beta}\, ,
 \label{eq:wmatrix-hubbard}
\end{align}
with the on-site interaction $U$ and the spin projection labeled by greek indices. 
Recall that we allow for a time dependence of the interaction to capture the adiabatic-switching protocol of initial correlations, see Sec.~\ref{ss:diagonal-kbe}, as well as nonequilibrium processes such as an interaction quench, cf. the discussion of \refeqn{eq:h} above.
The kinetic energy matrix is replaced by a hopping Hamiltonian,
\begin{align}
    h^{(0)}_{ij} = - \delta_{\left<i,j\right>} J \, ,
\end{align}
which includes hopping processes between nearest-neighbor sites $\left<i,j\right>$ with amplitude $J$.
Thus, the total Hamiltonian is given by
\begin{align}
 \chat{H}(t) = -J\sum_{\left<i,j\right>} \sum_\alpha \cop{i}{\alpha}\aop{j}{\alpha} + U(t) \sum_i \chat{n}_{i}^{\uparrow} \chat{n}_{i}^{\downarrow} + \chat{F}(t)\, ,
\label{eq:h-hubbard}
\end{align}
with a general single-particle excitation $\chat{F}(t)$.
Extensions to more complicated models, going beyond the nearest-neighbor single-band case are straightforward, see e.g. Ref.~\cite{joost_pss_18}, but will not be considered here.

In this work we are assuming spin symmetry of the system. In that case  a single spin-component of the single- and correlated two-particle Green functions is sufficient to completely  describe the dynamics of the system: $G^{<,\uparrow}$ and $\mathcal{G}^{\uparrow\downarrow\uparrow\downarrow}$, respectively. All other components can be expressed via these functions.
The time-diagonal EOM for the single-particle Green function, \refeqn{eq:eom_gone}, takes the following form
\begin{align}\label{eq:eom_g1_hubbard}
\i\hbar \frac{\d}{\d t} G^{<,\uparrow}_{ij}(t) &= \left[h^{\tn{HF},\uparrow}, G^{<,\uparrow}\right]_{ij}(t) + \left[I+I^\dagger\right]^{\uparrow}_{ij}(t)\, ,\quad \\
 I^{\uparrow}_{ij}(t) &= -\i\hbar U(t) \mathcal{G}^{\uparrow\downarrow\uparrow\downarrow}_{iiji}(t) \,. \label{eq:colint_hubbard}
\end{align}
The Hartree--Fock Hamiltonian in \refeqn{eq:eom_g1_hubbard} becomes, in the Hubbard basis, [cf. \refeqn{eq:h_HF}]:
\begin{align}
    h^{\tn{HF},\uparrow}_{ij}(t) = h^{(0)}_{ij} -\i\hbar \delta_{ij} U(t) G^{<,\downarrow}_{ii}(t) \, ,
\end{align}
with $G^{<,\downarrow} = G^{<,\uparrow}$, due to spin symmetry.
The equation for the time-diagonal two-particle Green function, \refeqn{eq:g12-equation-as}, now reads
\begin{align}
 \i\hbar  \frac{\d}{\d t} \mathcal{G}^{\uparrow\downarrow\uparrow\downarrow}_{ijkl}(t) &- \Big[ h^{(2),\tn{HF}}_{\uparrow\downarrow}(t),\mathcal{G}^{\uparrow\downarrow\uparrow\downarrow}(t) \Big]_{ijkl} = \Psi^{\uparrow\downarrow\uparrow\downarrow}_{ijkl}(t) \nonumber \\
 &+L^{\uparrow\downarrow\uparrow\downarrow}_{ijkl}(t) + \Pi^{\uparrow\downarrow\uparrow\downarrow}_{ijkl}(t) + \Lambda^{\tn{ph},\uparrow\downarrow\uparrow\downarrow}_{ijkl}(t) 
 \, , \label{eq:G1-G2_DSL_compact_Hubbard}
\end{align}
with the two-particle Hartree--Fock Hamiltonian
\begin{align}
 h^{(2),\tn{HF}}_{ijkl,\uparrow\downarrow}(t) = \delta_{jl} h^{\tn{HF},\uparrow}_{ik}(t) + \delta_{ik}h^{\tn{HF},\downarrow}_{jl}(t)\, , \label{eq:twopart_hamiltonian_hubbard}
\end{align}
the inhomogeneity
\begin{align}
\Psi^{\uparrow\downarrow\uparrow\downarrow}_{ijkl}(t) &\coloneqq \left(\i\hbar\right)^2 U(t) \sum_{p}
  \left[ G^{>,\uparrow}_{ip}(t) G^{>,\downarrow}_{jp}(t) G^{<,\uparrow}_{pk}(t) G^{<,\downarrow}_{pl}(t)\right. \nonumber \\
  &\qquad \left.-G^{<,\uparrow}_{ip}(t) G^{<,\downarrow}_{jp}(t) G^{>,\uparrow}_{pk}(t) G^{>,\downarrow}_{pl}(t) \right] \, ,
\label{eq:phi-hubbard}  
\end{align}
the particle--particle $T$-matrix (ladder) term
\begin{align} \label{eq:lambdapp_hubb}
 &L^{\uparrow\downarrow\uparrow\downarrow}_{ijkl}(t) = (\i\hbar)^2 U(t)\times
 \\&\,
  \sum_{p} \Big[ G^{>,\uparrow}_{ip} (t) G^{>,\downarrow}_{jp} (t) - G^{<,\uparrow}_{ip} (t) G^{<,\downarrow}_{jp} (t) \Big] \mathcal{G}^{\uparrow\downarrow\uparrow\downarrow}_{ppkl}(t) \, , \nonumber
\end{align}
the particle--hole $T$-matrix (ladder) term
\begin{align} \label{eq:lambdaph_hubb}
 &\Lambda^{\tn{ph},\uparrow\downarrow\uparrow\downarrow}_{ijkl}(t) = (\i\hbar)^2 U(t)\times
 \\&\,
  \sum_{p} \Big[ G^{>,\uparrow}_{ip} (t) G^{<,\downarrow}_{pl} (t) - G^{<,\uparrow}_{ip} (t) G^{>,\downarrow}_{pl} (t) \Big] \mathcal{G}^{\uparrow\downarrow\uparrow\downarrow}_{pjkp}(t) \, , \nonumber
\end{align}
and the $GW$ contribution
\begin{align}
 &\Pi^{\uparrow\downarrow\uparrow\downarrow}_{ijkl}(t) = - (\i\hbar)^2 U(t)\times \label{eq:pihubbardtrans}
 \\&\,
  \sum_{p} \Big[ G^{>,\downarrow}_{jp} (t) G^{<,\downarrow}_{pl} (t) - G^{<,\downarrow}_{jp} (t) G^{>,\downarrow}_{pl} (t) \Big] \mathcal{G}^{\uparrow\uparrow\uparrow\uparrow}_{ipkp}(t) \, , \nonumber
\end{align}
where $\mathcal{G}^{\uparrow\uparrow\uparrow\uparrow}_{ijkl} = \mathcal{G}^{\uparrow\downarrow\uparrow\downarrow}_{ijkl} - \mathcal{G}^{\uparrow\downarrow\uparrow\downarrow}_{ijlk}$, due to spin symmetry.
We emphasize that  conservation of particles and energy that is inherent to the two-time selfenergy approximations listed in Sec.~\ref{s:sigmas}, including the DSL approximation, also holds for the HF-GKBA \cite{hermanns_prb14,bonitz_qkt} and the respective time-linear versions, for an analysis see Sec.~\ref{ss:econs-cc}. 

\begin{figure}[t]
\includegraphics[width=\columnwidth]{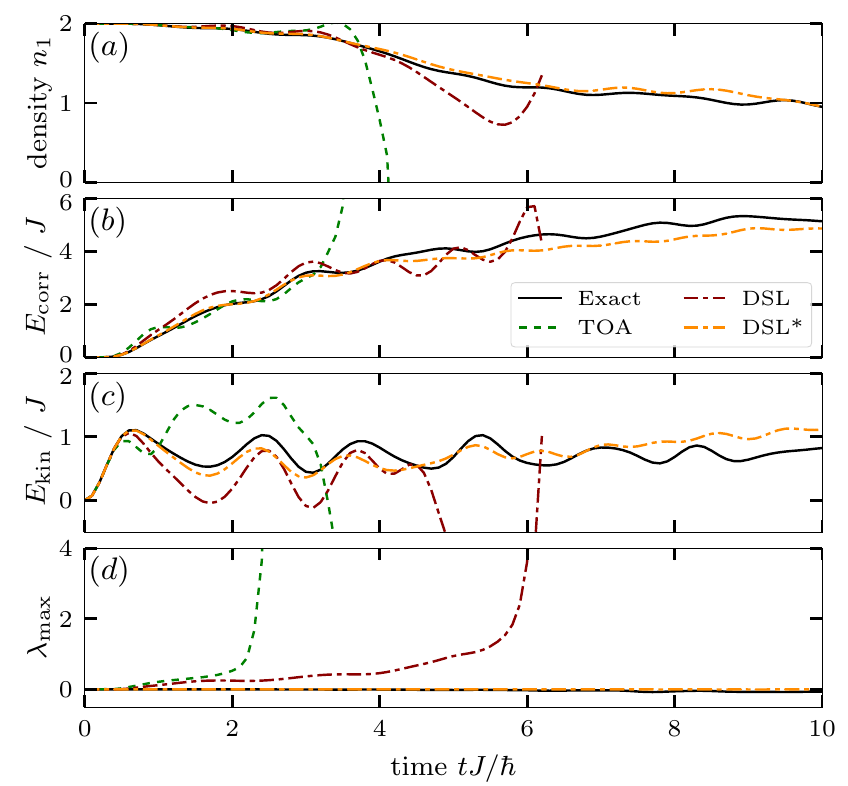}
\caption{Illustration of the instability of the G1--G2 scheme for a half-filled 6-site Hubbard system at moderate coupling, $U/J=4$. Sites 1--3 are initially doubly occupied and sites 4--6 are empty. At time $t=0$ the confinement potential is removed (confinement quench). In contrast to the exact (full black line) and DSL* solutions (dash-dotted yellow line), the TOA (dashed green line) and DSL solutions (dash-dotted red line) become unstable already at time $tJ/\hbar\approx 4$ and $tJ/\hbar\approx 6$, respectively.  (a): occupation $n$ of site 1. (b) and (c): correlation and kinetic energy, respectively. (d): largest
eigenvalue of the two-particle Green function.}
\label{fig:instability1}
\end{figure}

However, despite fulfilling these important conservation laws and advantageous scaling behavior which allows for long propagation times, in many cases, the G1--G2 scheme exhibits unstable dynamics. A typical example is shown in Fig.~\ref{fig:instability1}. There the time evolution of various observables in a moderately coupled 6-site Hubbard chain with $U/J=4$, following a confinement quench, is presented. We compare solutions of the G1--G2 equations for the TOA (green) and the DSL (brown) approximation to the exact dynamics (black). While the latter shows a stable dynamics, the G1--G2 solutions become unstable already after a short time interval of about 6 (4) time units for DSL (TOA). Note that this is not a consequence of the numerical scheme (such as a too large time step). 

The reason for this behavior is the violation of 
contraction consistency and N-representability~\cite{lacknerphd,coleman2000reduced,mazzioti_reduced_2007}, see our discussion in Sec.~\ref{ss:econs-cc}.
It has been shown, however, that contraction consistency can be enforced \textit{a posteriori} upon any reconstruction functional of the three-particle RDO \cite{lackner_propagating_2015,lacknerphd}, and (ensemble) N-representability can be partially restored ~\cite{lackner_propagating_2015,lackner_pra_17,lacknerphd}.  We present, in the following, both concepts for the two-particle Green function.

For this purpose it is convenient to introduce the full two- and three-particle Green functions which include the Hartree--Fock contribution alongside the correlation part. Due to spin symmetry, again, only a single spin component has to be considered for each quantity:
\begin{align}
    G^{(2),\uparrow\downarrow\uparrow\downarrow}_{ijkl} = G^{<,\uparrow}_{ik}G^{<,\downarrow}_{jl} + \mathcal{G}^{\uparrow\downarrow\uparrow\downarrow}_{ijkl}
\end{align}
and
\begin{align}
    G^{(3),\uparrow\uparrow\downarrow\uparrow\uparrow\downarrow}_{ijklpq} &= G^{<,\uparrow}_{il}G^{<,\uparrow}_{jp}G^{<,\downarrow}_{kq} - G^{<,\uparrow}_{ip}G^{<,\uparrow}_{jl}G^{<,\downarrow}_{kq}
    \\&+ G^{<,\uparrow}_{il} \mathcal{G}^{\uparrow\downarrow\uparrow\downarrow}_{jkpq} - G^{<,\uparrow}_{ip} \mathcal{G}^{\uparrow\downarrow\uparrow\downarrow}_{jklq} \\&+ G^{<,\uparrow}_{jp} \mathcal{G}^{\uparrow\downarrow\uparrow\downarrow}_{iklq} - G^{<,\uparrow}_{jl} \mathcal{G}^{\uparrow\downarrow\uparrow\downarrow}_{ikpq}
    \\& + G^{<,\downarrow}_{kq} \mathcal{G}^{\uparrow\uparrow\uparrow\uparrow}_{ijlp} + \mathcal{G}^{(3),\uparrow\uparrow\downarrow\uparrow\uparrow\downarrow}_{ijklpq}\,.
\end{align}
Applying above-mentioned methods to guarantee contraction consistency and partially restore (ensemble) N-representability to DSL leads to a stable variant which we call DSL*, which is also included in Fig.~\ref{fig:instability1} and shows much better agreement with the exact solution. Details of this approach are outlined in Sec.~\ref{ss:cc}, and further numerical results will be presented in Sec.~\ref{s:numerics}.

\subsection{Enforcing contraction consistency}\label{ss:cc}

The general definition of the reduced $s$-particle density operators in terms of the full $N$-particle density operator, \refeqn{eq:fs-def}, implies trace relations between the reduced density operators of different orders. Consequently, similar consistency relations have to be fulfilled by the time-diagonal Green functions. In particular, the two- and three-particle Green functions have to satisfy
\begin{align}
    \frac{N}{2} G^{\uparrow\uparrow}_{ij} &= - \mathrm{i}\hbar \sum_p G^{(2),\uparrow\downarrow\uparrow\downarrow}_{ipjp}\\
    G^{\uparrow\uparrow}_{ij} &= - \mathrm{i}\hbar \sum_p G^{(2),\uparrow\downarrow\uparrow\downarrow}_{ippj} \,
\end{align}
and
\begin{align}\label{eq:3to2_traces}
    \left(\frac{N}{2}-1\right) G^{(2),\uparrow\downarrow\uparrow\downarrow}_{ijkl} &= - \mathrm{i}\hbar \sum_p G^{(3),\uparrow\uparrow\downarrow\uparrow\uparrow\downarrow}_{ipjkpl}\\
    \frac{N}{2} G^{(2),\uparrow\uparrow\uparrow\uparrow}_{ijkl} &= - \mathrm{i}\hbar \sum_p G^{(3),\uparrow\uparrow\downarrow\uparrow\uparrow\downarrow}_{ijpklp}\\
    G^{(2),\uparrow\uparrow\uparrow\uparrow}_{ijkl} &= - \mathrm{i}\hbar \sum_p G^{(3),\uparrow\uparrow\downarrow\uparrow\uparrow\downarrow}_{ipjklp}\\
    G^{(2),\uparrow\uparrow\uparrow\uparrow}_{ijkl} &= - \mathrm{i}\hbar \sum_p G^{(3),\uparrow\uparrow\downarrow\uparrow\uparrow\downarrow}_{ijpkpl} \,,
\end{align}
respectively.
Despite the clear physical nature of the DSL approximation, which was established in Sec.~\ref{sss:dsl-comparison}, it is known to violate the trace consistency relations between the three- and the two-particle Green function since the neglected three-particle correlation part is not trace free, i.e. $\mathrm{Tr}\,\mathcal{G}^{(3)} \neq 0$, see the discussion in Sec.~\ref{ss:econs-cc}.

A method to restore contraction consistency of the approximate three-particle reduced density matrix that is independent of the applied reconstruction functional was presented by Lackner \textit{et al.}~\cite{lackner_propagating_2015}.
The general idea is to construct an additional correction term for the two-particle EOM in such a way that it restores the trace of the neglected three-particle correlation part $\mathcal{G}^{(3)}$. This scheme is based on the unitary decomposition, cf. Ref.~\cite{mazzioti_reduced_2007}, which will be illustrated with the following simple example for a two-particle quantity. Let $Y_{ijkl}$ be a two-particle quantity with two contraction relations to the single-particle level, $X_{ij} = \sum_{k}Y_{kikj}$ and $X_{ij} = \sum_{k}Y_{ikjk}$. Now one can construct a second two-particle quantity $Y^\tn{CC}$, which possesses the same trace relations as $Y$, without knowledge of the latter. This is done by expanding the single-particle property $X$ using Kronecker deltas to obtain
\begin{align}
   Y^\tn{CC}_{ijkl} = \frac{1}{N} \delta_{ik} X_{jl} + \frac{1}{N} \delta_{jl} X_{ik} - \frac{1}{N^2} \delta_{ik} \delta_{jl} \sum_p X_{pp}\,,
\end{align}
where $N$ is the basis size. The first (second) term on the right hand side directly reproduces the first (second) trace relation while the third term negates the respective redundant contribution for each case.\\

In the G1--G2 scheme contraction consistency is enforced by adding a similar correction term to the EOM of the two-particle correlation Green function 
\begin{align}
 \i\hbar  \frac{\d}{\d t} \mathcal{G}^{\uparrow\downarrow\uparrow\downarrow}_{ijkl}(t) &- \Big[ h^{(2),\tn{HF}}_{\uparrow\downarrow}(t),\mathcal{G}^{\uparrow\downarrow\uparrow\downarrow}(t) \Big]_{ijkl} = \Psi^{\uparrow\downarrow\uparrow\downarrow}_{ijkl}(t) \nonumber \\
 &+L^{\uparrow\downarrow\uparrow\downarrow}_{ijkl}(t) + \Pi^{\uparrow\downarrow\uparrow\downarrow}_{ijkl}(t) + \Lambda^{\tn{ph},\uparrow\downarrow\uparrow\downarrow}_{ijkl}(t) + C^{\tn{CC}}_{ijkl}(t)
 \, , \label{eq:G1-G2_DSL_compact_Hubbard_CC}
\end{align}
with the symmetrized correction
\begin{align}
 C^{\tn{CC}}_{ijkl}(t) = C_{ijkl}(t) + C_{jilk}(t) + C^*_{klij}(t) + C^*_{lkji}(t) \,,
\end{align}
where in the Hubbard model the correction part is given by
\begin{align}
 C_{ijkl}(t) = \pm \mathrm{i}\hbar U \left( G^{(3),\tn{CC}}_{iijkil}(t) +  G^{(3),\tn{CC}}_{jiilik}(t) \right)\,.
\end{align}
In the case of a general basis, \refeqn{eq:g12-equation-as}, it has the form
\begin{align}
 C_{ijkl}(t) = \pm \mathrm{i}\hbar \sum_{pqr} w_{ipqr} G^{(3),\tn{CC}}_{qrjkpl}(t)\,.
\end{align}
Following Lackner et al.~\cite{lackner_propagating_2015} the contraction consistent correction to the three-particle Green function (with arbitrary symmetry) $G^{(3),\tn{CC}}$ can be constructed as
\begin{align}
    &G^{(3),\tn{CC}}_{i_1i_2i_3j_1j_2j_3} = \sum_{k=1}^6 \sum_{\tau\in S_3} a^k_\tau\, \delta_{i_{\tau(1)}j_1}\, \delta_{i_{\tau(2)}j_2}\, \delta_{i_{\tau(3)}j_3}\, {}^kM^{(0)}\\
    &\qquad+ \sum_{k=1}^{18} \sum_{\substack{\sigma,\tau\in S_3\\ \sigma(1) < \sigma(2)}} b^k_{\tau\sigma}\, \delta_{i_{\tau(1)}j_{\sigma(1)}}\, \delta_{i_{\tau(2)}j_{\sigma(2)}}\, {}^kM^{(1)}_{i_{\tau(3)}j_{\sigma(3)}}\\
    &\qquad+ \sum_{k=1}^{9} \sum_{\sigma,\tau\in S_3} c^k_{\tau\sigma}\, \delta_{i_{\tau(1)}j_{\sigma(1)}} {}^kM^{(2)}_{i_{\tau(2)}i_{\tau(3)}j_{\sigma(2)}j_{\sigma(3)}}\,,\label{eq:CC_reconstruction}
\end{align}
where $S_3$ denotes the permutation group of three elements. The general structure of this reconstruction follows the above example for $Y^\tn{CC}$ with the terms in the last line mainly ensuring the correct trace relations and the terms in the first two lines negating abundant contributions. The main contribution to the correction are the zero-, one- and two-particle quantities $M^{(0)}$, $M^{(1)}$ and $M^{(2)}$, respectively, which are expanded to three-particle quantities using Kronecker deltas. They are constructed by (partial) traces over $M^{(3)} = \mathcal{G}^{(3)} = G^{(3)} - G^{(3),\tn{DSL}} $ which are known functions of $G^{(2)}$, to guarantee that trace consistency is restored. Their specific form and more details on the coefficients $a$, $b$ and $c$ are given in Appendix~\ref{app:cc}.

\subsection{Purification}

Beyond contraction consistency the general trace relation, \refeqn{eq:fs-def}, leads to the issue of $N$-representability~\cite{coleman2000reduced,mazzioti_reduced_2007}. Since the $N$-particle density matrix is directly connected to an $N$-particle wave function, the reduced density matrices have to obey certain positivity conditions demanding positive semidefiniteness of the respective density matrices~\cite{mazzioti_reduced_2007}. For the Green functions the according conditions on the single-particle level become
 \begin{align}
  \pm \tn{i}\hbar G^<_{ij} = \pm \tn{i}\hbar G^{\tn{p}}_{ij} &= \bra{\Psi} \hat{c}^\dagger_i \hat{c}_j \ket{\Psi} \succeq 0\\
  \pm \tn{i}\hbar G^{\tn{h}}_{ij} &= \bra{\Psi} \hat{c}_i \hat{c}^\dagger_j \ket{\Psi} \succeq 0\,.
 \end{align}
 On the two-particle level there are three necessary but not sufficient conditions that have to be fulfilled by the two-particle, two-hole and particle--hole Green function, respectively
 \begin{align}\label{eq:2p_positivity}
      \left(\tn{i}\hbar\right)^2 G^{(2)}_{ijkl} = \left(\tn{i}\hbar\right)^2 G^{\tn{pp}}_{ijkl} &= \bra{\Psi} \hat{c}^\dagger_i \hat{c}^\dagger_j \hat{c}_l \hat{c}_k \ket{\Psi} \succeq 0\\
      \left(\tn{i}\hbar\right)^2 G^{\tn{hh}}_{ijkl} &= \bra{\Psi} \hat{c}_i \hat{c}_j \hat{c}^\dagger_l \hat{c}^\dagger_k \ket{\Psi} \succeq 0\\
      \left(\tn{i}\hbar\right)^2 G^{\tn{ph}}_{ijkl} &= \bra{\Psi} \hat{c}^\dagger_i \hat{c}_j \hat{c}^\dagger_l  \hat{c}_k \ket{\Psi} \succeq 0\,.
\end{align}
A similar issue is known for two-time Green functions where the self-energy has to be positive semidefinite to guarantee a positive spectral function and a stable numerical propagation~\cite{stefanucci_diagrammatic_2014,uimonen_diagrammatic_2015}. Note that without the factor of $\left(\tn{i}\hbar\right)^2$ all two-(quasi-)particle Green functions have to be negative semidefinite. In Fig.~\ref{fig:instability1}(d) the instability of TOA and DSL is accompanied by an increasing, positive largest eigenvalue, indicating that N-representability is violated in these cases.\\
Like contraction consistency the issue of $N$-representability is well-studied for two-particle density matrices and the developed method to enforce the few necessary positivity conditions is called purification~\cite{mazzioti_reduced_2007}. In contrast to the procedure for restoring contraction consistency, purification does not lead to an additional term in the equation of motion but instead is applied subsequently, after the numerical propagation step is completed~\footnote{For completeness, we mention recent developments~\cite{schmelcher}, in which the purification procedure is integrated into a modified equation of motion.}.
As a consequence, the various purification schemes developed for the equilibrium setting~\cite{mazzioti_reduced_2007} can lead to the violation of conservation laws when applied for nonequilibrium dynamics. Therefore, the goal of the procedure used in this work is to enforce the $N$-representability conditions while still preserving contraction consistency, and the conservation of particles and energy. A detailed description of the purification procedure is given in Appendix~\ref{app:purification}. 

\section{Numerical results}\label{s:numerics}
In our recent benchmark study \cite{schluenzen_prb17} against DMRG results we could verify  that HF-GKBA simulations with rather sophisticated selfenergies, including TPP and TOA, are a powerful numerical method: 
these results are very close to the ``exact'' DMRG calculations, even for moderate to strong interactions ($U\sim4J$). Moreover, NEGF simulations are not restricted to one-dimensional systems like DMRG but are easily extended to higher dimensions. Despite this remarkable potential, in practice the method turned out to be restricted to rather short time scales because of both, its cubic scaling with the propagation time and its inherent (not numerical) instability observed for long propagations.
The G1--G2 scheme, being an exact reformulation of the HF-GKBA inherits all positive properties observed in our previous paper. On top of that it provides, in principle, access to much longer time scales due to its reduced numerical complexity (linear time scaling). To achieve these long times, we have to remove intrinsic instabilities by enforcing contraction consistency and performing a purification. In this paper we apply this procedure to our most advanced approximation---DSL. These results will be called DSL$^*$ below. We also present comparisons with TOA results because its validity range is independent of the particle density (filling factor) \cite{schluenzen_jpcm_19}.

The numerical results presented below confirm both statements: The results of the G1--G2--DSL$^*$ simulations are, for short times, qualitatively comparable with the best results we could achieve previously using TPP and TOA with the standard HF-GKBA, e.g. in Ref.~\cite{schluenzen_prb17}.
Here we present new numerical tests 
for small to medium size 1D Hubbard clusters (because here exact and DMRG benchmarks are available), for small to moderate coupling strength. Moreover, we study different excitation conditions by varying how far the system is driven out of equilibrium.

\subsection{Influence of deviation from equilibrium}
For the detailed characterization of the time-dependent dynamics in excited lattice systems, we use the total energy and its individual contributions,
\begin{align}
    E(t) &= E_\tn{kin}(t) + E_\tn{int}(t)\,,\quad \tn{with} \label{eq:etot-def}\\
    E_\tn{int}(t) &= E_\tn{HF}(t) + E_\tn{corr}(t)\, . \label{eq:int_energy}
\end{align}
Here, the kinetic part, the HF energy, and the correlation energy are explicitly given as 
\begin{align}
    E_\tn{kin}(t) &= - J \sum_{ij} \delta_{\left<i,j\right>} n_{ji}(t)\, , \label{eq:kinetic_energy} \\
    E_\tn{HF}(t) &= U(t) \sum_{i}n^\uparrow_{ii}(t)n^\downarrow_{ii}(t) \, , \label{eq:HF_energy}\\
    E_\tn{corr}(t) &= \left(\i\hbar\right)^2 U(t) \sum_{i} \mathcal{G}_{iiii}^{\uparrow\downarrow\uparrow\downarrow}(t)\, . \label{eq:corr_energy}
\end{align}

As a first test setup we consider a finite six-site Hubbard chain that is excited by an instantaneous interaction quench. This type of excitation retains spatial homogeneity and keeps the system close to  equilibrium. 
\begin{figure}[ht]
\includegraphics[width=\columnwidth]{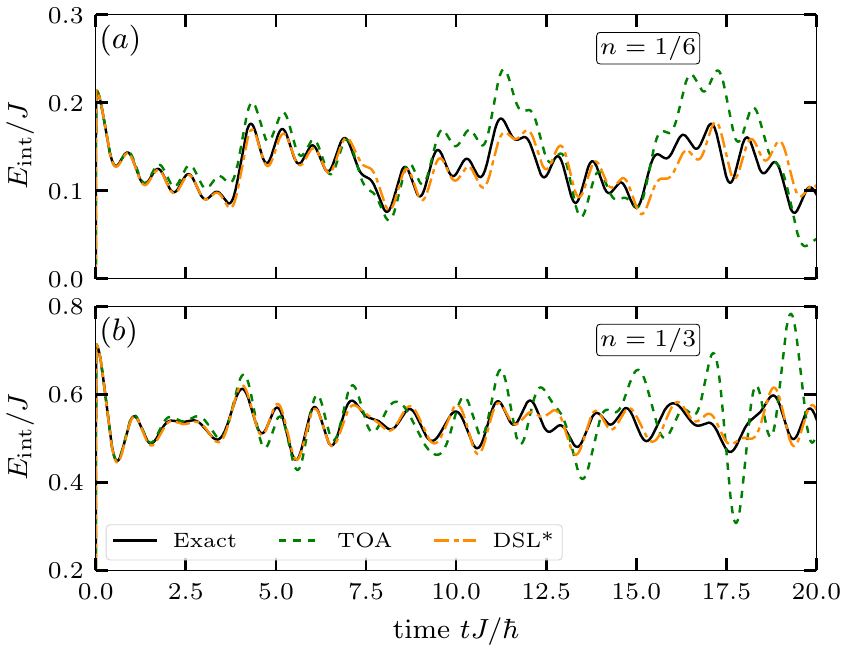}
\caption{Evolution of the interaction energy following an interaction quench. The ideal groundstate is prepared for a six-site Hubbard chain after which the interaction is instantaneously switched from $U=0$ to $U=J$ at $t=0$. Panel (a) corresponds to a filling of $n=1/6$ and (b) to $n=1/3$.
Exact results are shown in black. The green (orange) curves correspond to the TOA (DSL*) results.}
\label{fig:quench}
\end{figure}
A particularly  interesting observable for this setup is the time-dependent interaction energy, $E_\tn{int}$, of the system [cf. Eq.~\eqref{eq:int_energy}], which is shown in \fref{fig:quench} for two different electronic filling ratios [\fref{fig:quench}(a): $n=1/6$, \fref{fig:quench}(b): $n=1/3$]. The exact benchmark results (black) reveal a sudden jump to an energy maximum at $t=0$, followed by unsteady, intermittent oscillations. This behavior results from the energy transfer between $E_\tn{int}$ and the kinetic energy, overlayed by finite-size effects. Previous simulations with the TOA selfenergy (green) capture the general trends of the dynamics, but fail to predict the correct oscillation amplitudes, eventually leading to unreasonably pronounced energy peaks (cf. $15\leq tJ/\hbar \leq 20$ for both fillings). With our new DSL* scheme (orange), we find a close agreement with the exact benchmark data with only slight derivations towards the end of the propagation. The approach produces consistently accurate results for both considered filling ratios.

\begin{figure}[ht]
\includegraphics[width=\columnwidth]{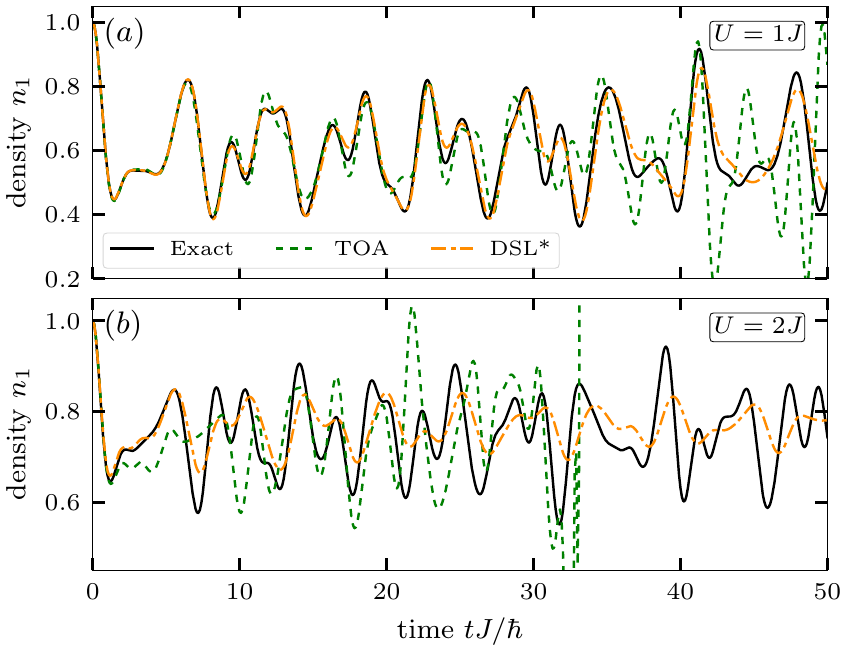}
\caption{Density dynamics on the first site of a six-site Hubbard system following a sudden switch on of an external potential, for (a) $U=1$ and (b) $U=2$. The dynamics start from the interacting groundstate (generated by the adiabatic-switching and, at $t=0$, a constant local potential of $w_0=J$ is applied to the first site.
Exact results are shown black. The green (orange) curves correspond to the TOA (DSL*) results.}
\label{fig:switch_on}
\end{figure}

Next,  in \fref{fig:switch_on} we consider a setup with a stronger spatial perturbation.
Here, the finite Hubbard cluster is propagated from the half-filled interacting groundstate and excited by a sudden increase of the on-site potential at the first site according to [cf. Eq.~\eqref{eq:h-hubbard}]
\begin{align}
 \chat{F}(t) = w_0 \Theta\left(t\right) \left(\chat{n}_1^\uparrow + \chat{n}_1^\downarrow\right) \, ,
\label{eq:h-ex}
\end{align}
with $w_0=J$. This procedure induces density oscillations throughout the finite system. The density on the first (excited) site is shown for $U=1J$ [\fref{fig:switch_on}(a)] and $U=2J$ [\fref{fig:switch_on}(b)]. Clearly, the DSL* results (orange) again exhibit a far superior accuracy in comparison to the exact benchmark data (black) than the TOA calculations (green). For $U=2J$ however, we observe growing deviations that indicate that the omitted many-body correlations gain importance in the intermediate- to strong-coupling regime. It should be noted that the TOA calculation for this case becomes unstable and reaches unphysical values, which is successfully overcome by the DSL* approach. 

\begin{figure}[ht]
\includegraphics[width=\columnwidth]{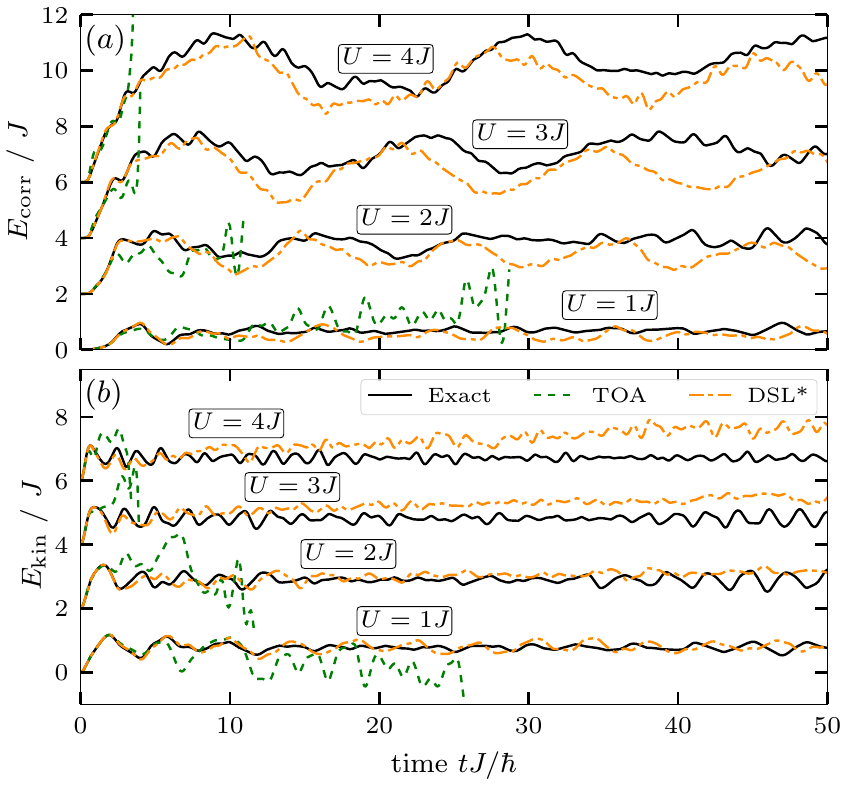}
\caption{Same setup as in Fig.~\ref{fig:instability1} but for various interactions from $U=J$ to $U=4J$. (a) and (b): correlation and kinetic energy of the system, respectively. Each curve beyond $U=J$ is shifted vertically by $2J$ with respect to the previous one. The TOA results (dashed green line) become increasingly unstable (diverge earlier) for increasing interaction strength.}
\label{fig:akbari_corr_kin}
\end{figure}

We now analyze the performance of our approach under the influence of strong nonequilibrium conditions. To this end, we return to the confinement setup discussed in \fref{fig:instability1}. We again consider a six-site Hubbard chain with all particles initially confined in the left half of the system, but now for different choices of the interaction strength. Interesting methodological insights can be gained from the individual energy contributions [cf. \eqrefrno{eq:kinetic_energy}{eq:corr_energy}{eq:HF_energy}] which are shown in \fref{fig:akbari_corr_kin}.
For the correlation energy [\fref{fig:akbari_corr_kin}(a)] the exact calculations (black) show an initial energy increase followed by broad oscillations that become more pronounced, but significantly slower for larger values of $U$. While the TOA calculations (green) exhibit severe instability problems especially for strong coupling, the DSL* results (orange) show the correct trends of the dynamics. However, our approach seems to consistently overestimate the oscillation frequencies. At the same time, the exact kinetic energy [\fref{fig:akbari_corr_kin}(b)] exhibits high-frequency oscillations around a characteristic constant base level. While our DSL* calculations capture the oscillations, they show an unreasonable drift towards increasing base levels. Both effects are possibly caused by the applied contraction-consistency and purification procedures. For the description of such complex states featuring an intricate interplay between strong nonequilibrium effects and higher-order many-body correlations these small gradual corrections inevitably start to alter the intrinsic consistency of the propagations. On the other hand, these side effects appear on a tolerable scale, especially in comparison with the numerically unstable TOA results.

As a final example, we study the relaxation of a system being initially in a charge-density wave (CDW) state. In this extreme nonequilibrium setup the electrons are initially confined in an alternating chain of empty and fully occupied lattice sites. To establish the properties of such a state in a Hubbard system, we use a larger chain of $L=20$ lattice sites. At this scale, benchmark data by exact-diagonalization methods are no longer available. For this reason, we use quasi-exact DMRG results for comparison (see Ref.~\cite{schluenzen_prb17} for details of the approach).
\begin{figure}[ht]
\includegraphics[width=\columnwidth]{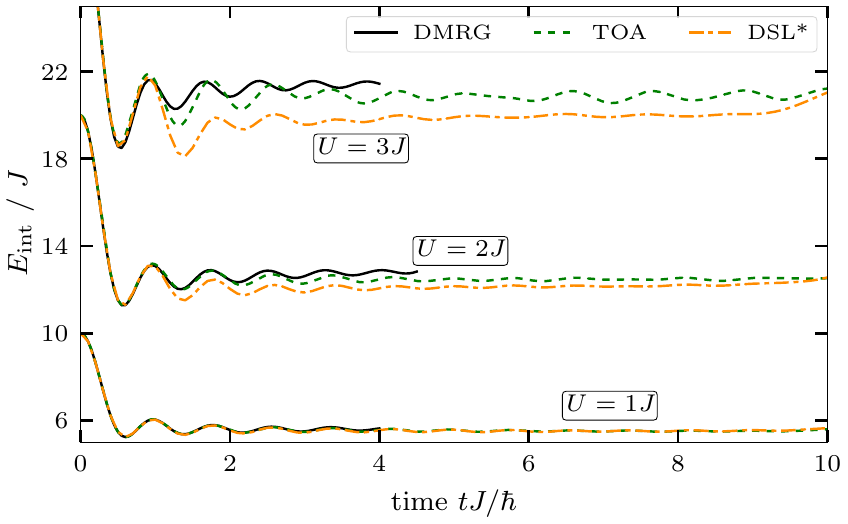}
\caption{Relaxation of a CDW state of doublons. Shown is the interaction energy for $U/J = 1, 2, 3$ and $L = N = 20$. The present DSL results with contraction consistency and purifciation (DSL*, orange) are compared to the DMRG benchmark (black) and to G1--G2--TOA (green) simulations.
}
\label{fig:totDocA}
\end{figure}
In \fref{fig:totDocA} we show results for the interaction energy for three different coupling strengths.
The overall behavior features a sudden decrease of $E_\tn{int}$ as part of the energy is transferred to particle motion, i.e. $E_\tn{kin}$, after which small oscillations appear, decaying into a prethermalized steady energy level. Again, we find increasing deviations between the DMRG data (black) and the DSL* scheme (orange) with larger interaction strengths due to the higher complexity of the dynamical many-body state. Surprisingly, we observe that the TOA approach (green) exhibits a better agreement with the DMRG results. At the moment no strict explanation for this behavior is known. A possible reason could be the high symmetry between the dynamics of the electrons and holes in the system. Due to cancellation effects of higher-order scattering diagrams the TOA scheme might already be quite accurate while the DSL* approximation involves partial resummations of the respective higher-order terms. 

We conclude that, overall, the DSL* scheme provides excellent results in the regime of weak to moderate coupling and significantly improves the reach of the NEGF method in comparison to previous approaches. Under strong nonequilibrium conditions (such as a CDW state) we find a slightly decreased accuracy which we hypothesize, is due to the increased complexity of the corresponding nonequilibrium many-body states.  

\section{Discussion and Outlook}\label{s:discussion}
In this paper we discussed the nonequilibrium dynamically screened ladder approximation (DSL) which fully selfconsistently combines particle--particle, particle--hole ladder ($T$-matrix) diagrams and polarization (bubble) diagrams. This allows to take dynamical screening and strong-coupling effects into account simultaneously. This extends earlier ground-state and equilibrium results that have been obtained in the frame of the Bethe--Salpeter equation by Zimmermann \textit{et al.} \cite{zimmermann_pss_78}, Kremp \textit{et al.} \cite{kremp-springer} and others to systems driven ouf of equilibrium by an external excitation to ultrafast processes and strong fields \cite{bonitz_99_cpp}. The most important application are correlated electrons and excitons in graphene and TMDC monolayers and bilayers exposed to laser fields.

A direct extension of the Bethe--Salpeter approach of Green functions theory  turned out to be unsuccessful so far; the question of a nonequilibrium DSL selfenergy is still open. However, the independent approach of reduced density operators (RDO) or, equivalently, density matrices (TD2RDM) provided a direct solution for DSL. Performing a cluster expansion of the RDO and neglecting three-particle correlations, $g^\pm_{123}\to 0$, we obtained the equation of motion for the pair-correlation operator and, correspondingly, for the two-particle Green function,  ${\cal G}_{12}$. The result are the equations of the G1--G2 scheme  on the DSL level \cite{schluenzen_19_prl, joost_prb_20}. In contrast to earlier versions of the equations, the present result fully includes exchange effects, also in the particle--hole $T$-matrix and $GW$ diagrams.

Our DSL-G1--G2 simulations for finite Hubbard systems confirmed the previously reported linear scaling with the simulation duration \cite{schluenzen_19_prl} paving the way to long simulations that are needed to simulate pump--probe experiments and to produce accurate energy spectra. However, when the coupling strength was increased the simulations were found to become unstable even though all conservation laws are accurately satisfied. The reason turned out to be the missing contraction consistency between the two-particle and three-particle reduced density operator. Enforcing contraction consistency by including additional contributions from the three-particle correlations improved the behavior of the solution. However for a stabilization of the simulations it was necessary to apply, in addition, a purification scheme that eliminates positive eigenvalues of the two-particle Green function, thereby partially restoring N-representability. Our solution improves previous purification schemes \cite{lackner_propagating_2015,lackner_pra_17} 
by maintaining total-energy conservation. 

The resulting DSL* simulations were tested against exact benchmark results (CI and DMRG) and showed very good accuracy. Best performance was observed in weak to intermediate nonequilibrium situations, whereas for extreme nonequilibrium conditions, such as a charge-density-wave initial state, the accuracy was slightly worse.

For future developments of NEGF theory it will be very interesting to derive a selfenergy that is equivalent to the DSL-G1--G2 approximation. Furthermore, for the improvement of the G1--G2 scheme it will be important to test more advanced approximations of reduced-density-operator theory that go beyond DSL and partially include three-particle correlation effects, i.e., contributions to $g_{123}^\pm$, such as the 
     the Nakatsuji--Yasuda approximation \cite{nakatsuji_prl_96,yasuda_pra_97}, approximations by Maziotti
     \cite{maziotti_pra_99,MAZZIOTTI_cpl_00},
     the selfconsistent RPA \cite{schuck_epja_16}, or the Fadeev approximation \cite{pavlyukh_prb_21}.

\section*{Acknowledgements}
This work was supported 
by grant shp00026 for CPU time at 
the Norddeutscher Verbund f\"ur Hoch- und H\"ochstleistungsrechnen (HLRN).
WWTF Grant No. MA14-002, the International Max Planck
Research School of Advanced Photon Science (IMPRSAPS), and the FWF doctoral school Solids4Fun.

\appendix

\section{Derivation of the (anti-)symmetrized equation for the pair correlation operator
}\label{app:g12-equation}

In this appendix we provide the main steps that lead to equation  (\refeq{eq:g12-equation-as}).
We start by summarizing the main properties of the permutation operators, $\hat P_{ij}$ and (an\-ti-)symmetrization operators $\lambda^\pm$ that will be needed for the derivations below.
\begin{enumerate}
\item The permutation operator obeys,
\begin{align}
    \hat P_{ij}^2 &= \hat 1,
\nonumber\\
\mbox{Tr}_{j} \hat P_{ij} &= \mbox{Tr}_{i} \hat P_{ij} = \hat 1,
\label{eq:trace_pij}\\
\hat P_{ij} A_{ij}  &= A_{ji} \hat P_{ij}\,,
\label{eq:pij-commutator}
\end{align}
where $A_{ij}$ is an arbitrary two-particle operator.
\item Pair permutations of different particles don't commute, i.e., $\hat P_{ij} \hat P_{jk}\ne \hat P_{jk} \hat P_{ij}$. For the case of three-particle states, the three different permutations, labeled by $\alpha, \beta, \gamma$ (denoting $12, 13$ and $23$), have the properties
\begin{align}
    \hat P_\beta \hat P_\gamma = \hat P_\alpha \hat P_\beta
    \label{eq:3permutation1}\,,\\
    \hat P_\gamma \hat P_\beta  = \hat P_\beta \hat P_\alpha \,.
    \label{eq:3permutation2}
\end{align}
\item Important properties of $\lambda^\pm$ are
\begin{align}
    (\hat 1-\epsilon \hat P_{ij})\lambda^\pm_{12} &= 0,\quad \epsilon\ne 0\,,
    \nn\\
    \lambda^\pm_{12}(\hat 1+\epsilon \hat P_{12}+\epsilon \hat P_{13})&=(\hat 1+\epsilon \hat P_{12}+\epsilon \hat P_{13})\lambda^\pm_{12}
        \nn\\
    \hat P_{23}\lambda^\pm_{13} &= \lambda^\pm_{12}\hat P_{23}    
        \nn\\
    (\lambda^\pm_{1\dots s})^2 &= s! \lambda^\pm_{1\dots s}\,,
    \label{eq:lambdapm-quadrat}
\end{align}
\end{enumerate}
All these relations are easily proven  by direct calculation.

Now we turn to hierarchy equations. Recall the (anti-)symmetrized first hierarchy equation, \refeqn{eq:bbgky1-as} of the main text:
\begin{align}
    \i\hbar \frac{\d}{\d t} &F_{1} - \left[{H}_{1}+U_1^{\rm HF},F_{1}\right] =
    \mbox{Tr}_2 [V_{12},g^\pm_{12}]=I_{1,\rm DO}\,,\quad
    \label{app:bbgky1-as}
\end{align}
with 
 the (anti-)symmetrized version of the collision integral and the Hartree--Fock potential energy operator,
\begin{align}
    U_1^{\rm HF} &= \mbox{Tr}_2 V_{12}^\pm F_2\,,
    \label{aoo:uhf-def}\\
    V_{12}^\pm &= V_{12}\lambda_{12}^\pm\,.
    \label{app:vpm-def}
\end{align}
For the derivation of the equation for $g_{12}^\pm$ we start from the second hierarchy equation, \refeqn{eq:bbgky_finite}, where for all density operators we use their (anti-)symmetrized versions and insert the cluster expansion, \refeqn{g12-def}, for $F_{12}$ and $F_{123}$,
\begin{align}
\nn
&\i\hbar \frac{\d}{\d t}\Big(F_{1}F_{2}\lambda^{\pm}_{12}+g^\pm_{12}\Big)
-[H^0_{12}+V_{12},F_{1}F_{2}\lambda^{\pm}_{12}+g^\pm_{12}]
\\
&=
\mbox{Tr}_{3}\Big\{[V_{13}+V_{23},F_{1}F_{2}F_{3}]
+ [V_{13}+V_{23},F_{1}g_{23}]
\nonumber\\
&\quad +
[V_{13}+V_{23},F_{2}g_{13}]
+ [V_{13}+V_{23},F_{3}g_{12}]
\nonumber\\
&\quad +
[V_{13}+V_{23},g_{123}]\Big\}(1+\epsilon \hat P_{13}+\epsilon \hat P_{23})\lambda^{\pm}_{12}\,.
\label{app:g12-as-1}
\end{align}
From this we subtract the equation of motion of $F_{1}F_{2}\lambda^{\pm}_{12}$ that directly follows from \refeqn{app:bbgky1-as} by multiplying with $F_2\lambda_{12}^\pm$. Now we group terms and perform the following transformations:
\begin{enumerate}
    \item The commutator $[V_{12},F_1F_2\lambda^\pm_{12}]$, in the second term on the left of Eq.~(\refeq{eq:g12-as-1}), is not cancelled by subtracting the equation of motion for $F_1F_2\lambda_{12}^\pm$. It contributes to the inhomogeneity, $\Psi^\pm_{12,\rm DO}$, \refeqn{eq:psi-pm-do}. The remaining Pauli blocking terms in $\Psi^\pm_{12,\rm DO}$ arise from the terms involving the product $F_1F_2F_3$, on the r.h.s. of Eq.~(\refeq{eq:g12-as-1}). Consider the transformation for a typical term:
    \begin{align}
        \mbox{Tr}_3 V_{13}F_1 F_2 F_3  \hat P_{23}\lambda^\pm_{12} &= 
F_2 \mbox{Tr}_3 V_{13}F_1 F_3  \hat P_{23}\lambda^\pm_{12}
\nonumber\\
F_2 \mbox{Tr}_3 \hat P_{23} V_{12}F_1 F_2  \lambda^\pm_{12} &=
F_2 \mbox{Tr}_3 \hat P_{23} V_{12}\lambda^\pm_{12}F_1 F_2  =
\nonumber\\
& =F_2 V^\pm_{12}F_1 F_2\,,
\label{eq:pauli_fff1}
    \end{align}
    where, in the last step, we used property (\refeq{eq:trace_pij}). There is a analogous term following from interchanging $1\leftrightarrow 2$.
    Collecting all terms, and also the first part of the commutator, $[V_{12},F_1F_2\lambda^\pm_{12}]$, we obtain
\begin{align}
 & V_{12}F_1F_2\lambda^\pm_{12} + 
 \nn\\
 & \mbox{Tr}_3 (V_{13}+V_{23})F_1 F_2\lambda^\pm_{12} F_3 (1+\epsilon \hat P_{13} + \epsilon \hat P_{23}) \nn\\
&= (H_1^{\rm HF}+H_2^{\rm HF})F_1 F_2\lambda^\pm_{12} + \hat V^\pm_{12}F_1 F_2\,,
    \label{eq:pauli_fff}
\end{align}
where $\hat V$ was defined in \refeqn{eq:vhat}. Similarly, the second part of the commutators is transformed into 
\begin{align}
& F_1F_2\lambda^\pm_{12}V_{12} + 
\nn\\
&\mbox{Tr}_3 F_1 F_2 F_3\lambda^\pm_{12} (1+\epsilon P_{13} + \epsilon P_{23})(V_{13}+V_{23}) 
\nn\\
&= F_1 F_2\lambda^\pm_{12}(H_1^{\rm HF}+H_2^{\rm HF}) + F_1 F_2\hat V^{\pm\dagger}_{12}\,,
    \nn
\end{align}
where the Hartree--Fock terms are cancelled when the equation for $F_1 F_2\lambda^\pm_{12}$ is subtracted.
    
\item We now consider the ladder term, \refeqn{eq:l-pm-do}. In addition to the spinless contribution, $L^0_{\rm DO}$, it also involves Pauli blocking terms that arise from terms on the r.h.s. of Eq.~(\refeq{eq:g12-as-1}) of the following form  
\begin{align}
    \mbox{Tr}_{3}V_{13}F_{2}g_{13}\hat P_{23}\lambda^\pm_{12}
    &= F_2\mbox{Tr}_{3}\hat P_{23} V_{12}F_{2}g_{12}\lambda^\pm_{12}=
    \nn\\
    &=F_{2}V_{12}g^\pm_{12}\,,
\end{align}
where we used relations (\refeq{eq:pij-commutator}) and (\refeq{eq:trace_pij}). Collecting all terms that contribute to the first parts of the commutators, we obtain
\begin{align}
    V_{12}g^\pm_{12} &+\epsilon \mbox{Tr}_{3}
    \left\{
    V_{13}F_{2}g_{13}\hat P_{23} +
    V_{23}F_{1}g_{23}\hat P_{13}
    \right\}
    \lambda^\pm_{12} =
    \nn\\
    &= \hat V_{12}g_{12}^\pm\,.
\end{align}
Similarly, the second parts of the commutators lead to 
\begin{align}
    g^\pm_{12}V_{12} &+\epsilon \mbox{Tr}_{3}
    \left\{
    F_{2}g_{13}V_{13}\hat P_{23} +
    F_{1}g_{23}V_{23}\hat P_{13}
    \right\}
    \lambda^\pm_{12} =
    \nn\\
    &= g_{12}^\pm\hat V_{12}^\dagger\,,
\end{align}
which gives the complete ladder term, \refeqn{eq:l-pm-do}.
\item The three-particle correlation term on the r.h.s. of \refeqn{eq:g12-equation-as} follows by applying the three-particle (anti-)symmetrization operator directly to $g_{123}$ which yields $g_{123}^\pm$ instead of $g_{123}$. But otherwise this term has the same form as for the spinless case, cf. \refeqn{eq:g12-equation}.
\item We now consider all the remaining terms that contain products of one-particle and two-particle density operators,
\begin{align}
    \mbox{Tr}_3
    V_{13}F_3g_{12}(\epsilon \hat P_{13}+\epsilon \hat P_{23})\lambda^\pm_{12} +\quad
    \label{app-rest}\\
    \mbox{Tr}_3 V_{23}F_2g_{13}(1+\epsilon \hat P_{13}+\epsilon \hat P_{12}+\hat P_{13}\hat P_{12}) + 1 \leftrightarrow 2
    \,,
    \label{app-rest2}
\end{align}
and, similar for the second parts of the commutators. The one in the second parentheses leads to the classical polarization term, \refeqn{eq:pi-rdo-01}. When, in addition, the term $\epsilon \hat P_{13}$ is included, we obtain $\mbox{Tr}_3 V_{23}F_2g_{13}(1+\epsilon \hat P_{13})=\mbox{Tr}_3 V_{23}F_2g^\pm_{13}$---the classical result with $g$ replaced by $g^\pm$.

Let us now analyze the remaining six terms in \refeqn{app-rest}. We start by transforming the first two contributions from the first term (of the total of four terms):
\begin{align}
    \mbox{Tr}_3V_{13}F_3g_{12}\epsilon \hat P_{13} &= \mbox{Tr}_3V_{13}\epsilon \hat P_{13}F_1g_{23} \,,
     \nn\\
    \mbox{Tr}_3V_{13}F_3g_{12}\epsilon \hat P_{23}\epsilon \hat P_{12} &= \mbox{Tr}_3V_{13}\epsilon \hat P_{13} F_1g_{23}\epsilon \hat P_{23}\,,
\nn
\end{align}
where, in the second expression, we used $\hat P_{23}\hat P_{12}=\hat P_{13}\hat P_{23}$. Adding these two expressions and the classical polarization term discussed above (where we exchange $1\leftrightarrow 2$) yields
\begin{align}
    \Pi^{\pm(1)}_{12,\rm DO} = \mbox{Tr}_{3}[V_{13}^{\pm},F_1]g_{23}^{\pm}\,,
    \label{app:pi1_do}
\end{align}
thus the previous result was ``upgraded'' once more, by replacing $V_{12} \to V_{12}^\pm$.
This is the result that was derived in Refs.~\cite{dufty-etal.97} and ~\cite{bonitz_qkt} and describes polarization effects including exchange effects (in the Green functions language this corresponding to $GW$ with exchange). 

There are four terms left from \refeqn{app-rest} which we transform such that an (anti-)symmetric pair correlation operator is produced (note that we have to retain the index change $1 \leftrightarrow 2$ in the second term),
\begin{align}
 R^{(1)x}    = &   \mbox{Tr}_3V_{13}F_3g_{12}(\hat P_{13}\hat P_{12}+\epsilon \hat P_{23}) + 
 \nn\\
      &+   \mbox{Tr}_3V_{13}F_1g_{23}(\hat P_{23}\hat P_{12}+\epsilon \hat P_{12})\,.
    \label{app:p1x}
\end{align}
We now transform the two parantheses,
\begin{align}
    \hat P_{13}\hat P_{12} &+\epsilon \hat P_{23} = 
    \epsilon \hat P_{13}\epsilon \hat P_{12} + \epsilon^3 \hat P^2_{13}\hat P_{23} = \nn\\ 
    = (\epsilon \hat P_{13} &+\epsilon^2 \hat P_{13}\hat P_{23})\epsilon \hat P_{12} = \epsilon \hat P_{13}\lambda^\pm_{23}\cdot \epsilon \hat P_{12}\,,
    \nn\\
    \hat P_{23}\hat P_{12} &+\epsilon \hat P_{12} = \lambda^\pm_{23}\cdot \epsilon\hat P_{12}\,,    
\nn
\end{align}
and insert these results into \refeqn{app:p1x}. Using \refeqn{eq:pij-commutator}, an overall factor $\epsilon \hat P_{12}$ can be taken out to the right, with the result
\begin{align}
   R^{(1)x}    = & \big\{
   \mbox{Tr}_3V_{13}\epsilon \hat P_{13}F_1g_{23}\lambda_{23}^\pm + 
 \nn\\
      &+   \mbox{Tr}_3V_{13}F_1g_{23}\lambda_{23}^\pm
      \big\}\cdot \epsilon\hat P_{12}
      \nn\\
      =& \mbox{Tr}_3V^\pm_{13}F_1g^\pm_{23}\cdot\epsilon\hat P_{12}\,.
    \label{app:p1x2}
\end{align}
Together with the second part of the commutator, this yields exactly the polarization term, \refeqn{app:pi1_do} times $\epsilon\hat P_{12}$. Analogously, the symmetric term arising from exchanging $1\leftrightarrow 2$ yields $    \Pi^{\pm(2)}_{12,\rm DO}\cdot \epsilon\hat P_{12}$. Thus, combining these terms with the two terms of the form of \refeqn{app:pi1_do} yields \begin{align}
    \tn{Eq.~}\eqref{app-rest} = \left(\Pi^{\pm(1)}_{12,\rm DO}+\Pi^{\pm(2)}_{12,\rm DO}\right)\lambda_{12}^\pm\,.
\end{align}
\end{enumerate}
Gathering all terms we obtain 
the final result which coincides with \refeqn{eq:g12-equation-as} of the main text and is reproduced here:
\begin{align}
\i\hbar \frac{\d}{\d t}
g^\pm_{12} - [{\bar H}^0_{12}, g^\pm_{12}]
=& \Psi^\pm_{12,\rm DO} + L_{12,\rm DO}
\label{app:g12-equation-as}\\
&+ P^\pm_{12, \rm DO} +
\mbox{Tr}_{3}[V^{(12),3},g^\pm_{123}]
\,,\nn
\\
\Psi^\pm_{12,\rm DO} =& 
\hat{V}^\pm_{12}F_{1}F_{2} - F_{1}F_{2}\hat{V}^{\pm\dagger}_{12}\,,
\label{app:psi-pm-do}
\\
L_{12,\rm DO} =& \hat{V}_{12}\,g^\pm_{12} - g^\pm_{12}\,\hat{V}^\dagger_{12}\,,
\label{app:l-pm-do}
\\
P^\pm_{12, \rm DO} =&
\left(\Pi_{12,\rm DO}^{\pm (1)} + 
\Pi_{12,\rm DO}^{\pm (2)}\right)\lambda_{12}^\pm \,,
\label{app:ppm-do}
\\
\Pi_{12,\rm DO}^{\pm (1)} =& \mbox{Tr}_{3}
    [V^\pm_{13},F_1 g^\pm_{23}]\,,
\label{app:ppm-do1}
\end{align}
where $\Pi_{12,\rm DO}^{\pm (2)}$ follows from $\Pi_{12,\rm DO}^{\pm (1)}$ by exchanging $(1 \leftrightarrow 2)$.

\section{Energy conservation of the G1--G2 scheme}\label{app:econs}
Here we discuss the question of total-energy conservation of the G1--G2 equations that were shown, in the main text, to coincide with the related reduced-density-operator results. We consider two cases. The first is the system of two equations for the one-particle and two-particle density operators, $F_1$ and $F_{12}$, that follows from the BBGKY hierarchy,~\refeqn{eq:bbgky_finite}, and the second is the case that contraction consistency between the one-particle and two-particle density operators is imposed first.
\subsection{Total-energy conservation of the Reduced-density-operator approach}
The energy per particle of a pair is given by 
\begin{align}
    \hat H_{12} = \frac{\hat H_{1} + \hat H_{2}}{2} + \frac{1}{2}\hat V_{12}\,,
\label{app:h12-def}
\end{align}
which yields the expectation value of the total energy of the $N$-particle system
\begin{align}
    \langle \hat H \rangle = \mbox{Tr}_1 \hat H_1 F_{1} + \frac{1}{2}\mbox{Tr}_{12} \hat V_{12} F_{12}
    \label{app:etot}\,,
\end{align}
being completely determined by the single-particle and two-particle reduced density operators, as introduced in Sec.~\ref{ss:rdm-definition}. Energy conservation is readily derived from \refeqn{app:etot} by time differentiation \cite{bonitz_qkt} and using the equations of motion \refeqn{eq:bbgky_finite} of $F_1$ and $F_{12}$:
\begin{align}
    \i\hbar \frac{\d}{\d t} \mbox{Tr}_1 \hat H_1 F_{1} &=\mbox{Tr}_1 \hat H_1 [\hat H_1,F_{1}] + \mbox{Tr}_{12} \hat H_1 [\hat V_{12},F_{12}] = \nonumber\\\label{app:ddte1}
    &=\mbox{Tr}_{12} \hat H_1 [\hat V_{12},F_{12}]\,,\\
    \i\hbar \frac{\d}{\d t} \frac{1}{2}\mbox{Tr}_{12} \hat V_{12} F_{12} &=
    \frac{1}{2}\mbox{Tr}_{12} \hat V_{12} [2\hat H_1, F_{12}] + \nonumber\\
    &+\frac{1}{2}\mbox{Tr}_{12} \hat V_{12} [\hat V_{12}, F_{12}]+\nonumber\\
    &+ \frac{1}{2}\mbox{Tr}_{123} \hat V_{12} [\hat V_{13}+\hat V_{23}, F_{123}]=\nonumber\\
    = -\mbox{Tr}_{12} \hat H_1 [\hat V_{12},F_{12}] &+ \frac{1}{2}\mbox{Tr}_{123} \hat V_{12} [\hat V_{13}+\hat V_{23}, F_{123}]\,,
    \label{app:ddte2}
    \end{align}
where the cyclic invariance of the trace has been used. As a result we obtain, by adding \refeqns{app:ddte1}{app:ddte2},
\begin{align}
    \i\hbar \frac{\d}{\d t} \langle\hat H\rangle &=
    \frac{1}{2}\mbox{Tr}_{123} \hat V_{12} [\hat V_{13}+\hat V_{23}, F_{123}]\,.
    \label{app:ddte12}
\end{align}
Thus total energy is conserved if three-particle density operator is symmetric in the particle indices, cf.~\refeqn{eq:f123-symmetry}
 and Refs.~\cite{dufty-etal.97, bonitz_qkt,akbari_prb_12}. This is fulfilled, of course, for the exact solution whereas for approximations it imposes a (rather weak) symmetry constraint. 

\subsection{Total-energy conservation in case of contraction consistency between $F_{12}$ and $F_{1}$}
The difference compared to the former case is that the single-particle density operator is now not an independent quantity but depends on the two-particle density operator via the trace consistency condition 
\begin{align}
F_1 &= \frac{1}{N-1}\mbox{Tr}_2 F_{12}\,.
\end{align}
Therefore, the total energy \refeqn{app:etot} is now expressed via $F_{12}$ alone
\begin{align}
    \langle \hat H \rangle^c &=  \frac{1}{2}\mbox{Tr}_{12} \hat H^{c}_{12} F_{12}\,,\\
    \hat H^{c}_{12} & = \frac{\hat H_1+\hat H_2}{N-1} +  \hat V_{12}
    \label{app:etot-cc}\,.
\end{align}
Now, for the time derivative follows
\begin{align}
    \i\hbar \frac{\d}{\d t} \langle\hat H\rangle^c &= \mbox{Tr}_{12} \frac{2}{N-1} \hat{H}_1 \left[\hat{H}_1,F_{12}\right] + \mbox{Tr}_{12} \hat{V}_{12} \left[ \hat{H}_1 , F_{12}\right]
    \nonumber\\
    & + \mbox{Tr}_{12} \frac{\hat{H}_1}{N-1} \left[\hat{V}_{12},F_{12}\right] + \mbox{Tr}_{12} \hat{V}_{12} \left[\hat{V}_{12},F_{12}\right]
     \nonumber\\
    & + \mbox{Tr}_{123} \frac{\hat{H}_1}{N-1} \left[\hat{V}_{12}+\hat{V}_{23},F_{123}\right]
    \nonumber\\
    &+ 
    \frac{1}{2}\mbox{Tr}_{123} \hat V_{12} \left[\hat V_{13}+\hat V_{23}, F_{123}\right]\,,
    \label{app:ddte12-cc}
\end{align}
where in the last line we can identify the contribution of \refeqn{app:ddte12} again. Assuming the aforementioned symmetry of $F_{123}$ and using the cyclic property of the trace we simplify \refeqn{app:ddte12-cc} to 
\begin{align}
    \i\hbar \frac{\d}{\d t} \langle\hat H\rangle^c &= \mbox{Tr}_{12} \frac{N-2}{N-1} \left[\hat{V}_1,\hat{H}_1\right] F_{12} \nonumber\\
    & + \mbox{Tr}_{123} \frac{\hat{H}_1}{N-1} \left[\hat{V}_{12}+\hat{V}_{23},F_{123}\right]\,.
    \label{app:ddte12-cc_reduced}
\end{align}
By using \refeqn{eq:f123-symmetry} for the second line of \refeqn{app:ddte12-cc_reduced}, we end up with the following expression,
\begin{align}
    \i\hbar \frac{\d}{\d t} \langle\hat H\rangle^c =& \frac{1}{N-1} \mbox{Tr}_{12} \Bigg\{ \left[\hat{V}_{12},\hat{H}_1\right] \times \nonumber \\
    &\times \Big[(N-2) F_{12} - \mbox{Tr}_3 F_{123}\Big]\Bigg\}\,.
\end{align}
This implies that total-energy conservation in the case of contraction consistency between $F_{12}$ and $F_1$ requires fulfilling the additional condition of contraction consistency between $F_{123}$ and $F_{12}$, i.e. [cf. \refeqn{eq:cc-f12-f123}], 
\begin{align}
    \mbox{Tr}_3 F_{123} = \left(N-2\right)F_{12}\, .
\end{align}
This result is in line with previous findings, cf. Refs.~\cite{lackner_propagating_2015,lacknerphd}. 

\section{Enforcing Contraction Consistency between $G^{(3)}$ and $G^{(2)}$}\label{app:cc}

In the following we outline the derivation of contraction consistency and unitary decomposition of three-particle matrices and reproduce the procedure presented in detail in Refs.~\cite{lackner_propagating_2015,lacknerphd}.
We point out, that slight deviations of Tabs.~\ref{tab:parameters}, \ref{tab:parameter_b}, \ref{tab:parameter_c} and \refeqn{eq:parameters} to Ref.~\cite{lacknerphd} are due to typos in Ref.~\cite{lacknerphd}.\\
The zero-, one- and two-particle quantities $M^{(0)}$, $M^{(1)}$ and $M^{(2)}$ are crucial for the construction of the three-particle Green function correction, \refeqn{eq:CC_reconstruction}. They are defined as (partial) traces over the three-particle quantity $M^{(3)}$ which is defined as the difference between the correct $G^{(3)}$ and the used approximation
\begin{align*}
    M^{(3)} = G^{(3)} - G^{(3),\tn{DSL}} (= \mathcal{G}^{(3)})\,.
\end{align*}
The notation $\tn{Tr}{\left(\substack{1,4\\2,5\\3,6}\right)}$ used in the following relations denotes that contractions are performed over indices in the same line. For the sake of readability the indices of the $M$ quantities are neglected.
\begin{align*}
    {}^1M^{(2)} &= \tn{Tr}_{(3,6)} M^{(3)}\qquad {}^2M^{(2)} = \tn{Tr}_{(2,6)} M^{(3)}\\
    {}^3M^{(2)} &= \tn{Tr}_{(1,6)} M^{(3)}\qquad {}^4M^{(2)} = \tn{Tr}_{(3,5)} M^{(3)}\\
    {}^5M^{(2)} &= \tn{Tr}_{(2,5)} M^{(3)}\qquad {}^6M^{(2)} = \tn{Tr}_{(1,5)} M^{(3)}\\
    {}^7M^{(2)} &= \tn{Tr}_{(3,4)} M^{(3)}\qquad {}^8M^{(2)} = \tn{Tr}_{(2,4)} M^{(3)}\\
    {}^9M^{(2)} &= \tn{Tr}_{(1,4)} M^{(3)}
\end{align*}
\begin{align*}
    {}^1M^{(1)} &= \tn{Tr}{\left(\substack{2,5\\3,6}\right)} M^{(3)}\qquad {}^2M^{(1)} = \tn{Tr}{\left(\substack{1,5\\3,6}\right)} M^{(3)}\\
    {}^4M^{(1)} &= \tn{Tr}{\left(\substack{1,5\\2,6}\right)} M^{(3)}  \qquad {}^4M^{(1)} = \tn{Tr}{\left(\substack{2,6\\3,5}\right)} M^{(3)}\\
    {}^5M^{(1)} &= \tn{Tr}{\left(\substack{1,6\\3,5}\right)} M^{(3)}\qquad {}^6M^{(1)} = \tn{Tr}{\left(\substack{1,6\\2,5}\right)} M^{(3)}\\
    {}^7M^{(1)} &= \tn{Tr}{\left(\substack{2,4\\3,6}\right)} M^{(3)}  \qquad {}^8M^{(1)} = \tn{Tr}{\left(\substack{1,4\\3,6}\right)} M^{(3)}\\
    {}^9M^{(1)} &= \tn{Tr}{\left(\substack{1,4\\2,6}\right)} M^{(3)}\qquad {}^{10}M^{(1)} = \tn{Tr}{\left(\substack{2,6\\3,4}\right)} M^{(3)}\\
    {}^{11}M^{(1)} &= \tn{Tr}{\left(\substack{1,6\\3,4}\right)} M^{(3)}  \qquad {}^{12}M^{(1)} = \tn{Tr}{\left(\substack{1,6\\2,4}\right)} M^{(3)}\\
    {}^{13}M^{(1)} &= \tn{Tr}{\left(\substack{2,4\\3,5}\right)} M^{(3)}\qquad {}^{14}M^{(1)} = \tn{Tr}{\left(\substack{1,4\\3,5}\right)} M^{(3)}\\
    {}^{15}M^{(1)} &= \tn{Tr}{\left(\substack{1,4\\2,5}\right)} M^{(3)}  \qquad {}^{16}M^{(1)} = \tn{Tr}{\left(\substack{2,5\\3,4}\right)} M^{(3)}\\
    {}^{17}M^{(1)} &= \tn{Tr}{\left(\substack{1,5\\3,4}\right)} M^{(3)}\qquad {}^{18}M^{(1)} = \tn{Tr}{\left(\substack{1,5\\2,4}\right)} M^{(3)}
\end{align*}
\begin{align*}
    {}^1M^{(0)} &= \tn{Tr}{\left(\substack{1,4\\2,5\\3,6}\right)} M^{(3)}\qquad
    {}^2M^{(0)} = \tn{Tr}{\left(\substack{1,4\\2,6\\3,5}\right)} M^{(3)}\\
    {}^3M^{(0)} &= \tn{Tr}{\left(\substack{1,5\\2,4\\3,6}\right)} M^{(3)}\qquad
    {}^4M^{(0)} = \tn{Tr}{\left(\substack{1,5\\2,6\\3,4}\right)} M^{(3)}\\
    {}^5M^{(0)} &= \tn{Tr}{\left(\substack{1,6\\2,4\\3,5}\right)} M^{(3)}\qquad
    {}^6M^{(0)} = \tn{Tr}{\left(\substack{1,6\\2,5\\3,4}\right)} M^{(3)}
\end{align*}
In practice the traces over $M^{(3)}$ are taken separately for its constituents $G^{(3)}$ and $G^{(3),\tn{DSL}}$. For the former this can be done analytically using the relations \refeqn{eq:3to2_traces}, for the latter the traces are performed numerically over 
\begin{align}
    G^{(3),\uparrow\uparrow\downarrow\uparrow\uparrow\downarrow}_{ijklpq} &= G^{<,\uparrow}_{il}G^{<,\uparrow}_{jp}G^{<,\downarrow}_{kq} - G^{<,\uparrow}_{ip}G^{<,\uparrow}_{jl}G^{<,\downarrow}_{kq}
    \\&+ G^{<,\uparrow}_{il} \mathcal{G}^{\uparrow\downarrow\uparrow\downarrow}_{jkpq} - G^{<,\uparrow}_{ip} \mathcal{G}^{\uparrow\downarrow\uparrow\downarrow}_{jklq} \\&+ G^{<,\uparrow}_{jp} \mathcal{G}^{\uparrow\downarrow\uparrow\downarrow}_{iklq} - G^{<,\uparrow}_{jl} \mathcal{G}^{\uparrow\downarrow\uparrow\downarrow}_{ikpq}
    \\& + G^{<,\downarrow}_{kq} \mathcal{G}^{\uparrow\uparrow\uparrow\uparrow}_{ijlp}\,.
\end{align}

The weight parameters $a_\tau^k$, $b_{\tau,\sigma}^k$ and $c_{\tau,\sigma}^k$ of \refeqn{eq:CC_reconstruction} can be expressed by a reduced amount of auxiliary parameters $\alpha$, $\beta$ and $\gamma$ as listed in Tables~\ref{tab:parameter_a}, \ref{tab:parameter_b} and \ref{tab:parameter_c}, respectively. For the index permutation the following shortened notation is used:
\begin{align*}
    \mathbf{1}&=(1,2,3) \qquad \mathbf{2}=(1,3,2) \qquad \mathbf{3}=(2,1,3)\\
    \mathbf{4}&=(2,3,1) \qquad \mathbf{5}=(3,1,2) \qquad \mathbf{6}=(3,2,1)
\end{align*}
For each $\alpha$, $\beta$ and $\gamma$ the respective parameter value can be calculated utilizing Table~\ref{tab:parameters} and \refeqn{eq:parameters}. Note that \refeqn{eq:parameters} differs from the one given in Ref.~\cite{lacknerphd} by a factor of $1/36$.

\onecolumngrid

\begin{align}\label{eq:parameters}
    X = \frac{1}{36}\left(\frac{A_1}{N_{\tn{b}}-4} + \frac{A_2}{N_{\tn{b}}+4} + \frac{B_1}{N_{\tn{b}}-3} + \frac{B_2}{N_{\tn{b}}+3} + \frac{C_1}{N_{\tn{b}}-2} + \frac{C_2}{N_{\tn{b}}+2} + \frac{D_1}{N_{\tn{b}}-1} + \frac{D_2}{N_{\tn{b}}+1} + \frac{E_1}{N_{\tn{b}}} + \frac{E_2}{N^2_{\tn{b}}}\right)
\end{align}

 \begin{table}[h]
  \begin{center}
  \setlength{\tabcolsep}{4pt}
    \begin{tabular}{l|cccccccccccccccccccc} 
      & {$\alpha_1$} & {$\alpha_2$} & {$\alpha_3$} & {$\beta_1$} & {$\beta_2$} & {$\beta_3$} & {$\beta_4$} & {$\beta_5$} & {$\beta_6$} & {$\beta_7$} & {$\gamma_1$} & {$\gamma_2$} & {$\gamma_3$} & {$\gamma_4$} & {$\gamma_5$} & {$\gamma_6$} & {$\gamma_7$} & {$\gamma_8$} & {$\gamma_9$} & {$\gamma_{10}$}\\
      \hline
      {$A_1$} &  1 &   1 &  1 & -1 & -1 & -1 & -1 &  1 &  1 &  1 & -3 &             3 &            3 &             -2 &             -2 &             -2 &             -2 &             2 &  2 &  2\\
      {$A_2$} &  1 &   1 &  1 &  1 &  1 &  1 &  1 &  1 &  1 &  1 &  3 &             3 &            3 &              2 &              2 &              2 &              2 &             2 &  2 &  2\\
      {$B_1$} &  2 &  -4 &  4 &  0 &  3 & -3 &  0 & -1 &  1 & -2 &  6 &            -6 &           -6 &             -2 &              4 &              1 &              4 &            -4 & -1 &  2\\
      {$B_2$} &  2 &  -4 &  4 &  0 & -3 &  3 &  0 & -1 &  1 & -2 & -6 &            -6 &           -6 &              2 &             -4 &             -1 &             -4 &            -4 & -1 &  2\\
      {$C_1$} &  1 &   4 &  4 &  1 & -2 & -2 &  4 & -2 & -2 &  1 & -3 & $\frac{3}{2}$ &            6 & -$\frac{5}{2}$ &  $\frac{1}{2}$ &  $\frac{5}{4}$ & -$\frac{7}{4}$ & $\frac{1}{2}$ & -1 &  2\\
      {$C_2$} &  1 &   4 &  4 & -1 &  2 &  2 & -4 & -2 & -2 &  1 &  3 & $\frac{3}{2}$ &            6 &  $\frac{5}{2}$ & -$\frac{1}{2}$ & -$\frac{5}{4}$ &  $\frac{7}{4}$ & $\frac{1}{2}$ & -1 &  2\\
      {$D_1$} & -2 &   4 &  4 &  2 & -1 & -1 & -4 &  1 &  1 & -2 &  0 &             0 &            0 &              4 &             -2 &              1 &             -2 &             2 & -1 & -4\\
      {$D_2$} & -2 &   4 &  4 & -2 &  1 &  1 &  4 &  1 &  1 & -2 &  0 &             0 &            0 &             -4 &              2 &             -1 &              2 &             2 & -1 & -4\\
      {$E_1$} & -4 & -10 & 10 &  0 &  0 &  0 &  0 &  2 & -2 &  4 &  0 &             3 &           -6 &              0 &              0 &              0 &              0 &            -1 &  2 & -4\\
      {$E_2$} &  0 &   0 &  0 &  0 &  0 &  0 &  0 &  0 &  0 &  0 &  0 &             0 &            0 &             -6 &              6 &              3 &              3 &             0 &  0 &  0\\
    \end{tabular}
  \end{center}
  \caption{List of parameters A--E to calculate the values of $\alpha$, $\beta$ and $\gamma$ for Tables~\ref{tab:parameter_a}--\ref{tab:parameter_c} using \refeqn{eq:parameters}. Note that the table differs from the one given in Ref.~\cite{lacknerphd}.
}
  \label{tab:parameters}
\end{table}

\twocolumngrid

 \begin{table}[t]
  \begin{center}
    \begin{tabular}{l|cccccc} 
    $a^k_\tau$ & {$1$} & {$2$} & {$3$} & {$4$} & {$5$} & {$6$} \\
      \hline
        $\mathbf{1}$ &$\alpha_3$& $\alpha_1$& $\alpha_1$& $\alpha_2$& $\alpha_2$& $\alpha_1$\\
        $\mathbf{2}$ &$\alpha_1$& $\alpha_3$& $\alpha_2$& $\alpha_1$& $\alpha_1$& $\alpha_2$\\
        $\mathbf{3}$ &$\alpha_1$& $\alpha_2$& $\alpha_3$& $\alpha_1$& $\alpha_1$& $\alpha_2$\\
        $\mathbf{4}$ &$\alpha_2$& $\alpha_1$& $\alpha_1$& $\alpha_3$& $\alpha_2$& $\alpha_1$\\
        $\mathbf{5}$ &$\alpha_2$& $\alpha_1$& $\alpha_1$& $\alpha_2$& $\alpha_3$& $\alpha_1$\\
        $\mathbf{6}$ &$\alpha_1$& $\alpha_2$& $\alpha_2$& $\alpha_1$& $\alpha_1$& $\alpha_3$
    \end{tabular}
  \end{center}
  \caption{Parameters $a^k_\tau$ which enter \refeqn{eq:CC_reconstruction}. The values of $\alpha$ can be determined using \refeqn{eq:parameters} and Table~\ref{tab:parameters}.}
  \label{tab:parameter_a}
\end{table}
 \begin{table}[t]
  \begin{center}
  \setlength{\tabcolsep}{1pt}
    \begin{tabular}{l|cccccccccccccccccc} 
    $b^k_{\tau,\sigma}$  & {$1$} & {$2$} & {$3$} & {$4$} & {$5$} & {$6$} & {$7$} & {$8$} & {$9$} & {$10$} & {$11$} & {$12$} & {$13$} & {$14$} & {$15$} & {$16$} & {$17$} & {$18$}\\
      \hline
        $\mathbf{1,1}$   & $\beta_4$& $\beta_5$& $\beta_5$& $\beta_3$& $\beta_2$& $\beta_6$& $\beta_5$& $\beta_4$& $\beta_2$& $\beta_6$& $\beta_5$& $\beta_3$& $\beta_3$& $\beta_6$& $\beta_6$& $\beta_1$& $\beta_3$& $\beta_7$\\
        $\mathbf{1,2}$   & $\beta_5$& $\beta_3$& $\beta_4$& $\beta_5$& $\beta_6$& $\beta_2$& $\beta_2$& $\beta_6$& $\beta_5$& $\beta_4$& $\beta_3$& $\beta_5$& $\beta_6$& $\beta_1$& $\beta_3$& $\beta_6$& $\beta_7$& $\beta_3$\\
        $\mathbf{1,3}$   & $\beta_5$& $\beta_4$& $\beta_2$& $\beta_6$& $\beta_5$& $\beta_3$& $\beta_4$& $\beta_5$& $\beta_5$& $\beta_3$& $\beta_2$& $\beta_6$& $\beta_6$& $\beta_3$& $\beta_3$& $\beta_7$& $\beta_6$& $\beta_1$\\
        $\mathbf{1,4}$   & $\beta_3$& $\beta_5$& $\beta_6$& $\beta_2$& $\beta_4$& $\beta_5$& $\beta_6$& $\beta_2$& $\beta_3$& $\beta_5$& $\beta_5$& $\beta_4$& $\beta_1$& $\beta_6$& $\beta_7$& $\beta_3$& $\beta_3$& $\beta_6$\\
        $\mathbf{1,5}$   & $\beta_2$& $\beta_6$& $\beta_5$& $\beta_4$& $\beta_3$& $\beta_5$& $\beta_5$& $\beta_3$& $\beta_4$& $\beta_5$& $\beta_6$& $\beta_2$& $\beta_3$& $\beta_7$& $\beta_6$& $\beta_3$& $\beta_1$& $\beta_6$\\
        $\mathbf{1,6}$   & $\beta_6$& $\beta_2$& $\beta_3$& $\beta_5$& $\beta_5$& $\beta_4$& $\beta_3$& $\beta_5$& $\beta_6$& $\beta_2$& $\beta_4$& $\beta_5$& $\beta_7$& $\beta_3$& $\beta_1$& $\beta_6$& $\beta_6$& $\beta_3$\\
        $\mathbf{2,1}$   & $\beta_5$& $\beta_2$& $\beta_4$& $\beta_6$& $\beta_5$& $\beta_3$& $\beta_3$& $\beta_6$& $\beta_6$& $\beta_1$& $\beta_3$& $\beta_7$& $\beta_5$& $\beta_4$& $\beta_2$& $\beta_6$& $\beta_5$& $\beta_3$\\
        $\mathbf{2,2}$   & $\beta_4$& $\beta_6$& $\beta_5$& $\beta_2$& $\beta_3$& $\beta_5$& $\beta_6$& $\beta_1$& $\beta_3$& $\beta_6$& $\beta_7$& $\beta_3$& $\beta_2$& $\beta_6$& $\beta_5$& $\beta_4$& $\beta_3$& $\beta_5$\\
        $\mathbf{2,3}$   & $\beta_2$& $\beta_5$& $\beta_5$& $\beta_3$& $\beta_4$& $\beta_6$& $\beta_6$& $\beta_3$& $\beta_3$& $\beta_7$& $\beta_6$& $\beta_1$& $\beta_4$& $\beta_5$& $\beta_5$& $\beta_3$& $\beta_2$& $\beta_6$\\
        $\mathbf{2,4}$   & $\beta_6$& $\beta_4$& $\beta_3$& $\beta_5$& $\beta_5$& $\beta_2$& $\beta_1$& $\beta_6$& $\beta_7$& $\beta_3$& $\beta_3$& $\beta_6$& $\beta_6$& $\beta_2$& $\beta_3$& $\beta_5$& $\beta_5$& $\beta_4$\\
        $\mathbf{2,5}$   & $\beta_5$& $\beta_3$& $\beta_2$& $\beta_5$& $\beta_6$& $\beta_4$& $\beta_3$& $\beta_7$& $\beta_6$& $\beta_3$& $\beta_1$& $\beta_6$& $\beta_5$& $\beta_3$& $\beta_4$& $\beta_5$& $\beta_6$& $\beta_2$\\
        $\mathbf{2,6}$   & $\beta_3$& $\beta_5$& $\beta_6$& $\beta_4$& $\beta_2$& $\beta_5$& $\beta_7$& $\beta_3$& $\beta_1$& $\beta_6$& $\beta_6$& $\beta_3$& $\beta_3$& $\beta_5$& $\beta_6$& $\beta_2$& $\beta_4$& $\beta_5$\\
        $\mathbf{4,1}$   & $\beta_3$& $\beta_6$& $\beta_6$& $\beta_1$& $\beta_3$& $\beta_7$& $\beta_5$& $\beta_2$& $\beta_4$& $\beta_6$& $\beta_5$& $\beta_3$& $\beta_2$& $\beta_5$& $\beta_5$& $\beta_3$& $\beta_4$& $\beta_6$\\
        $\mathbf{4,2}$   & $\beta_6$& $\beta_1$& $\beta_3$& $\beta_6$& $\beta_7$& $\beta_3$& $\beta_4$& $\beta_6$& $\beta_5$& $\beta_2$& $\beta_3$& $\beta_5$& $\beta_5$& $\beta_3$& $\beta_2$& $\beta_5$& $\beta_6$& $\beta_4$\\
        $\mathbf{4,3}$   & $\beta_6$& $\beta_3$& $\beta_3$& $\beta_7$& $\beta_6$& $\beta_1$& $\beta_2$& $\beta_5$& $\beta_5$& $\beta_3$& $\beta_4$& $\beta_6$& $\beta_5$& $\beta_2$& $\beta_4$& $\beta_6$& $\beta_5$& $\beta_3$\\
        $\mathbf{4,4}$   & $\beta_1$& $\beta_6$& $\beta_7$& $\beta_3$& $\beta_3$& $\beta_6$& $\beta_6$& $\beta_4$& $\beta_3$& $\beta_5$& $\beta_5$& $\beta_2$& $\beta_3$& $\beta_5$& $\beta_6$& $\beta_4$& $\beta_2$& $\beta_5$\\
        $\mathbf{4,5}$   & $\beta_3$& $\beta_7$& $\beta_6$& $\beta_3$& $\beta_1$& $\beta_6$& $\beta_5$& $\beta_3$& $\beta_2$& $\beta_5$& $\beta_6$& $\beta_4$& $\beta_4$& $\beta_6$& $\beta_5$& $\beta_2$& $\beta_3$& $\beta_5$\\
        $\mathbf{4,6}$   & $\beta_7$& $\beta_3$& $\beta_1$& $\beta_6$& $\beta_6$& $\beta_3$& $\beta_3$& $\beta_5$& $\beta_6$& $\beta_4$& $\beta_2$& $\beta_5$& $\beta_6$& $\beta_4$& $\beta_3$& $\beta_5$& $\beta_5$& $\beta_2$\\
    \end{tabular}
  \end{center}
  \caption{Parameters $b^k_\tau$ which enter \refeqn{eq:CC_reconstruction}. The values of $\beta$ can be determined using \refeqn{eq:parameters} and Table~\ref{tab:parameters}. Note that the table differs from the one given in Ref.~\cite{lacknerphd}.}
  \label{tab:parameter_b}
\end{table}
 \begin{table}[t]
  \begin{center}
  \setlength{\tabcolsep}{1pt}
    \begin{tabular}{l|ccccccccc} 
    $c^k_{\tau,\sigma}$ & {$1$} & {$2$} & {$3$} & {$4$} & {$5$} & {$6$} & {$7$} & {$8$} & {$9$}\\
      \hline
                     $\mathbf{1,1}$ &$\gamma_7$& $\gamma_1$& $\gamma_6$& $\gamma_1$& $\gamma_8$& $\gamma_3$& $\gamma_6$& $\gamma_3$& $\gamma_{10}$\\
                     $\mathbf{1,2}$ &$\gamma_1$& $\gamma_7$& $\gamma_2$& $\gamma_8$& $\gamma_1$& $\gamma_5$& $\gamma_3$& $\gamma_6$& $\gamma_4$\\
                     $\mathbf{1,3}$ &$\gamma_1$& $\gamma_6$& $\gamma_1$& $\gamma_8$& $\gamma_3$& $\gamma_8$& $\gamma_3$& $\gamma_{10}$& $\gamma_3$\\
                     $\mathbf{1,4}$ &$\gamma_7$& $\gamma_2$& $\gamma_7$& $\gamma_1$& $\gamma_5$& $\gamma_1$& $\gamma_6$& $\gamma_4$& $\gamma_6$\\
                     $\mathbf{1,5}$ &$\gamma_6$& $\gamma_1$& $\gamma_7$& $\gamma_3$& $\gamma_8$& $\gamma_1$& $\gamma_{10}$& $\gamma_3$& $\gamma_6$\\
                     $\mathbf{1,6}$ &$\gamma_2$& $\gamma_7$& $\gamma_1$& $\gamma_5$& $\gamma_1$& $\gamma_8$& $\gamma_4$& $\gamma_6$& $\gamma_3$\\
                     $\mathbf{2,1}$ &$\gamma_1$& $\gamma_8$& $\gamma_3$& $\gamma_7$& $\gamma_1$& $\gamma_6$& $\gamma_2$& $\gamma_5$& $\gamma_4$\\
                     $\mathbf{2,2}$ &$\gamma_8$& $\gamma_1$& $\gamma_5$& $\gamma_1$& $\gamma_7$& $\gamma_2$& $\gamma_5$& $\gamma_2$& $\gamma_9$\\
                     $\mathbf{2,3}$ &$\gamma_8$& $\gamma_3$& $\gamma_8$& $\gamma_1$& $\gamma_6$& $\gamma_1$& $\gamma_5$& $\gamma_4$& $\gamma_5$\\
                     $\mathbf{2,4}$ &$\gamma_1$& $\gamma_5$& $\gamma_1$& $\gamma_7$& $\gamma_2$& $\gamma_7$& $\gamma_2$& $\gamma_9$& $\gamma_2$\\
                     $\mathbf{2,5}$ &$\gamma_3$& $\gamma_8$& $\gamma_1$& $\gamma_6$& $\gamma_1$& $\gamma_7$& $\gamma_4$& $\gamma_5$& $\gamma_2$\\
                     $\mathbf{2,6}$ &$\gamma_5$& $\gamma_1$& $\gamma_8$& $\gamma_2$& $\gamma_7$& $\gamma_1$& $\gamma_9$& $\gamma_2$& $\gamma_5$\\
                     $\mathbf{3,1}$ &$\gamma_1$& $\gamma_8$& $\gamma_3$& $\gamma_6$& $\gamma_3$& $\gamma_{10}$& $\gamma_1$& $\gamma_8$& $\gamma_3$\\
                     $\mathbf{3,2}$ &$\gamma_8$& $\gamma_1$& $\gamma_5$& $\gamma_3$& $\gamma_6$& $\gamma_4$& $\gamma_8$& $\gamma_1$& $\gamma_5$\\
                     $\mathbf{3,3}$ &$\gamma_8$& $\gamma_3$& $\gamma_8$& $\gamma_3$& $\gamma_{10}$& $\gamma_3$& $\gamma_8$& $\gamma_3$& $\gamma_8$\\
                     $\mathbf{3,4}$ &$\gamma_1$& $\gamma_5$& $\gamma_1$& $\gamma_6$& $\gamma_4$& $\gamma_6$& $\gamma_1$& $\gamma_5$& $\gamma_1$\\
                     $\mathbf{3,5}$ &$\gamma_3$& $\gamma_8$& $\gamma_1$& $\gamma_{10}$& $\gamma_3$& $\gamma_6$& $\gamma_3$& $\gamma_8$& $\gamma_1$\\
                     $\mathbf{3,6}$ &$\gamma_5$& $\gamma_1$& $\gamma_8$& $\gamma_4$& $\gamma_6$& $\gamma_3$& $\gamma_5$& $\gamma_1$& $\gamma_8$\\
                     $\mathbf{4,1}$ &$\gamma_7$& $\gamma_1$& $\gamma_6$& $\gamma_2$& $\gamma_5$& $\gamma_4$& $\gamma_7$& $\gamma_1$& $\gamma_6$\\
                     $\mathbf{4,2}$ &$\gamma_1$& $\gamma_7$& $\gamma_2$& $\gamma_5$& $\gamma_2$& $\gamma_9$& $\gamma_1$& $\gamma_7$& $\gamma_2$\\
                     $\mathbf{4,3}$ &$\gamma_1$& $\gamma_6$& $\gamma_1$& $\gamma_5$& $\gamma_4$& $\gamma_5$& $\gamma_1$& $\gamma_6$& $\gamma_1$\\
                     $\mathbf{4,4}$ &$\gamma_7$& $\gamma_2$& $\gamma_7$& $\gamma_2$& $\gamma_9$& $\gamma_2$& $\gamma_7$& $\gamma_2$& $\gamma_7$\\
                     $\mathbf{4,5}$ &$\gamma_6$& $\gamma_1$& $\gamma_7$& $\gamma_4$& $\gamma_5$& $\gamma_2$& $\gamma_6$& $\gamma_1$& $\gamma_7$\\
                     $\mathbf{4,6}$ &$\gamma_2$& $\gamma_7$& $\gamma_1$& $\gamma_9$& $\gamma_2$& $\gamma_5$& $\gamma_2$& $\gamma_7$& $\gamma_1$\\
                     $\mathbf{5,1}$ &$\gamma_6$& $\gamma_3$& $\gamma_{10}$& $\gamma_1$& $\gamma_8$& $\gamma_3$& $\gamma_7$& $\gamma_1$& $\gamma_6$\\
                     $\mathbf{5,2}$ &$\gamma_3$& $\gamma_6$& $\gamma_4$& $\gamma_8$& $\gamma_1$& $\gamma_5$& $\gamma_1$& $\gamma_7$& $\gamma_2$\\
                     $\mathbf{5,3}$ &$\gamma_3$& $\gamma_{10}$& $\gamma_3$& $\gamma_8$& $\gamma_3$& $\gamma_8$& $\gamma_1$& $\gamma_6$& $\gamma_1$\\
                     $\mathbf{5,4}$ &$\gamma_6$& $\gamma_4$& $\gamma_6$& $\gamma_1$& $\gamma_5$& $\gamma_1$& $\gamma_7$& $\gamma_2$& $\gamma_7$\\
                     $\mathbf{5,5}$ &$\gamma_{10}$& $\gamma_3$& $\gamma_6$& $\gamma_3$& $\gamma_8$& $\gamma_1$& $\gamma_6$& $\gamma_1$& $\gamma_7$\\
                     $\mathbf{5,6}$ &$\gamma_4$& $\gamma_6$& $\gamma_3$& $\gamma_5$& $\gamma_1$& $\gamma_8$& $\gamma_2$& $\gamma_7$& $\gamma_1$\\
                     $\mathbf{6,1}$ &$\gamma_2$& $\gamma_5$& $\gamma_4$& $\gamma_7$& $\gamma_1$& $\gamma_6$& $\gamma_1$& $\gamma_8$& $\gamma_3$\\
                     $\mathbf{6,2}$ &$\gamma_5$& $\gamma_2$& $\gamma_9$& $\gamma_1$& $\gamma_7$& $\gamma_2$& $\gamma_8$& $\gamma_1$& $\gamma_5$\\
                     $\mathbf{6,3}$ &$\gamma_5$& $\gamma_4$& $\gamma_5$& $\gamma_1$& $\gamma_6$& $\gamma_1$& $\gamma_8$& $\gamma_3$& $\gamma_8$\\
                     $\mathbf{6,4}$ &$\gamma_2$& $\gamma_9$& $\gamma_2$& $\gamma_7$& $\gamma_2$& $\gamma_7$& $\gamma_1$& $\gamma_5$& $\gamma_1$\\
                     $\mathbf{6,5}$ &$\gamma_4$& $\gamma_5$& $\gamma_2$& $\gamma_6$& $\gamma_1$& $\gamma_7$& $\gamma_3$& $\gamma_8$& $\gamma_1$\\
                     $\mathbf{6,6}$ &$\gamma_9$& $\gamma_2$& $\gamma_5$& $\gamma_2$& $\gamma_7$& $\gamma_1$& $\gamma_5$& $\gamma_1$& $\gamma_8$
    \end{tabular}
  \end{center}
  \caption{Parameters $c^k_\tau$ which enter \refeqn{eq:CC_reconstruction}. The values of $\gamma$ can be determined using \refeqn{eq:parameters} and Table~\ref{tab:parameters}. Note that the table differs from the one given in Ref.~\cite{lacknerphd}.}
  \label{tab:parameter_c}
\end{table}

\section{Purification}\label{app:purification}
The following purification scheme is based on the one presented in Ref.~\cite{lackner_propagating_2015}. Consequently, of the three positivity conditions on the two-particle level, cf. \refeqn{eq:2p_positivity}, only the two-particle and two-hole condition will be considered. In a first step the full two-particle and two-hole Green functions are calculated from the two-particle correlation part and the single-particle Green function,
 \begin{align*}
    G^{(2),\uparrow\downarrow\uparrow\downarrow}_{ijkl} &= G^{<,\uparrow\uparrow}_{ik}G^{<,\downarrow\downarrow}_{jl} + \mathcal{G}^{\uparrow\downarrow\uparrow\downarrow}_{ijkl}\\
    Q^{(2),\uparrow\downarrow\uparrow\downarrow}_{ijkl} &= G^{(2),\uparrow\downarrow\uparrow\downarrow}_{ijkl} + \frac{1}{\left(\tn{i}\hbar\right)^2} \delta_{ik} \delta_{jl} \\&\quad+ \frac{1}{\tn{i}\hbar} \delta_{ik} G^{<,\downarrow\downarrow}_{jl} + \frac{1}{\tn{i}\hbar} \delta_{jl} G^{<,\uparrow\uparrow}_{ik} \,.
 \end{align*}
The following steps are outlined for the two-particle Green function only but have to be performed for the two-hole Green function in the same way. To purify its eigenvalues the two-particle Green function can be mapped to a $N^2_{\tn{b}}\times N^2_{\tn{b}}$ matrix, $G^{(2)}_{ij,kl} \rightarrow G^{(2)}_{x,y}$, in order to perform the eigendecomposition
 \begin{align*}
    \bm{G^{(2)}} = \bm{V} \bm{\lambda} \bm{V^\dagger}\,,
 \end{align*}
where the bold quantities $\bm{\lambda}$ and $\bm{V}$ are matrices containing the eigenvalues (on the diagonal) and the eigenvectors of $\bm{G^{(2)}}$, respectively. Next, $\bm{G^{(2)}_{\tn{pos}}}$, the unphysical (positive) part of the two-particle Green function containing only the positive eigenvalues, is constructed via
 \begin{align*}
    \bm{G^{(2)}_{\tn{pos}}} = \bm{V} \bm{\lambda_{\tn{pos}}} \bm{V^\dagger}\,.
 \end{align*}
In principle, subtracting this quantity from the full two-particle Green function would be sufficient to purify its eigenvalues. However, doing so would violate contraction consistency, as well as conservation of energy. Contraction consistency between the two-particle and the one-particle level can be ensured by calculating the contraction-free component of $\bm{G^{(2)}_{\tn{pos}}}$. Starting with the symmetrized and antisymmetrized auxilary quantities
 \begin{align*}
    A^{(1)}_{ij} &= \frac{1}{2} \sum_p \left(G^{(2),\tn{pos}}_{ipjp} - G^{(2),\tn{pos}}_{ippj}\right)\,, \qquad A^{(0)} = \tn{Tr}A^{(1)}\\
    S^{(1)}_{ij} &= \frac{1}{2} \sum_p \left(G^{(2),\tn{pos}}_{ipjp} + G^{(2),\tn{pos}}_{ippj}\right)\,, \qquad S^{(0)} = \tn{Tr}S^{(1)}
 \end{align*}
 and
 \begin{align*}
    A^{(2)}_{ijkl} &= \delta_{ik} \frac{A^{(1)}_{jl}}{N_{\tn{b}}-2} + \delta_{jl} \frac{A^{(1)}_{ik}}{N_{\tn{b}}-2} - \delta_{jk} \frac{A^{(1)}_{il}}{N_{\tn{b}}-2} - \delta_{il} \frac{A^{(1)}_{jk}}{N_{\tn{b}}-2} \\ &\quad- \delta_{ik} \delta_{jl} \frac{A^{(0)}}{\left(N_{\tn{b}}-1\right)\left(N_{\tn{b}}-2\right)} + \delta_{il} \delta_{jk} \frac{A^{(0)}}{\left(N_{\tn{b}}-1\right)\left(N_{\tn{b}}-2\right)}\\
    S^{(2)}_{ijkl} &= \delta_{ik} \frac{S^{(1)}_{jl}}{N_{\tn{b}}+2} + \delta_{jl} \frac{S^{(1)}_{ik}}{N_{\tn{b}}+2} + \delta_{jk} \frac{S^{(1)}_{il}}{N_{\tn{b}}+2} + \delta_{il} \frac{S^{(1)}_{jk}}{N_{\tn{b}}+2} \\ &\quad - \delta_{ik} \delta_{jl} \frac{S^{(0)}}{\left(N_{\tn{b}}+1\right)\left(N_{\tn{b}}+2\right)} - \delta_{il} \delta_{jk} \frac{S^{(0)}}{\left(N_{\tn{b}}+1\right)\left(N_{\tn{b}}+2\right)}
 \end{align*}
 the contraction-free part of the positive two-particle Green function is given by
 \begin{align*}
    \bm{G^{(2)}_{\tn{pos,CC}}} = \bm{G^{(2)}_{\tn{pos}}} - \bm{A^{(2)}} - \bm{S^{(2)}}\,.
 \end{align*}
 Ensuring contraction consistency from the two-particle to the single-particle level also guarantees conservation of all single-particle observables. Therefore, violations to the conservation of energy can only originate from the correlation energy, which for a general diagonal basis is given by
 \begin{align*}
    E_{\tn{corr}} = \left(\tn{i}\hbar\right)^2 \sum_{pq} V_{pq} \left( 2\,\mathcal{G}^{\uparrow\downarrow\uparrow\downarrow}_{pqpq} - \mathcal{G}^{\uparrow\downarrow\uparrow\downarrow}_{pqqp} \right)\,.
 \end{align*}
 Note, that in the Hubbard basis this reduces to \refeqn{eq:corr_energy}.
 Thus, conservation of total energy can be ensured by setting
 
  \begin{align*}
    G^{(2),\tn{pos}}_{ijkl} = 0 \qquad \tn{for } (i=k \tn{ and } j=l) \tn{ or } (i=l \tn{ and } j=k)
 \end{align*}
 before calculating $\bm{G^{(2)}_{\tn{pos,CC}}}$ so that the parts of $\bm{G^{(2)}}$ that enter the correlation energy ($\sim N^2_{\tn{b}}$ of the $N^4_{\tn{b}}$ entries) are not modified by the purification procedure. After repeating the above procedure for the two-hole Green function the purified two-particle Green function can be constructed as
  \begin{align*}
    \bm{G^{(2)}_{\tn{pur}}} = \bm{G^{(2)}} - \bm{G^{(2)}_{\tn{pos,CC}}} - \bm{Q^{(2)}_{\tn{pos,CC}}}\,.
 \end{align*}
 Due to the modifications done to $\bm{G^{(2)}_{\tn{pos,CC}}}$ and $\bm{Q^{(2)}_{\tn{pos,CC}}}$ the above step in general does not eliminate all positive eigenvalues of the two-particle Green function completely. While the presented purification scheme can be repeated iteratively to further converge the result, in practice it was found that one iteration is enough to ensure a stable propagation in all considered cases. In a final step the correlation part of the purified two-particle Green function is reconstructed by
\begin{align}
 \mathcal{G}^{\uparrow\downarrow\uparrow\downarrow}_{ijkl} &= G^{(2),\uparrow\downarrow\uparrow\downarrow}_{ijkl} - G^{<,\uparrow\uparrow}_{ik}G^{<,\downarrow\downarrow}_{jl}\,.
\end{align}

\input{main-g3x.bbl}

\end{document}

%% file: main-g3x.bbl
%